\documentclass[aps, prd, onecolumn, tightenlines, notitlepage, superscriptaddress, nofootinbib, preprintnumbers, floatfix,showkeys,11pt]{revtex4-2}
\usepackage[normalem]{ulem}
\usepackage{amstext}
\usepackage{amssymb}
\usepackage{amsmath}
\usepackage{graphicx}
\graphicspath{{plots/}}
\usepackage{url}
\usepackage{color}
\usepackage{caption}
\usepackage{subcaption}
\usepackage{placeins} %%for FloatBarrier
\usepackage{ulem}
\usepackage[utf8]{inputenc}
\pdfoutput=1
\usepackage{textcomp}
\usepackage{comment}
\usepackage{yfonts}
\usepackage{epsfig,amsfonts,mathrsfs,amsmath,amssymb,graphicx,color,slashed,multirow}
\usepackage{amsmath,latexsym,amssymb,graphicx,slashed,color,enumerate,url,cancel,gensymb}
\usepackage{textcomp}

\usepackage[x11names]{xcolor}
\usepackage[colorlinks]{hyperref}

\usepackage{booktabs}
\usepackage{adjustbox}

\definecolor{vdrgreen}{rgb}{0.0, 0.7, 0.0}

\def\cevns{CE$\nu$NS }
\def\eves{E$\nu$ES }

\definecolor{indianred}{rgb}{0.8, 0.36, 0.36}
\definecolor{blue(ncs)}{rgb}{0.0, 0.53, 0.74}

\AtBeginDocument{\hypersetup{citecolor=indianred,linkcolor=indianred,urlcolor=indianred}}

\usepackage{float}

\usepackage[T1]{fontenc}
\usepackage{ae,aecompl}
\graphicspath{{Figures/}}
\usepackage{appendix}



\newcommand{\AddrIISERB}{Department of Physics, Indian Institute of Science Education and Research - Bhopal, \\ 
Bhopal Bypass Road, Bhauri, Bhopal 462066, India}

\newcommand{\AddrAHEP}{%
  AHEP Group, Institut de F\'{i}sica Corpuscular --
  CSIC/Universitat de Val\`{e}ncia \\ % 
  C/Catedr\'atico Jos\'e Beltr\'an, 2 E-46980 Paterna, Spain}

\usepackage{orcidlink}
\begin{document}

\title{\LARGE \textcolor{indianred}{Probing conventional and new physics at the ESS with coherent elastic neutrino-nucleus scattering}}
\author{Ayan Chattaraj~\orcidlink{0009-0001-3561-7049}}
\email{ayan23@iiserb.ac.in}
\affiliation{\AddrIISERB}
\author{Anirban Majumdar~\orcidlink{0000-0002-1229-7951}}\email{anirban19@iiserb.ac.in}
\affiliation{\AddrIISERB}
\author{Dimitrios K. Papoulias~\orcidlink{0000-0003-0453-8492}}\email{dipapou@ific.uv.es}
\affiliation{\AddrAHEP}
\author{Rahul Srivastava~\orcidlink{0000-0001-7023-5727}}
\email{rahul@iiserb.ac.in}
\affiliation{\AddrIISERB}

%%%%%%%%%%%
\begin{abstract}
We explore the potential of the European Spallation Source (ESS) in probing physics within and beyond the Standard Model  (SM), based on future measurements of coherent elastic neutrino-nucleus scattering (CE$\nu$NS).  We consider two SM physics cases, namely the weak mixing angle and the nuclear radius. Regarding physics beyond the SM, we focus on neutrino generalized interactions (NGIs) and on various aspects of sterile neutrino and sterile neutral lepton phenomenology. For this,  we explore the violation of lepton unitarity, active-sterile oscillations as well as interesting upscattering channels such as the sterile dipole portal and the production of sterile neutral leptons via NGIs.  The projected ESS sensitivities are estimated by performing  a statistical analysis considering the various CE$\nu$NS detectors and expected backgrounds.  We find that the enhanced statistics achievable in view of the highly intense ESS neutrino beam, will offer a drastic improvement in the current constraints obtained from existing CE$\nu$NS measurements. Finally, we discuss how the ESS has the potential to provide the leading CE$\nu$NS-based  constraints, complementing  also further experimental probes and astrophysical observations. 

\end{abstract}
\maketitle
%%%%%%%%
\section{Introduction}
\label{sec:intro}
Coherent elastic neutrino-nucleus scattering (CE$\nu$NS) was theoretically proposed over five decades ago by Freedman~\cite{Freedman:1973yd} and later revisited by Drukier and Stodolsky~\cite{Drukier:1984vhf}. In this process, a neutrino interacts coherently with the entire nucleus via the exchange of a virtual Standard Model (SM) $Z$ boson, leading to nuclear ground-state-to-ground-state transitions.  The contributions from all nucleons add up coherently, resulting in an enhanced cross section that scales approximately with the square of the neutron number of the target nucleus. For this to occur, the momentum transfer $\mathfrak{q}$ must remain comparable to or smaller than the inverse of the nuclear radius. Notably, \cevns constitutes the dominant neutrino scattering process, significantly surpassing other relevant neutrino processes such as the elastic neutrino-electron scattering (E$\nu$ES). Despite its large  cross section, the experimental observation of \cevns remained elusive for decades due to the extremely small recoil energies imparted to the nucleus in the interaction. 

The first experimental observation of \cevns was achieved by the COHERENT Collaboration at Oak Ridge National Laboratory, which utilized neutrinos from pion-decay-at-rest ($\pi$-DAR) produced at the Spallation Neutron Source (SNS) facility. This milestone was accomplished using a CsI detector~\cite{COHERENT:2017ipa, COHERENT:2021xmm}, it was later confirmed with a liquid argon detector~\cite{COHERENT:2020iec}, and more recently with a germanium detector~\cite{Adamski:2024yqt}. By exploiting reactor antineutrinos, a suggestive evidence for \cevns was  reported by the Dresden-II Collaboration, also using a germanium target~\cite{Colaresi:2022obx}. Very recently, the upgraded version of the CONUS experiment~\cite{CONUS:2021dwh, CONUSCollaboration:2024oks}, referred to as the CONUS+ experiment, has observed reactor antineutrino-induced \cevns signals with a statistical significance of $3.7\sigma$ using a germanium detector~\cite{Ackermann:2025obx}. Moreover, the XENONnT~\cite{XENON:2024ijk} and PandaX-4T~\cite{PandaX:2024muv} Collaborations, using liquid xenon detectors, reported their first \cevns measurements induced by solar $^8$B neutrinos.  Beyond these, several experimental efforts aim to exploit diverse neutrino sources and detection technologies to further advance \cevns studies. The Coherent Captain Mills experiment at Los Alamos National Laboratory plans to utilize a liquid argon target for \cevns detection~\cite{CCM:2021leg}. Reactor-based experiments such as CONNIE~\cite{CONNIE:2016nav}, $\nu$Gen~\cite{nGeN:2022uje}, NUCLEUS~\cite{NUCLEUS:2019igx}, RICOCHET~\cite{Billard:2016giu}, MINER~\cite{MINER:2016igy}, $\nu$IOLETA~\cite{Fernandez-Moroni:2020yyl}, TEXONO~\cite{Wong:2005vg,TEXONO:2024vfk}, NUXE~\cite{Ni:2021mwa}, CHILLAX~\cite{Bernard:2022zyf}, RED-100~\cite{Akimov:2017hee, Akimov:2022xvr, Akimov:2024lnl}, Scintillating Bubble Chamber~\cite{SBC:2021yal} etc. are also under development for \cevns measurements with high precision (for a review see~\cite{Abdullah:2022zue}). 

The current data have already marked a significant milestone in the field of low-energy neutrino physics, prompting several studies to explore both SM and beyond the SM physics. For instance, \cevns measurements allow for precision tests on fundamental parameters of SM electroweak theory such as the weak mixing angle at low energies~\cite{Papoulias:2019txv,Khan:2019cvi,Cadeddu:2019eta,Miranda:2020tif,Cadeddu:2021ijh, DeRomeri:2022twg, Majumdar:2022nby, AtzoriCorona:2023ktl, AtzoriCorona:2024vhj, Maity:2024aji, DeRomeri:2024iaw}. They also constitute powerful probes of nuclear structure parameters like the neutron root mean square (rms) radius~\cite{Cadeddu:2017etk,Cadeddu:2019eta,Canas:2019fjw,Coloma:2020nhf, DeRomeri:2022twg,Sierra:2023pnf},  which remains poorly constrained for most nuclei. Concerning physics beyond the SM,  \cevns data has been analyzed to set stringent constraints on nonstandard neutrino interactions (NSIs)~\cite{Barranco:2005yy,Scholberg:2005qs,Liao:2017uzy,Giunti:2019xpr,Miranda:2020tif,Denton:2020hop,Galindo-Uribarri:2020huw,Khan:2021wzy, DeRomeri:2022twg, AristizabalSierra:2024nwf, Li:2024iij}, neutrino generalized interactions (NGIs)~\cite{Dutta:2015nlo,Lindner:2016wff,AristizabalSierra:2018eqm,Flores:2021kzl, Majumdar:2022nby, DeRomeri:2022twg}, light mediators~\cite{Farzan:2018gtr,Denton:2018xmq,Flores:2020lji,Cadeddu:2020nbr, Bertuzzo:2021opb, AristizabalSierra:2022axl, DeRomeri:2022twg, Majumdar:2024dms,Xia:2024ytb, DeRomeri:2024iaw, Blanco-Mas:2024ale}, heavy scalar and vector mediators~\cite{Billard:2018jnl,Arcadi:2019uif, DeRomeri:2022twg}, sterile neutrino oscillation~\cite{Dutta:2015nlo,Kosmas:2017zbh,Miranda:2020syh, DeRomeri:2022twg}, upscattering production of sterile neutral leptons~\cite{Miranda:2021kre, DeRomeri:2022twg, Candela:2023rvt, Candela:2024ljb}, deviations from lepton unitarity~\cite{Miranda:2020syh}, and neutrino electromagnetic properties~\cite{Cadeddu:2019eta, AtzoriCorona:2022qrf, DeRomeri:2022twg, DeRomeri:2024hvc}. 

The European Spallation Source (ESS), currently under development in Lund, Sweden, will utilize the world’s most powerful superconducting proton linear accelerator in combination with an advanced hydrogen moderator to produce the most intense neutron beams  in the world, paving the way for a diverse experimental program~\cite{Abele:2022iml}. The facility will operate with a 2 GeV proton beam and an unprecedented beam power of 5 MW, resulting to a proton-on-target (POT) number of $2.8\times10^{23}$ over 208 days, (equivalent to a calendar year) of operation. In addition to its primary focus on neutron science, the ESS will generate an intense, pulsed neutrino flux offering unique opportunities for \cevns measurements with high statistics.  Compared to existing $\pi$-DAR facilities, the ESS is expected to achieve an order-of-magnitude enhancement in neutrino flux compared to the SNS. This significant increase, combined with the proposed cutting-edge detector technologies, positions the ESS to enable precision studies of \cevns with unparalleled statistical accuracy. Six advanced detector technologies have been proposed~\cite{Baxter:2019mcx}, including detectors based on CsI, Xe, Ge, Si, Ar, and $\mathrm{C_3F_8}$.  The resulting experimental reach is anticipated to provide deep insights into both SM parameters and physics beyond the SM, offering enhanced sensitivity far beyond current experimental capabilities. Surprisingly, only a limited number of phenomenological studies have investigated the potential of ESS in probing SM and BSM physics parameters. These, mainly include projections for a precise determination of the weak mixing angle~\cite{Baxter:2019mcx} and constraints on neutrino electromagnetic properties~\cite{Baxter:2019mcx, Parada:2024opw}. Moreover, the ESS sensitivity on NSIs using \cevns  has been analyzed in Ref.~\cite{Chatterjee:2022mmu}, highlighting its capability to probe a large portion of a previously unexplored region of the parameter space.  

Motivated by the latter we are intended to explore the discovery potential of ESS by performing a thorough study investigating several physics scenarios, for the first time. We begin by evaluating the projected sensitivities focusing first  on low-energy SM precision measurements at the ESS, e.g.,  the determination of the weak mixing angle and the nuclear neutron rms radius. Then, concerning new physics we  focus our attention in scenarios beyond conventional vector-type NSIs, in an effort to extend previous works. To this purpose, we instead consider the broader  framework of NGIs, and explore both the light and heavy mediator regimes taking into account several Lorentz-invariant structures. Furthermore, we study potential signatures due to the violation of lepton unitarity and the existence  of  sterile neutrinos and sterile neutral leptons. For the latter, we examine two different possibilities: i) short-baseline active-sterile oscillations, and ii) sterile neutral lepton production in the context of the sterile dipole portal which is possible via the upscattering of active neutrinos on nuclei. Going one step further, we explore sterile neutral leptons by systematically analyzing various Lorentz-invariant interactions including scalar, pseudoscalar, vector, axial vector, and tensor ones. For all the aforementioned scenarios, our results are presented for the individual detectors as well as in terms of a combined analysis.

The remainder of this paper is organized as follows. In Sec.~\ref{Sec:Theory}, we outline the theoretical framework, providing a brief description of the \cevns cross sections for the different scenarios within and beyond the SM. Section~\ref{Sec:Events_simulation} describes the adopted methodology for event simulation and statistical analysis used in the present work in order to estimate the projected sensitivities. This section also includes a detailed overview of the  specifications regarding the different \cevns detector technologies proposed at the ESS. In Sec.~\ref{Sec:Results}, we present and discuss the results of our sensitivity projections. Finally, we conclude with a summary of our findings in Sec.~\ref{Sec:Conclusions}.

%%%%%%%%%%%%%%%%%%%%%%%%%%%%%%%%%%%%%%%%%%%%
%%%%%%%%%%%%%%%%%%%%%%%%%%%%%%%%%%%%%%%%%%%

\section{\label{Sec:Theory} Theoretical framework}
In this section, we present the \cevns cross sections corresponding to the various physics scenarios within and beyond the SM.
\subsection{The standard \cevns cross section}
\label{subsec:CEvNS_SM}

At tree level, within the SM  \cevns is a flavor-blind neutral current process, mediated by the   $Z$ boson~\cite{Tomalak:2020zfh}. For incoming neutrino energies much below the $Z$ boson mass ($E_\nu \ll M_Z$), the quark-level four-Fermi interaction Lagrangian governing this process can be written as~\cite{Barranco:2005yy}
\begin{equation}
\label{equn:CEvNS_SM_Lagrangian}
\mathscr{L}_\mathrm{SM}^{\nu q} \subset -\frac{G_F}{\sqrt{2}} \sum_{\substack{q=u,d \\ \alpha=e,\mu,\tau}} \left[\bar{\nu}_\alpha \gamma^\rho (1-\gamma^5) \nu_\alpha \right] \left[\bar{q} \gamma_\rho (g_{q}^{V \, \text{(SM)}} - g_{q}^{A\,\text{(SM)}} \gamma^5) q \right] \,,
\end{equation}
where $G_F=1.1663787\times 10^{-5} \mathrm{~GeV}^{-2}$ denotes the Fermi constant~\cite{ParticleDataGroup:2022pth}, $q = \{u, d\}$ represents the up and down quarks, and $g_{q}^{V\, \text{(SM)}}$ and $g_{q}^{A\, \text{(SM)}}$ represent the SM vector and axial vector couplings of the quarks to the $Z$ boson. The values of these couplings are given by
\begin{equation}
\label{equn:SM_Quark_Couplings}
\begin{aligned}
    g_{u}^{V\, \text{(SM)}} &= \frac{1}{2} - \frac{4}{3} \sin^2 \theta_W, \,\,\,\,\, & g_{d}^{V\, \text{(SM)}} &= -\frac{1}{2} + \frac{2}{3} \sin^2 \theta_W, \\
    g_{u}^{A\, \text{(SM)}}&= \frac{1}{2}, & g_{d}^{A\, \text{(SM)}} &= -\frac{1}{2}\,.
\end{aligned}
\end{equation}
These couplings depend on the weak mixing angle, which constitutes a fundamental parameter of the Salam-Weinberg electroweak theory~\cite{Weinberg:1967tq, Salam:1968rm} and has been precisely measured at the $Z$ pole  to be $\sin^2\theta_W(M_Z) = 0.23121 \pm 0.00004$. At low energies, however, such as those relevant to \cevns (i.e. for $\mathfrak{q} \rightarrow 0$), the weak mixing angle remains poorly constrained by experiments. Indeed, despite numerous experimental efforts~\cite{ParticleDataGroup:2022pth}, existing low-energy measurements suffer by substantial uncertainties. Theoretically, its value is extrapolated via renormalization group equations (RGE) in the context of $\overline{\text{MS}}$ scheme, as $\sin^2\theta_W(\mathfrak{q}=0) = 0.23857 \pm 0.00005$~\cite{ParticleDataGroup:2022pth, Erler:2019hds}. Therefore, a precise low-energy determination of the weak mixing angle will constitute a  critical test of the SM.  CE$\nu$NS, being a low-energy process, provides a valuable avenue for probing the weak mixing angle at low momentum transfer.  

For a neutrino with energy $E_\nu$ scattering off a nucleus of mass $m_\mathcal{N}$, the differential cross section with respect to the nuclear recoil energy, \( T_\mathcal{N} \), is expressed as~\cite{Freedman:1973yd, Hoferichter:2020osn}
\begin{equation}
\label{Eq:CEvNS_SM_xSec}
\begin{split}
\left[\frac{\mathrm{d}\sigma}{\mathrm{d}T_\mathcal{N}}\right]_\mathrm{SM}^{\mathrm{CE}\nu\mathrm{NS}} = \frac{G_F^2 m_\mathcal{N}}{\pi} &\Big[\mathcal{F}_W^2(\mathfrak{q}^2) \, (Q_{W}^V)^2 \left(1 - \frac{m_\mathcal{N} T_\mathcal{N}}{2 E_{\nu}^2}-\frac{T_\mathcal{N}}{E_\nu}\right)\\
+&\mathcal{F}_A(\mathfrak{q}^2)\left(1 + \frac{m_\mathcal{N} T_\mathcal{N}}{2 E_{\nu}^2}-\frac{T_\mathcal{N}}{E_\nu}\right)\Big] \,.
\end{split}
\end{equation}
The SM vector charge, $Q_{W}^V$, is defined as,
\begin{eqnarray}
Q_{W}^V = \mathbb{Z}\,(2g_{u}^{V \, \text{(SM)}} + g_{d}^{V \, \text{(SM)}}) + \mathbb{N}\,(g_{u}^{V \, \text{(SM)}} + 2g_{d}^{V \, \text{(SM)}}) \,,
\label{eq:weak_charge_SM}
\end{eqnarray}
with $\mathbb{Z}$ and $\mathbb{N}$ being the proton and neutron numbers of the nucleus, respectively. Given that the nucleus is a composite object, a nuclear form factor, \( \mathcal{F}_W(\mathfrak{q}^2) \), accounting for the spatial distribution of the nucleons within the nucleus, is also incorporated. In this analysis, we adopt the Helm parametrization~\cite{PhysRev.104.1466}
\begin{eqnarray}
\label{Eq:Helm_FF}
\mathcal{F}_W(\mathfrak{q}^2) = 3 \frac{j_1(\mathfrak{q} R_0)}{\mathfrak{q} R_0} e^{-\frac{\mathfrak{q}^2 s^2}{2}} \,,
\end{eqnarray}
where $j_1$ is the spherical Bessel function of the first order, $s = 0.9$ fm~\cite{Friedrich:1982esq} represents the surface thickness, and $R_0^2 = \frac{5}{3} (R^2 - 3 s^2)$, with $R$ denoting the rms nuclear radius which is parametrized as $R=1.23\mathbb{A}^{1/3}$ fm. At this point it should be mentioned that when different charge radii are assumed for protons ($R_p$) and neutrons ($R_n$), the corresponding form factors for protons ($\mathcal{F}_p$) and neutrons ($\mathcal{F}_n$) do not factorize and the quantity $\mathcal{F}_W^2 (Q_W^V)^2$ in Eq.~(\ref{Eq:CEvNS_SM_xSec}) must be replaced by
\begin{equation}
    \mathcal{F}_W^2(\mathfrak{q}^2) (Q_W^V)^2 \to \left[\mathbb{Z}\,(2g_{u}^{V \, \text{(SM)}} + g_{d}^{V \, \text{(SM)}}) \mathcal{F}_p(\mathfrak{q}^2) + \mathbb{N}\,(g_{u}^{V \, \text{(SM)}} + 2g_{d}^{V \, \text{(SM)}})\mathcal{F}_n(\mathfrak{q}^2)\right]^2 \, .
    \label{eq:exact_Qw}
    \end{equation}
In this case, $\mathcal{F}_p$ and  $\mathcal{F}_n$ correspond to the individual Helm form factors following the definition of Eq.~(\ref{Eq:Helm_FF}) with $R_0^2= \frac{5}{3}(R_{p}^2 - 3s^2)$ for protons and $R_0^2= \frac{5}{3}(R_{n}^2 - 3s^2)$  for neutrons. Notice that by using Eq.~\eqref{eq:exact_Qw} the difference in the predicted rates is very small, and hence we use it only when probing  $R_n$. For the other cases explored below we have verified that the resulting limits remain unaffected.

The SM axial vector contribution to the \cevns cross section is subdominant in comparison to the vector component, being highly suppressed by the nuclear spin~\cite{Barranco:2005yy}. However, in our analysis, we have incorporated this component, with the corresponding axial vector form factor taken to be\footnote{We neglect subdominant contributions from strangeness and two-body currents.}~\cite{Markisch:2018ndu, Hoferichter:2020osn}
\begin{equation}
    \mathcal{F}_A(\mathfrak{q}^2) = \frac{2\pi}{2J+1} \cdot g_A^2 \cdot S_{11}^\mathcal{T}(\mathfrak{q}^2)\,,
\end{equation}
where $J$ denotes the total angular momentum of the nucleus in its ground state (see Table~\ref{Tab:Detectors_Specs}), while  the  axial vector coupling of the nucleon is parameterized as $g_A = \Delta_u^p - \Delta_d^p$.  Here, $\Delta_q^p$ parametrizes the contribution of the quark spin content of the nucleon. Assuming isospin symmetry, the relevant values are taken from lattice QCD studies~\cite{Lin:2018obj} and read $\Delta_u^p = \Delta_d^n= 0.777$ and $\Delta_d^p = \Delta_u^n = -0.438$.  The transverse spin structure functions, $S_{ij}^\mathcal{T}$, are evaluated following the methodology outlined in Ref.~\cite{Hoferichter:2020osn}, with  only  $S_{11}^\mathcal{T}$ being relevant for the SM case. This is because of the assumed isospin symmetry together with the fact that $g_{u}^{A \, \text{(SM)}} = -g_{d}^{A \, \text{(SM)}}$ for which only the isovector part of the full spin-structure function survives. Moreover,  tiny contributions from the longitudinal multipoles are of the order $\mathcal{O}(T_\mathcal{N}/m_\mathcal{N})$ and hence safely ignored. It is important to note that, due to the spin-dependent nature of the axial vector component, this contribution is non-zero only for nuclei with non-zero total angular momentum. Therefore, among the various proposed ESS detectors considered here, only the CsI and $\mathrm{C}_3\mathrm{F}_8$ ones receive non-zero axial vector contributions. 
%%%%%%%%%%%
%%%%%%%%%%%%%%

\begin{table}[ht!]
\centering
\begin{tabular}{|c|c|c|c|c|c|}
\hline
\hspace{0.5cm}Target nucleus\hspace{0.5cm} & \hspace{0.5cm}$\mathbb{Z}$\hspace{0.5cm} & \hspace{0.5cm}$\mathbb{N}$\hspace{0.5cm} & \hspace{0.5cm}$m_\mathcal{N}$ (a.m.u)~\cite{nist_handbook}\hspace{0.5cm} & \hspace{0.5cm}$J^\pi$~\cite{nist_handbook}\hspace{0.5cm} & \hspace{0.5cm}$R_p$ (fm)~\cite{Mann:1973ais, Angeli:2013epw}\hspace{0.5cm} \\ 
\hline
\hline
$^{133}\mathrm{Cs}$  & $55$       & $78$         & $132.91$    & $7/2^+$  & $4.824$ \\ 
\hline
$^{127}\mathrm{I}$   & $53$       & $74$         & $126.90$    & $5/2^+$  & $4.766$ \\ 
\hline
$^{132}\mathrm{Xe}$  & $54$       & $78$         & $131.29$    & $0^+$    & $4.786$ \\ 
\hline
$^{72}\mathrm{Ge}$   & $32$       & $40$         & $71.92$     & $0^+$    & $4.050$ \\ 
\hline
$^{28}\mathrm{Si}$   & $14$       & $14$         & $27.98$     & $0^+$    & $3.150$ \\ 
\hline
$^{40}\mathrm{Ar}$   & $18$       & $22$         & $39.95$     & $0^+$    & $3.393$ \\ 
\hline
$^{12}\mathrm{C}$    & $6$        & $6$          & $12.0$      & $0^+$    & $2.471$ \\ 
\hline
$^{19}\mathrm{F}$    & $9$        & $10$         & $19.00$     & $1/2^+$  & $2.900$ \\ 
\hline
\end{tabular}
\caption{Key properties of the proposed ESS nuclear targets (see also  Table~\ref{Tab:detectors_details}): the number of protons ($\mathbb{Z}$), neutrons ($\mathbb{N}$), atomic mass ($m_\mathcal{N}$), nuclear ground-state ($J^\pi$), and proton rms radius ($R_p$).}
\label{Tab:Detectors_Specs}
\end{table}

For completeness, Table~\ref{Tab:Detectors_Specs} lists the physical properties of the target nuclei employed in these detectors, including their proton and neutron numbers, atomic masses, spin values, and proton rms radii.

\subsection{Neutrino Generalized Interactions}
\label{subsec:NGIs}

NSIs provide straightforward and theoretically motivated SM extensions. They can be naturally accommodated by extending the SM with an additional $U(1)$ gauge group, which introduces a new vector mediator. If such mediators exist, they could contribute significantly to \cevns processes,  resulting in observable distortions of the detectable rates, being particularly relevant at low-energy nuclear recoils. Several studies  examined NSI contributions to \cevns in both the light~\cite{Majumdar:2024dms, DeRomeri:2024dbv} and heavy~\cite{Chatterjee:2022mmu, Mustamin:2021mtq, Majumdar:2022nby} mediator regimes. To our knowledge, up to now there exists  only one study which explored the impact of NSI at the ESS, focusing mainly on the heavy mediator regime~\cite{Chatterjee:2022mmu}.

In addition to the conventional NSI, which usually encompasses vector ($V$) or axial vector ($A$) interactions arising from gauge extensions, a more generalized framework can be constructed to account for all Lorentz-invariant interactions that may indicate new physics. This scenario is known as NGI~\cite{Lindner:2016wff,AristizabalSierra:2018eqm}, which accounts also for scalar ($S$), pseudoscalar ($P$) and tensor ($T$) interactions. A few comments are in order: scalar and pseudoscalar interactions can emerge from Yukawa-type interaction terms if an additional scalar/pseudoscalar mediator is introduced alongside the existing SM particle content, while vector and axial vector interactions are commonly generated within gauge extensions of the SM. In contrast, tensor interactions represent an effective type of interaction, typically necessitating multiple mediators or composite particles within a UV complete theory to be fully realized. As such, tensor interactions often arise in low energy phenomenological processes, while a full UV complete model may require a more intricate theoretical foundation involving composite structures or the interplay of multiple mediator particles.

Following a model-independent approach, in this work, we add the most general effective  NGI operators to the SM Lagrangian below the electroweak scale,  as~\cite{AristizabalSierra:2018eqm}
\begin{equation}
\label{Eq:NGI_Lagrangian}
\mathscr{L}^{\nu q}_\mathrm{NGI} \supset -\frac{G_F}{\sqrt{2}} \sum_{\substack{X=S,P,V,A,T\\q=u,d} } \varepsilon_{\nu q}^X\left[\bar{\nu} \Gamma^X  \nu \right] \left[\bar{q} \Gamma_X q \right] \, .
\end{equation}
 This Lagrangian incorporates all possible Lorentz-invariant,  neutral-current interactions between neutrinos and first-generation quarks. Here, $\Gamma_X\equiv\{\mathbb{I}, i\gamma^5, \gamma_\rho, \gamma_\rho\gamma^5, \sigma_{\rho\delta}\}$ (with $\sigma_{\rho\delta}=\frac{i}{2}[\gamma_\rho,\gamma_\delta]$) correspond to $X\equiv\{S,P,V,A,T\}$. The relative strengths of the new physics interactions are encoded in the real, dimensionless coefficients $\varepsilon_{\nu q}^X$.  For the case of light mediators, where their  mass is comparable to or smaller than the typical momentum transfer in the experiment ($\sim\mathcal{O}(10)$ MeV for CE$\nu$NS), the interaction strength is parameterized as $\varepsilon_{\nu q}^X \sim \left[\sqrt{2}/G_F\right]\left[\textsl{g}_X^2/(\mathfrak{q}^2+M_X^2)\right]$. Here, $M_X$ represents the mediator mass of the type $X \equiv \{S,P,V,A,T\}$, and $\textsl{g}_X$ denotes the corresponding coupling strength  defined as  $\textsl{g}_X = \sqrt{g_\nu^X g_q^X}$, where   universal quark couplings $g_u^X=g_d^X=g_q^X$ have been assumed, while $g_\nu^X$ stands for the corresponding  neutrino  coupling. Notably, for mediators with masses much higher than the momentum transfer of the experiment, one can integrate out the mediator mass from the denominator of the interaction strength.

Moving from quark- to nuclear-level operators the resulting Lagrangian takes the form~\cite{AristizabalSierra:2018eqm}
\begin{equation}
\label{equn:NGI_nuclear_level_interaction}
\mathscr{L}^{\nu \mathcal{N}}_\mathrm{NGI} \supset -\sum_{\substack{X=S,P,V,A,T}} \frac{C_X}{\mathfrak{q}^2+M_X^2}\left[\bar{\nu} \Gamma^X \nu \right] \left[\bar{\mathcal{N}} \Gamma_X \mathcal{N} \right] \,,
\end{equation}
where $C_X$ represents the effective $\nu$--$\mathcal{N}$ coupling. Among the different interactions, scalar and vector interactions are spin-independent, whereas pseudoscalar, axial vector, and tensor interactions depend on nuclear spin. The contributions of different interactions to the \cevns differential cross section are expressed as~\cite{DeRomeri:2024iaw}
\begin{subequations}
\label{Eq.NGI_xSec}
\begin{align}
    \left[\frac{\mathrm{d}\sigma}{\mathrm{d}T_\mathcal{N}}\right]_S^\mathrm{CE\nu NS} &= \frac{m_\mathcal{N}^2T_\mathcal{N} C_S^2}{4\pi E_\nu^2 (M_S^2 + 2m_\mathcal{N} T_\mathcal{N})^2} \mathcal{F}_W^2(\mathfrak{q}^2)\,, \\[4pt]
%%%%%%%%%%%%%%%%%%%%%%%%%%%%%%%%%%%%%%%%%%%%%%%%%%%%%%%%%%
    \left[\frac{\mathrm{d}\sigma}{\mathrm{d}T_\mathcal{N}}\right]_V^\mathrm{CE\nu NS} &= \left(1+\frac{C_V}{\sqrt{2}G_FQ_{W}^V(M_V^2 + 2m_\mathcal{N} T_\mathcal{N})}\right)^2\left[\frac{\mathrm{d}\sigma}{\mathrm{d}T_\mathcal{N}}\right]_\mathrm{SM}^{\mathrm{CE}\nu\mathrm{NS}} \,, \\[4pt]
%%%%%%%%%%%%%%%%%%%%%%%%%%%%%%%%%%%%%%%%%%%%%%%%%%%%%%%%%%
    \left[\frac{\mathrm{d}\sigma}{\mathrm{d}T_\mathcal{N}}\right]_A^\mathrm{CE\nu NS} &= \frac{4 m_\mathcal{N}}{2J + 1} \frac{C_A^2}{(M_A^2 + 2 m_\mathcal{N} T_\mathcal{N})^2} \left[1 + \frac{m_\mathcal{N} T_\mathcal{N}}{2E_\nu^2}-\frac{T_\mathcal{N}}{E_\nu}\right] S_{00}^\mathcal{T}(\mathfrak{q}^2)\,, \\[4pt]
%%%%%%%%%%%%%%%%%%%%%%%%%%%%%%%%%%%%%%%%%%%%%%%%%%%%%%%%%%
    \left[\frac{\mathrm{d}\sigma}{\mathrm{d}T_\mathcal{N}}\right]_T^\mathrm{CE\nu NS} &= \frac{m_\mathcal{N} }{2J + 1} \frac{C_T^2}{(M_T^2 + 2 m_\mathcal{N} T_\mathcal{N})^2} \Bigg\lbrace \left[2 - \frac{m_\mathcal{N} T_\mathcal{N}}{E_\nu^2}-\frac{2T_\mathcal{N}}{E_\nu}\right] S_{00}^\mathcal{T}(\mathfrak{q}^2) + \left[1-\frac{T_\mathcal{N}}{E_\nu}\right]S_{00}^\mathcal{L}(\mathfrak{q}^2) \Bigg\rbrace ,
\end{align}
\end{subequations}
where $\mathcal{O}(T_\mathcal{N}/m_\mathcal{N})$ and higher order $\mathcal{O}(T^2_\mathcal{N})$ terms have been neglected.
The pseudoscalar interaction is neglected  since its contribution at low-momentum transfer ($\mathfrak{q}\rightarrow0$) is found to be negligible~\cite{Cerdeno:2016sfi,DeRomeri:2024iaw}.
It is important to note that the differential cross sections for $X = S, A, T$ interactions are provided for pure new physics contributions. In contrast, due to the interference term, the differential cross section for the vector interaction is expressed as the sum of the SM and BSM contributions. Even though the SM  \cevns cross section given in Eq.~\eqref{Eq:CEvNS_SM_xSec} includes an axial vector component, there is no interference between the SM and axial vector NGI contributions considered here. The absence of interference arises since the relevant spin-structure function for the SM \cevns case is $S_{11}^\mathcal{T}$, while for the NGI case the relevant spin structure function is $S_{00}^\mathcal{T}$. This is due to the fact that for the NGI scenarios, universal quark couplings $g_u^A=g_d^A$ are assumed.  The effective couplings for the different interactions are given as~\cite{Barranco:2011wx, Cirelli:2013ufw,Cerdeno:2016sfi, Hoferichter:2020osn, Demirci:2021zci}
\begin{subequations}
\label{eq:vector}
\begin{align}
C_S &= g_\nu^S\left( \mathbb{Z}\sum_{q = u,d}g_q^S\frac{m_p}{m_q}f_{T_q}^p + \mathbb{N}\sum_{q = u,d}g_q^S\frac{m_n}{m_q}f_{T_q}^n \right)\,,\label{Eq:Scalar_NGI_Coupling}\\
C_V &= \kappa g_\nu^V\left[\mathbb{Z}(2g_u^V+g_d^V)+ \mathbb{N}(g_u^V+2g_d^V)\right] = 3 \kappa \mathbb{A} g_\nu^V g_q^V \,,\label{Eq:Vector_NGI_Coupling}\\
C_A &= g_\nu^A g_q^A \left(\Delta_u^p + \Delta_d^p\right)\,,\label{Eq:Axial_Vector_NGI_Coupling}\\
C_T &= g_\nu^T g_q^T \left(\delta_u^p + \delta_d^p\right)\,,\label{Eq:Tensor_NGI_Coupling}
\end{align}
\end{subequations}
where $m_p$ and $m_n$ are the proton and neutron masses, $m_q$ are the quark masses, and  finally the hadronic structure parameters for the scalar interaction are~\cite{DelNobile:2021wmp}
\begin{equation*}
f_{T_u}^p = 0.026\, ,  ~~~~~ f_{T_d}^p =  0.038\, ,  ~~~~~ f_{T_u}^n = 0.018\, ,  ~~~~~ f_{T_d}^n =  0.056\, .
\end{equation*} 
For the vector-mediated interaction, we consider the anomaly-free 
${U}(1)_{B-L}$ gauge extension of the SM as the benchmark model. In this case, anomaly cancellation conditions dictate the charges assigned to leptons ($Q'_\ell = -1$) and quarks ($Q'_q = 1/3$) under the additional ${U}(1)'$ gauge symmetry. This results in an effective vector charge as defined in Eq.~\eqref{Eq:Vector_NGI_Coupling}, with $\kappa = Q'_\ell\cdot Q'_q = -1/3$. For axial vector and tensor-mediated cross sections the longitudinal and transverse spin structure functions, $S_{ij}^\mathcal{L}$ and $S_{ij}^\mathcal{T}$, are evaluated following the methodology specified in Ref.~\cite{Hoferichter:2020osn} (for more details see the Appendix B of Ref.~\cite{Candela:2024ljb}). Finally, for the tensor interaction, the hadronic structure functions read~\cite{DelNobile:2021wmp}
\begin{equation*}
\delta_u^p = \delta_d^n= 0.784, \qquad \delta_d^p = \delta_u^n= -0.204\,.
\end{equation*}
Again and similarly to the axial vector case, for tensor NGI the relevant spin-structure functions are $S_{00}^\mathcal{L}$ and $S_{00}^\mathcal{T}$.

As explained previously, among the different detectors analyzed in this study, only the CsI and $\mathrm{C}_3\mathrm{F}_8$ detectors, which contain spin-dependent isotopes, acquire non-vanishing sensitivity to axial vector and tensor-mediated \cevns processes.

\subsection{Phenomenology of sterile neutrinos and sterile neutral leptons}

Within the context of the SM, neutrinos are left-handed and  massless without any isosinglet ``right-handed'' counterparts. One of the simplest mechanisms to provide mass to neutrinos is the so-called ``seesaw'' mechanism, which introduces an arbitrary number of  new right-handed sterile neutral leptons for each left-handed neutrino~\cite{Minkowski:1977sc, Mohapatra:1979ia, Yanagida:1980xy, Magg:1980ut, Mohapatra:1986bd, Foot:1988aq}. These right-handed neutral leptons, typically  taken as singlets under the SM gauge group, are commonly referred to as sterile neutrinos or sterile neutral lepton in the literature\footnote{Note that, as far as Lorentz and SM gauge symmetries are concerned, there is no difference between right handed neutrinos, sterile neutrinos and sterile neutral leptons.  Hence these terms are often used interchangeably.}. Depending on their masses, the experimental signatures of these sterile neutral leptons can be  best probed in different ways. Therefore, in order to differentiate their role and implications in neutrino phenomenology, in this work we make the following distinction.
\begin{itemize}
 \item \textbf{Sterile Neutrinos ($\nu_s$):} We call them sterile neutrinos if they are sufficiently light. In that case, they may participate in neutrino oscillations in the short baseline experiments, leading to active-sterile oscillations~\cite{LSND:2001aii, MiniBooNE:2013uba, Miranda:2020syh, Capozzi:2023ltl}.  
 \item \textbf{Sterile Neutral Leptons ($N_R$):} When they are sufficiently heavier than active neutrinos, we simply call them sterile neutral leptons (SNL). Indeed, if their masses are significantly larger than the electroweak scale, they would contribute to weak interaction processes only through their mixing with the SM neutrinos, leading to a phenomenon known as lepton unitarity violation~\cite{Schechter:1980gr, Escrihuela:2015wra, Capozzi:2023ltl, CentellesChulia:2024sff}.
\end{itemize}

SNLs can be also produced via active-sterile neutrino Transition Magnetic Moments (TMM)~\cite{McKeen:2010rx}. Going one step further, one can also build up the framework for the production of a massive sterile neutral lepton through the upscattering of active neutrinos on nuclei, involving various Lorentz-invariant interactions (scalar, pseudoscalar, vector, axial vector, and tensor)~\cite{Brdar:2018qqj, Chang:2020jwl, Chen:2021uuw, Candela:2023rvt, Candela:2024ljb}. 

In this subsection, we will briefly discuss all the aforementioned scenarios, with a particular emphasis on their implications in the context of \cevns.

\subsubsection{Violation of lepton unitarity}\label{SubSec:Violation_of_lepton_unitarity}
We consider a scenario that extends the SM by incorporating $n$ heavy neutral leptons ($N_{R,i}$ ; $i = 1,2,\cdots n$) in addition to the three standard light neutrinos. These new heavy neutral leptons are assumed to be SM gauge singlets and much heavier than the electroweak scale. In such scenarios, the generalized unitary lepton mixing matrix can be expressed as~\cite{Schechter:1980gr, Nardi:1994iv}
\begin{equation}
\label{Eq:4X4_Unitary}
    U \equiv \begin{pmatrix}
        N_{3\times 3} & S_{3\times n}\\
        V_{n\times 3} & T_{n\times n}
    \end{pmatrix}\,,
\end{equation}
where $N_{3\times 3}$ describes the mixing among light neutrino states, $S$ and $V$ represent the mixing between light and heavy states, and $T$ denotes the mixing among the heavy states. The $N_{3\times 3}$ submatrix can be written as~\cite{Escrihuela:2015wra}
\begin{equation}
    N_{3\times 3} = N_{3\times 3}^\mathrm{NP} U_{3\times 3}^\mathrm{PMNS} \,,
\end{equation}
where $U_{3\times 3}^\mathrm{PMNS}$ is the standard $3 \times 3$ Pontecorvo-Maki-Nakagawa-Sakata (PMNS) unitary mixing matrix~\cite{Maki:1962mu, Bilenky:1978nj}, while $N_{3\times 3}^\mathrm{NP}$ accounts for new physics effects associated with unitarity violation. The latter matrix is parameterized in a lower-triangular form~\cite{Escrihuela:2015wra}
\begin{equation}
\label{equn:New_Physics_Matrix}
N_{3\times 3}^\mathrm{NP} \equiv \begin{pmatrix}
\alpha_{11} & 0 & 0 \\
\alpha_{12} & \alpha_{22} & 0 \\
\alpha_{13} & \alpha_{23} & \alpha_{33}
\end{pmatrix} \, ,
\end{equation}
where the diagonal elements $\alpha_{ii}$ are real, and the off-diagonal elements $\alpha_{ij}$ are generally small and complex. These elements satisfy the ``triangle inequalities''~\cite{Fernandez-Martinez:2016lgt,Escrihuela:2016ube}
\begin{equation}
\begin{aligned}
|\alpha_{21}| &\leq \sqrt{(1-\alpha_{11}^2)(1-\alpha_{22}^2)} \, , \\
|\alpha_{31}| &\leq \sqrt{(1-\alpha_{11}^2)(1-\alpha_{33}^2)} \, , \\
|\alpha_{32}| &\leq \sqrt{(1-\alpha_{22}^2)(1-\alpha_{33}^2)} \, .
\end{aligned}
\label{eq:triangle_ineq}
\end{equation}

In Eq.~\eqref{Eq:4X4_Unitary}, the submatrix $S \sim \mathcal{O}(\varepsilon)$, where $\varepsilon$ is the seesaw expansion parameter~\cite{Schechter:1981cv}. The unitarity condition of $U$ then implies $NN^\dagger + SS^\dagger = I $, leading to $NN^\dagger \sim 1 - \mathcal{O}(\varepsilon^2)$. Consequently, within the seesaw paradigm, we find $\alpha_{ii}^2 \sim 1 - \mathcal{O}(\varepsilon^2)$ and $|\alpha_{ij}|^2 \sim \mathcal{O}(\varepsilon^4) $~\cite{CentellesChulia:2024sff}.
For the most general charged-current interactions, the relevant rectangular sub-block is $K\equiv\begin{pmatrix}
    N_{3\times 3} & S_{3\times n}
\end{pmatrix}$, while for neutral-current interactions, $P = K^\dagger K$ is required~\cite{Schechter:1980gr}. If the energy of a given process is much lower than the masses of the heavy sterile states, these will not be produced in the short baseline experiments, like e.g. the ESS. For example, the heavy states will not take part in oscillation experiments. Then, effectively, only the first $3\times 3$ blocks of $K~(N)$ and $P~(N^\dagger N)$ will play a crucial role in the charged- and neutral-current weak interactions, respectively.

The SM \cevns cross section, discussed in Eq.~\eqref{Eq:CEvNS_SM_xSec}, is proportional to the Fermi constant $G_F$, which is usually extracted from the $\mu^-$ decay width. However, in the presence of non-unitarity (NU), the $W$-boson vertices are modified, and as a result the measured quantity would be the effective muon decay coupling $G_\mu$. These are related by~\cite{Miranda:2020syh}
\begin{equation}
    \frac{G_F^2}{G_\mu^2} = \frac{1}{(NN^\dagger)_{ee}(NN^\dagger)_{\mu\mu}} \,.
\end{equation}

Given that at the ESS the neutrinos are generated through charged-current interaction, and subsequently scatter with the nucleus via the neutral-current interaction channel, the zero-distance probability factor is given as~\cite{CentellesChulia:2024sff}
\begin{equation}
    \mathcal{P} = (NN^\dagger NN^\dagger NN^\dagger)_{\alpha \alpha} \,,
\end{equation}
where $\nu_\alpha$ is the incident neutrino flavor. The ratio of the expected events in the SM ($ R_\mathrm{SM} $) and NU cases ($ R_\mathrm{NU} $) is then expressed as $ R_\mathrm{NU}/R_\mathrm{SM} = \mathcal{P} \cdot \left[G_F^2 / G_\mu^2\right] $. Following Ref.~\cite{CentellesChulia:2024sff}, we expand $R_\mathrm{NU}$ in powers of $\varepsilon $, retaining terms up to $\mathcal{O}(\varepsilon^2) $:
\begin{equation}
\begin{aligned}
R_{\nu_\mu \mathcal{N}}^\mathrm{NU} &= (2\alpha_{22}^2 - \alpha_{11}^2) R_{\nu_\mu \mathcal{N}}^\mathrm{SM} \, , \\
R_{\nu_e \mathcal{N}}^\mathrm{NU} &= (2\alpha_{11}^2 - \alpha_{22}^2) R_{\nu_e \mathcal{N}}^\mathrm{SM} \, , \\
R_{\bar{\nu}_\mu \mathcal{N}}^\mathrm{NU} &= (2\alpha_{22}^2 - \alpha_{11}^2) R_{\bar{\nu}_\mu \mathcal{N}}^\mathrm{SM} \, .
\end{aligned}
\label{equn:NU_Oscillation_Probability}
\end{equation}
Thus, for our analysis, the relevant NU parameters affecting the neutrino signal are $ \alpha_{11} $ and $ \alpha_{22} $. Unlike neutrino oscillation experiments, the neutrino signal in this case is independent of CP-violating phases and standard oscillation phenomena.

\subsubsection{Sterile neutrino oscillations}

Although most theoretical neutrino mass mechanisms suggest the existence of heavy sterile neutral leptons, it is  possible that singlet neutrinos also exist in nature, being  light enough to participate in oscillations~\cite{Branco:2020yvs}. These are commonly referred to as light sterile neutrinos ($\nu_s$). In our study, we consider the simplest 3+1 scenario in which the standard lepton mixing  matrix is extended to incorporate the three standard generations of active neutrinos plus  one light sterile neutrino.  In experiments such as ESS, where neutrinos are detected via CE$\nu$NS, the oscillation effects differ from those observed in conventional oscillation experiments. The very short baseline of such facilities suppresses the oscillations between active neutrino flavors, while matter effects can also be  neglected due to the minimal path length. Moreover, while neutrinos are primarily produced via conventional charged-current processes, the \cevns detection relies on neutral-current interactions, unlike the case of dedicated oscillation experiments. Therefore, \cevns signals in ESS-like experiments can provide complementary constraints on light sterile neutrinos, particularly given their potential for a high-intensity pulsed neutrino source and high-statistics data.

In this context, the survival probabilities of electron neutrinos ($\mathcal{P}_{ee}$) and muon (anti)neutrinos ($\mathcal{P}_{\mu\mu}$) are expressed as~\cite{Miranda:2020syh}
\begin{subequations}
\begin{equation}
\mathcal{P}_{ee}(E_\nu) \approx 1 - \sin^2(2\theta_{14})\sin^2\left(\frac{\Delta m^2_{41}L}{4E_\nu}\right),
\end{equation}
\begin{equation}
\mathcal{P}_{\mu\mu}(E_\nu) \approx 1 - \sin^2(2\theta_{24})\sin^2\left(\frac{\Delta m^2_{42}L}{4E_\nu}\right),
\end{equation}
\end{subequations}
where $\theta_{14}$ and $\theta_{24}$ are the active-sterile mixing angles, and $\Delta m^2_{41} \approx \Delta m^2_{42} = \Delta m^2$ represent the active-sterile mass splittings, while $L$ is the baseline of the experiment. Here, we assume $\Delta m^2$ to be much larger than the solar ($\Delta m^2_{21}$) and atmospheric ($|\Delta m^2_{31}|$) mass splittings. The sterile neutrino oscillation effects modify the unoscillated flux distributions reaching the detector according to 
\begin{equation}
    \begin{split}
        &\frac{\mathrm{d}\Phi_{\nu_e}}{\mathrm{d}E_\nu} = \mathcal{P}_{ee}(E_\nu) \frac{\mathrm{d}\Phi^0_{\nu_e}}{\mathrm{d}E_\nu}, \\
        &\frac{\mathrm{d}\Phi_{\nu_\mu}}{\mathrm{d}E_\nu} + \frac{\mathrm{d}\Phi_{\bar{\nu}_\mu}}{\mathrm{d}E_\nu} = \mathcal{P}_{\mu\mu}(E_\nu) 
        \left(\frac{\mathrm{d}\Phi^0_{\nu_\mu}}{\mathrm{d}E_\nu} + \frac{\mathrm{d}\Phi^0_{\bar{\nu}_\mu}}{\mathrm{d}E_\nu}\right),
    \end{split}
\end{equation}
where $\mathrm{d}\Phi^0_{\nu_\alpha}/\mathrm{d}E_\nu$ represents the unoscillated flux distributions produced at the source, and $\mathrm{d}\Phi_{\nu_\alpha}/\mathrm{d}E_\nu$ corresponds to the flux arriving at the detector.

\subsubsection{Sterile dipole portal}

As discussed previously, one possibility of producing SNLs is through electromagnetic upscattering  of an active neutrino on nuclei, provided that there is a large TMM between the active and sterile states. When  muon or electron (anti)neutrinos interact electromagnetically with a target nucleus of charge $\mathbb{Z}e$, it is possible to generate SNL via the upscattering process $\nu_{e/\mu}~\mathcal{N} \rightarrow N_R~\mathcal{N}$. Neglecting the tiny contribution from the nuclear magnetic dipole moment, the differential cross section for this process is given as~\cite{McKeen:2010rx, Chen:2021uuw, Miranda:2021kre}
\begin{equation}
  \label{eq:xsec_dipole_portal}
\begin{aligned}
 \left[\frac{\mathrm{d}\sigma_{\nu_\alpha \mathcal{N}}}{\mathrm{d}T_\mathcal{N}}\right]^{\nu\mathcal{N}-N_R\mathcal{N}}_\mathrm{DP} &=
  \dfrac{\pi \alpha^2_\mathrm{EM}}{m_{e}^2} \, \mathbb{Z}^2 \mathcal{F}_{W}^2(\mathfrak{q}^2)
\left| \dfrac{\mu_{\nu_{\alpha}}}{\mu_{B}} \right|^2 \\
 & \times \left[\frac{1}{T_\mathcal{N}} - \frac{1}{E_\nu} 
    - \frac{m_{N_R}^2}{2E_\nu T_\mathcal{N} m_\mathcal{N}}
    \left(1- \frac{T_\mathcal{N}}{2E_\nu} + \frac{m_\mathcal{N}}{2E_\nu}\right)
    + \frac{m_{N_R}^4(T_\mathcal{N}-m_\mathcal{N})}{8E_\nu^2 T_\mathcal{N}^2 m_\mathcal{N}^2}
  \right]\,,
  \end{aligned}
\end{equation}
where $\alpha_\mathrm{EM}$ is the fine-structure constant, $m_{N_R}$ is the mass of SNL, and $\mu_{\nu_{\alpha}}$ is the effective neutrino magnetic moment\footnote{The effective neutrino magnetic moment for the active-sterile transition is written in terms of fundamental TMMs, CP phases and neutrino mixing parameters~\cite{Miranda:2021kre}. In this work, for simplicity we will focus on effective magnetic moments only.}, expressed in units of the Bohr magneton, $\mu_B$.
The kinematics of this process imposes an upper bound on the produced SNL mass $m_{N_R}$, as 
\begin{equation}
    \label{Eq:Sterile_kinematics_limit_on_m4}
    m_{N_R}^2 \lesssim 2m_\mathcal{N}T_\mathcal{N}\left(\sqrt{\frac{2}{m_\mathcal{N}T_\mathcal{N}}}E_\nu-1\right)\,,
\end{equation}
while for this scenario the minimum neutrino energy is obtained by inverting the last expression, and reads
\begin{equation}
    \label{Eq:Sterile_kinematics_min_Ev}
    E_\nu^\mathrm{min} \approx \sqrt{\frac{m_\mathcal{N}T_\mathcal{N}}{2}}\left(1+\frac{m_{N_R}^2}{2m_\mathcal{N}T_\mathcal{N}}\right)\,.
\end{equation}

\subsubsection{Upscattering Production of  Sterile Neutral Leptons via  NGIs}
In analogy to the dipole portal scenario discussed previously, SNLs can also be produced via the upscattering of active neutrinos on nuclei mediated by generalized Lorentz-invariant interactions. For this scenario, we adopt a nuclear-level Lagrangian similar to the one considered for the NGI framework in Eq.~\eqref{equn:NGI_nuclear_level_interaction}, but with the outgoing active neutrino replaced by a massive SNL ($N_R$). The modified Lagrangian is expressed as
\begin{equation}
\label{equn:Upscattering_Sterile_nuclear_level_interaction}
\mathscr{L}^{\nu \mathcal{N}-{N_R}\mathcal{N}}_\mathrm{SNL} \supset -\sum_{\substack{X=S,P,V,A,T\\\alpha=e,\mu,\tau}} \frac{C_X}{\mathfrak{q}^2+M_X^2}\left[\bar{{N_R}} \Gamma^X  \nu_{L, \alpha } \right] \left[\bar{\mathcal{N}} \Gamma_X \mathcal{N} \right] \,,
\end{equation}
where $\nu_{L, \alpha }$ is the left-handed neutrino field.

The corresponding \cevns cross sections by dropping  
$\mathcal{O}(T_\mathcal{N}/m_\mathcal{N})$ and higher order $\mathcal{O}(T^2_\mathcal{N})$ terms\footnote{For the full expressions see Ref.~\cite{Candela:2024ljb}.}, read~\cite{Candela:2023rvt,Candela:2024ljb}
%%%%
%%%%%%%
\begin{subequations}
\begin{align}
    \left[\frac{\mathrm{d}\sigma}{\mathrm{d}T_\mathcal{N}}\right]_S^{\nu\mathcal{N}-{N_R}\mathcal{N}} &= 
    \dfrac{m_\mathcal{N} C_S^2}{4\pi (M_S^2 + 2m_\mathcal{N} T_\mathcal{N})^2} 
    \mathcal{F}_W^2(\mathfrak{q}^2)  
    \left(\dfrac{m_\mathcal{N} T_\mathcal{N}}{E_\nu^2} + \dfrac{m_{N_R}^2}{2 E_\nu^2}\right)\,, \\[4pt]
    %%%%%%%%%%%%%%%%%%%%%%%%%%%%%%%%%%%%%%%%%%%%%%%%%%%%%%
    \left[\frac{\mathrm{d}\sigma}{\mathrm{d}T_\mathcal{N}}\right]_V^{\nu\mathcal{N}-{N_R}\mathcal{N}} &= 
    \dfrac{m_\mathcal{N} C_V^2}{2\pi (M_V^2 + 2m_\mathcal{N} T_\mathcal{N})^2} 
    \mathcal{F}_W^2(\mathfrak{q}^2) \nonumber \\[4pt]
    &\times \left[
        \left(1 - \dfrac{m_\mathcal{N} T_\mathcal{N}}{2 E_\nu^2} - \dfrac{T_\mathcal{N}}{E_\nu}\right) 
        - \dfrac{m_{N_R}^2}{4 E_\nu^2}\left(1 + \dfrac{2 E_\nu}{m_\mathcal{N}}\right)
    \right]\,,\\[4pt]
    %%%%%%%%%%%%%%%%%%%%%%%%%%%%%%%%%%%%%%%%%%%%%%%%%
    \left[\frac{\mathrm{d}\sigma}{\mathrm{d}T_\mathcal{N}}\right]_A^{\nu\mathcal{N}-{N_R}\mathcal{N}} &=  
    \dfrac{2 m_\mathcal{N}}{2J + 1} \dfrac{C_A^2}{(M_A^2 + 2 m_\mathcal{N} T_\mathcal{N})^2} \nonumber \\[4pt]
    &\times \Bigg\lbrace \left[
        \left(
            2 + \dfrac{m_\mathcal{N} T_\mathcal{N}}{E_\nu^2} - \dfrac{2 T_\mathcal{N}}{E_\nu}\right)  - \dfrac{m_{N_R}^2}{2 E_\nu^2} \left(
            1 + \dfrac{3 E_\nu}{m_\mathcal{N}} + \dfrac{m_{N_R}^2 }{m_\mathcal{N} T_\mathcal{N}}\right)
    \right] S_{00}^{\mathcal{T}}(\mathfrak{q}^2) \nonumber \\[4pt]
    &~~~~+ 
        \dfrac{m_{N_R}^2}{E_\nu^2} \left(
        2 + \dfrac{3 E_\nu}{m_\mathcal{N}} + \dfrac{m_{N_R}^2}{m_\mathcal{N} T_\mathcal{N}}\right)  S_{00}^\mathcal{L}(\mathfrak{q}^2) \Bigg\rbrace\,,\\[4pt]
    %%%%%%%%%%%%%%%%%%%%%%%%%%%%%%%%%%%%%%%%%%%%%%%%%%%%%%%
   \left[\frac{\mathrm{d}\sigma}{\mathrm{d}T_\mathcal{N}}\right]_T^{\nu\mathcal{N}-{N_R}\mathcal{N}} &= 
   \dfrac{m_\mathcal{N}}{2J + 1} \dfrac{C_T^2}{(M_T^2 + 2 m_\mathcal{N} T_\mathcal{N})^2} \nonumber \\[4pt]
    &\times \Bigg\lbrace \left[\left(2 - \dfrac{m_\mathcal{N} T_\mathcal{N}}{E_\nu^2} - \dfrac{2 T_\mathcal{N}}{E_\nu} \right) + \dfrac{m_{N_R}^2}{2 E_\nu^2} \left(1 + \dfrac{3 E_\nu}{m_\mathcal{N}} + \dfrac{m_{N_R}^2}{m_\mathcal{N} T_\mathcal{N}}\right)\right] S_{00}^\mathcal{T}(\mathfrak{q}^2) \nonumber \\[4pt]
    &~~~~+ \left[\left(1 - \dfrac{T_\mathcal{N}}{E_\nu}\right) - \dfrac{m_{N_R}^2}{2 E_\nu^2} \left(1 + \dfrac{3 E_\nu}{2 m_\mathcal{N}} + \dfrac{m_{N_R}^2}{2 m_\mathcal{N} T_\mathcal{N}}\right)\right] S_{00}^\mathcal{L}(\mathfrak{q}^2) \Bigg\rbrace\,,
\end{align}
\end{subequations}
where $m_{N_R}$ is the SNL mass. For the axial vector and tensor interactions, non-zero contributions are expected only for Ar and $\mathrm{C_3F_8}$ detectors, as previously discussed. Unlike the SM and NGI cases, in this scenario the axial vector cross section receives non-negligible contributions from the longitudinal multipoles as well, as previously noted in Ref.~\cite{Candela:2024ljb}. Finally, the kinematics of this process remains essentially identical to that discussed for the sterile neutrino dipole portal scenario in Eqs.~\eqref{Eq:Sterile_kinematics_limit_on_m4} and \eqref{Eq:Sterile_kinematics_min_Ev}. One may notice that in the limit $m_{N_R} \to 0$ the NGI cross sections given in Eq.~\eqref{Eq.NGI_xSec} are recovered.

\section{Events simulation and statistical analysis}
\label{Sec:Events_simulation}
In this section, we detail the methodology  for accurately simulating the \cevns signal expected at the proposed ESS detectors. We furthermore  describe the statistical analysis procedures employed in this work for obtaining the attainable ESS sensitivities for  the various SM and BSM parameters of interest.

Following Ref.~\cite{Baxter:2019mcx},  we evaluate the expected \cevns events in each nuclear recoil energy bin, $i$, for different interaction channels $x \equiv\{\mathrm{SM},\mathrm{~BSM}\}$ through the expression
\begin{equation}
\label{eq:Nevents_CEvNS}
\begin{split}
\left[R_{\nu_\alpha \mathcal{N}}^\kappa\right]^i &= t_\mathrm{run}N_T\mathcal{A} \int_{T_\mathcal{N}^i}^{T_\mathcal{N}^{i+1}} \hspace{-0.3cm} \mathrm{d}T_\mathcal{N}^{\mathrm{reco}}\, \int_0^{T_\mathcal{N}^{\mathrm{max}}} \hspace{-0.4cm}\mathrm{d}T_\mathcal{N} \, G(T_\mathcal{N}^{\mathrm{reco}}, T_\mathcal{N}) \int_{E_\nu^{\mathrm{min}}(T_\mathcal{N})}^{E_\nu^{\mathrm{max}}} \mathrm{d}E_\nu \, \frac{\mathrm{d} \Phi_{\nu_\alpha}(E_\nu)}{\mathrm{d} E_\nu} \left[\frac{\mathrm{d}\sigma}{\mathrm{d}T_\mathcal{N}}\right]^\mathrm{CE\nu NS}_x\,,
\end{split}
\end{equation}
where $t_\mathrm{run}$ is the experimental run time, and $\mathcal{A}$ represents the detector efficiency which is  taken  to be flat $80$\% above the nuclear recoil threshold following Ref.~\cite{Baxter:2019mcx}. Moreover, $N_T = m_\mathrm{det} N_A / M$ is the number of target nuclei in the detector, where $m_\mathrm{det}$ is the detector mass, $N_A$ is Avogadro's number, and $M$ is the molar mass of the target. The variables $T_\mathcal{N}$ and $T_\mathcal{N}^\mathrm{reco}$ represent the true and measured nuclear recoil energies, respectively. The minimum neutrino energy \footnote{For the case of upscattering  processes, the kinematically allowed minimum neutrino energy is given in Eq.~\eqref{Eq:Sterile_kinematics_min_Ev}.} required to induce a recoil energy $T_\mathcal{N}$ is given by $E_\nu^\mathrm{min}\approx\sqrt{m_\mathcal{N}T_\mathcal{N}/2}$, and the maximum recoil energy is $T_\mathcal{N}^\text{max} \approx 2(E_\nu^\text{max})^2 / m_\mathcal{N}$, while the maximum neutrino energy from the ESS flux is $E_\nu^\mathrm{max} = m_\mu / 2 \approx 52.8$ MeV. The neutrino fluxes consist of a prompt $\nu_\mu$ beam from $\pi^+$ decay at rest, and a delayed $\bar{\nu}_\mu$ and $\nu_e$ beams from $\mu^+$ decay at rest. The differential neutrino energy spectra are given by the Michel spectra~\cite{Michel:1949qe, Bouchiat:1957zz}
\begin{equation}
\begin{aligned} 
\frac{\mathrm{d} \Phi_{\nu_\mu}(E_\nu)}{\mathrm{d} E_\nu} & = \eta \, \delta\left(E_\nu-\frac{m_{\pi}^{2}-m_{\mu}^{2}}{2 m_{\pi}}\right) \quad &(\text{prompt})\, , \\ 
\frac{\mathrm{d} \Phi_{\bar{\nu}_\mu}(E_\nu)}{\mathrm{d} E_\nu} & = \eta \frac{64 E^{2}_\nu}{m_{\mu}^{3}}\left(\frac{3}{4}-\frac{E_\nu}{m_{\mu}}\right) \quad &(\text{delayed})\, ,\\ 
\frac{\mathrm{d} \Phi_{\nu_e}(E_\nu)}{\mathrm{d} E_\nu} & = \eta \frac{192 E^{2}_\nu}{m_{\mu}^{3}}\left(\frac{1}{2}-\frac{E_\nu}{m_{\mu}}\right) \quad &(\text{delayed}) \, ,
\end{aligned}
\label{Eq.:SNS_Flux}
\end{equation}
where the flux normalization $\eta = rN_\mathrm{POT}/4\pi L^2$, with $L = 20$ m being the ESS baseline, $r = 0.3$ the neutrino yield per flavor per Proton On Target (POT), and $N_\mathrm{POT}=2.8\times 10^{23}$ the number of POT accumulated over one calendar year (208 effective days).

\begin{table}[t]
\newcommand{\mc}[3]{\multicolumn{#1}{#2}{#3}}
\newcommand{\mr}[3]{\multirow{#1}{#2}{#3}}
\centering
\begin{tabular}{|c|c|c|c|c|c|}
\hline
\hspace{0.05cm}Detector\hspace{0.05cm} & \hspace{0.05cm}$m_\mathrm{det}$ (kg)\hspace{0.05cm} & \hspace{0.1cm}$T_\mathcal{N}^\mathrm{th}$ ($\mathrm{keV}_{nr}$)\hspace{0.1cm} & \hspace{0.1cm}$T_\mathcal{N}^\mathrm{max}$ ($\mathrm{keV}_{nr}$)\hspace{0.1cm} & \hspace{0.1cm}$\sigma_0$ ($\mathrm{keV}_{nr}$)\hspace{0.1cm} & \hspace{0.5cm}Steady-state background\hspace{0.5cm} \\ 
\hline
\hline
CsI              & $22.5$       & $1$         & $46.1$         & $0.3$   & $10\mathrm{~~counts/keV}_{nr}\mathrm{/kg/day}$ \\ 
\hline
Xe               & $20$         & $0.9$       & $45.6$         & $0.36$  & $10\mathrm{~~counts/keV}_{nr}\mathrm{/kg/day}$ \\ 
\hline
Ge               & $7$          & $0.6$       & $78.9$         & $0.09$  & $3\mathrm{~~counts/keV}_{nr}\mathrm{/kg/day}$ \\ 
\hline
Si               & $1$          & $0.16$      & $212.9$        & $0.096$ & $0.04375\mathrm{~~counts/keV}_{nr}\mathrm{/kg/day}$ \\ 
\hline
Ar               & $10$         & $0.1$       & $150$          & $0.04$  & $0.1\mathrm{~~counts/kg/day}$ \\ 
\hline
$\rm C_3F_8$     & $10$         & $2$         & $329.6$        & $0.8$   & $0.1\mathrm{~~counts/kg/day}$ \\ 
\hline
\end{tabular}
\caption{Summary of the key parameters for different detectors proposed for \cevns measurements at the ESS. The table includes the target nucleus, detector mass, nuclear recoil energy threshold, maximum nuclear recoil energy, resolution power at $T_\mathcal{N}^\mathrm{th}$, and steady-state background counts. For CsI, Xe, Ge, and Si detectors, flat background rates are provided ($\mathrm{counts/keV}_{nr}\mathrm{/kg/day}$), whereas for Ar and $\rm C_3F_8$ detectors, total background counts/(kg-day) are provided. It should be noted that the background rates listed do not include the reduction factor of $4 \times 10^{-2}$ due to ESS duty cycle. The information presented here is adapted from Ref.~\cite{Baxter:2019mcx}.}
\label{Tab:detectors_details}
\end{table}

\begin{figure}[ht!]
    \centering
    \includegraphics[width=0.49\linewidth]{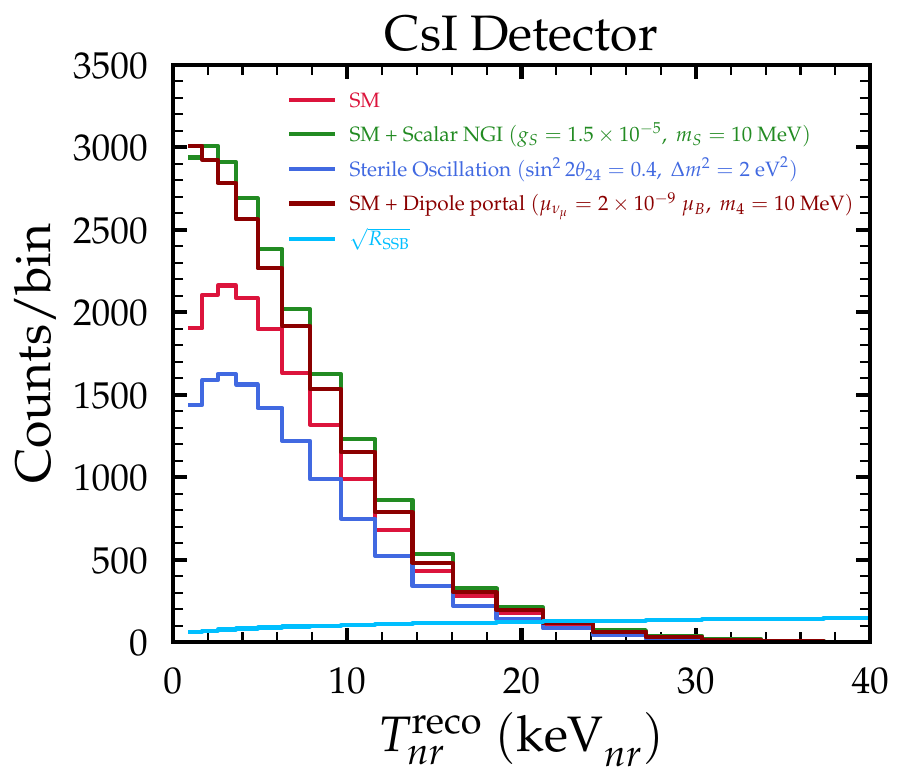}
    \includegraphics[width=0.49\linewidth]{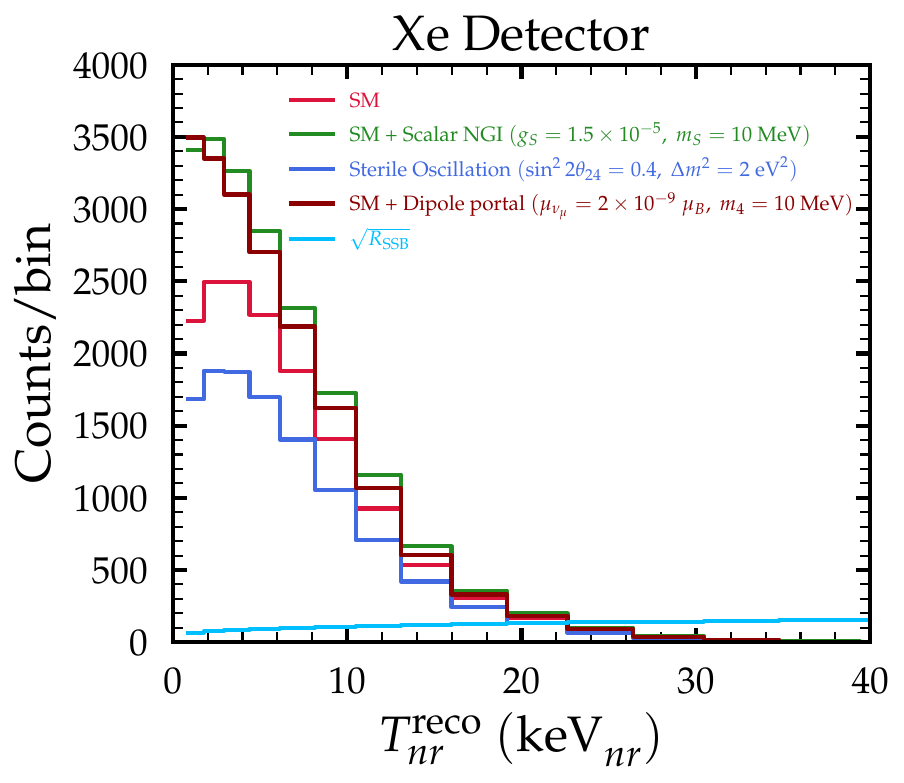}
    \includegraphics[width=0.49\linewidth]{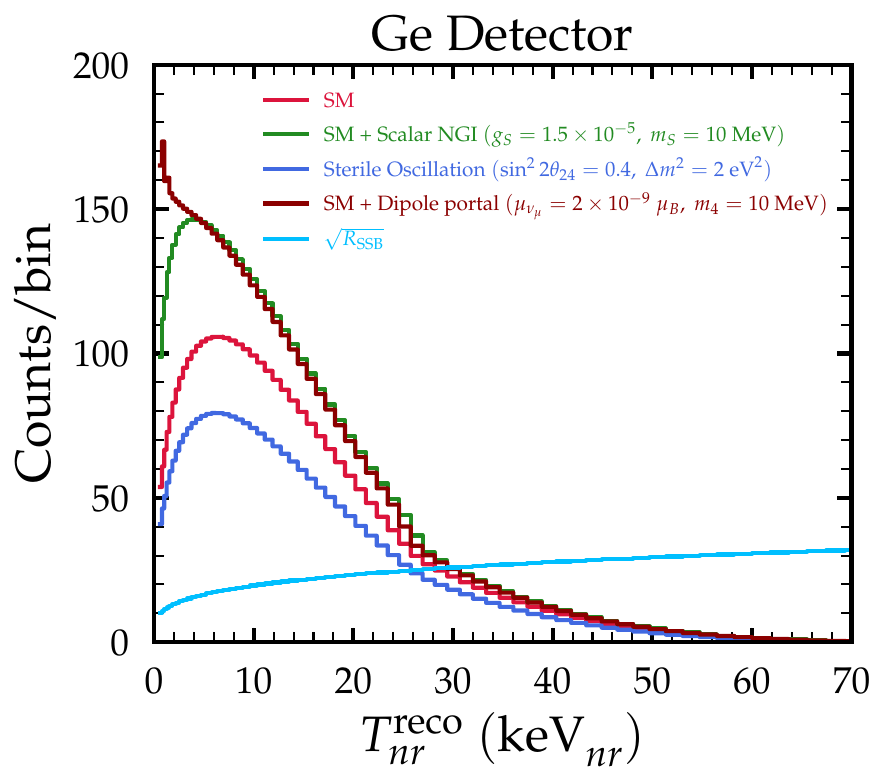}
    \includegraphics[width=0.49\linewidth]{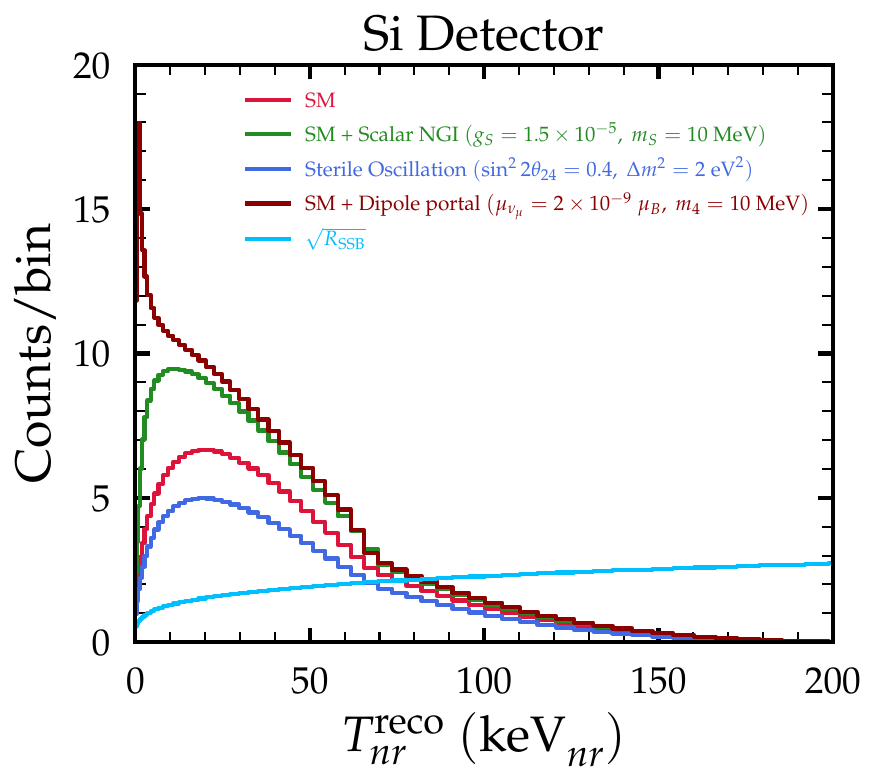}
    \caption{Expected signal and SSB event spectra as a function of the reconstructed nuclear recoil energy for CsI, Xe, Ge, and Si detectors. The results are obtained assuming 3 effective years of data taking time. The signal events are estimated for SM and various possible new physics contributions to the CE$\nu$NS. The cyan curve represents the square root of the expected SSB for visual clarity.}
    \label{fig:Events_Spectra}
\end{figure}

For realistic simulations, the event spectra are smeared using a Gaussian resolution function
\begin{equation}
    G(T_\mathcal{N}^{\mathrm{reco}}, T_\mathcal{N}) = \frac{1}{\sqrt{2\pi}~\sigma(T_\mathcal{N})} 
    \exp{\left(-\left[\frac{T_\mathcal{N}^{\mathrm{reco}} - T_\mathcal{N}}{\sqrt{2}~\sigma(T_\mathcal{N})}\right]^2 \right)}\,.
\end{equation}
The energy-dependent resolution power is parameterized as $\sigma(T_\mathcal{N}) = \sigma_0 \sqrt{T_\mathcal{N} / T_\mathcal{N}^\mathrm{th}}$, where $\sigma_0$ denotes the resolution power at the recoil energy threshold, $T_\mathcal{N}^\mathrm{th}$. The values of $\sigma_0$ for the various proposed detectors are provided in Ref.~\cite{Baxter:2019mcx} and are also listed in Table~\ref{Tab:detectors_details}.  As outlined in Ref.~\cite{Baxter:2019mcx}, the reconstructed nuclear recoil energy range is divided into bins for CsI, Xe, Ge, and Si detectors, with  the bin width taken to be twice the energy resolution at the bin center. However, for Ar and $\mathrm{C}_3\mathrm{F}_8$ detectors, no recoil energy binning is considered. In these cases, the total number of events is calculated by simply integrating the event rate between the threshold and the maximum nuclear recoil energies. 

The relevant specifications of the various proposed \cevns detectors at the ESS in Ref.~\cite{Baxter:2019mcx} are summarized in Table~\ref{Tab:detectors_details}, which includes information on detector-specific parameters such as mass, nuclear recoil thresholds, energy resolution, and background rates. These steady-state backgrounds (SSB), primarily arising from cosmic ray interactions, are anticipated to be the dominant background for \cevns searches at ESS, as discussed in Ref.~\cite{Baxter:2019mcx}. Other potential background sources, such as Beam-Related spallation Neutrons (BRN) and Neutrino-Induced Neutrons (NIN), are considered to be negligible due to their relatively small contributions.

In Fig.~\ref{fig:Events_Spectra}, we present the simulated event spectra for the CsI, Xe, Ge, and Si detectors, assuming 3  years of data taking  time.  The spectra show the number of signal events as a function of the reconstructed nuclear recoil energy, incorporating the SM prediction for \cevns as well as contributions from various new physics scenarios at future ESS. Notice that the SM case is found to be  in excellent agreement with Ref.~\cite{Chatterjee:2022mmu}. Additionally, the SSB is displayed for each detector with a cyan curve, by noting that for the sake of better visualization its square root shown.

For the spectral analysis, we employ the following Poissonian $\chi^2$ test statistic
\begin{equation}
\chi^2(\mathcal{S}) = 2\sum_{i} \left[R_\mathrm{th}^i(\mathcal{S};\alpha,\beta)- R_\mathrm{exp}^i+  R_\mathrm{exp}^i\ln \left(\frac{R_\mathrm{exp}^i}{R_\mathrm{th}^i(\mathcal{S}; \alpha,\beta)}\right) \right] + \left(\frac{\alpha}{\sigma_{\alpha}}\right)^2 + \left(\frac{\beta}{\sigma_{\beta}}\right)^2 \, ,
\label{Eq.:P_chi_2_Func}
\end{equation}
where $R_\mathrm{exp}^i$ represents the expected number of events in the $i$th energy bin, derived as the sum of the SM contributions from signal and SSB, i.e., $R_\mathrm{exp}^i = \sum_\alpha \left[R_{\nu_\alpha \mathcal{N}}^\mathrm{SM}\right]^i + \left[R_\mathrm{SSB}\right]^i$. The term $R_\mathrm{th}^i$ denotes the theoretically predicted events in the $i$th bin, incorporating both signal and background contributions as determined by the model under consideration. The latter quantity incorporates the nuisance parameters $\alpha$ and $\beta$, which account for signal and background systematic uncertainties, respectively, as
\begin{equation}
    R_\mathrm{th}^i(\mathcal{S}; \alpha, \beta) = (1 + \alpha)\sum_\alpha\left[R_{\nu_\alpha \mathcal{N}}^x \right]^i + (1 + \beta)\left[R_\mathrm{SSB}\right]^i \, .
\end{equation}
Here, $\mathcal{S}$ denotes the set of free parameters associated with the interaction channel $x$ being tested. As suggested in Ref.~\cite{Baxter:2019mcx}, a systematic uncertainty of $\sigma_\alpha = 10\%$ is adopted for the signal prediction, which accounts for cumulative uncertainties from neutrino flux estimation, nuclear form factors, and energy efficiency. Finally, the systematic uncertainty on background normalization is fixed to $\sigma_\beta = 1\%$ for all detectors considered in this analysis. For completeness, the effect of different detector-specific assumptions are explored in Appendix~\ref{Appendix_2}.

\section{Results}
\label{Sec:Results}
We now present the results on the projected ESS sensitivities considering 3 years of experimental run time for the various physics scenarios discussed in Sec.~\ref{Sec:Theory}. For visual clarity, whenever possible the latter are illustrated in the form of combined sensitivities, extracted from a simultaneous analysis of all the proposed ESS detectors. While it is not clear whether all the aforementioned detectors will be deployed at the ESS, in this work we intend to explore the full potential of future ESS measurements given the provided information~\cite{Baxter:2019mcx}.  For completeness, the projected sensitivities obtained for the individual detectors are shown in Appendix~\ref{Appendix_1}. Let us also stress that, as can be seen from the Appendix, for most of the cases the constraints coming out from the individual detector analyses are quite similar, hence the combined analysis presented below is not driven by a particular nuclear target.

\begin{figure}[t!]
    \centering
    \includegraphics[width=0.49\linewidth]{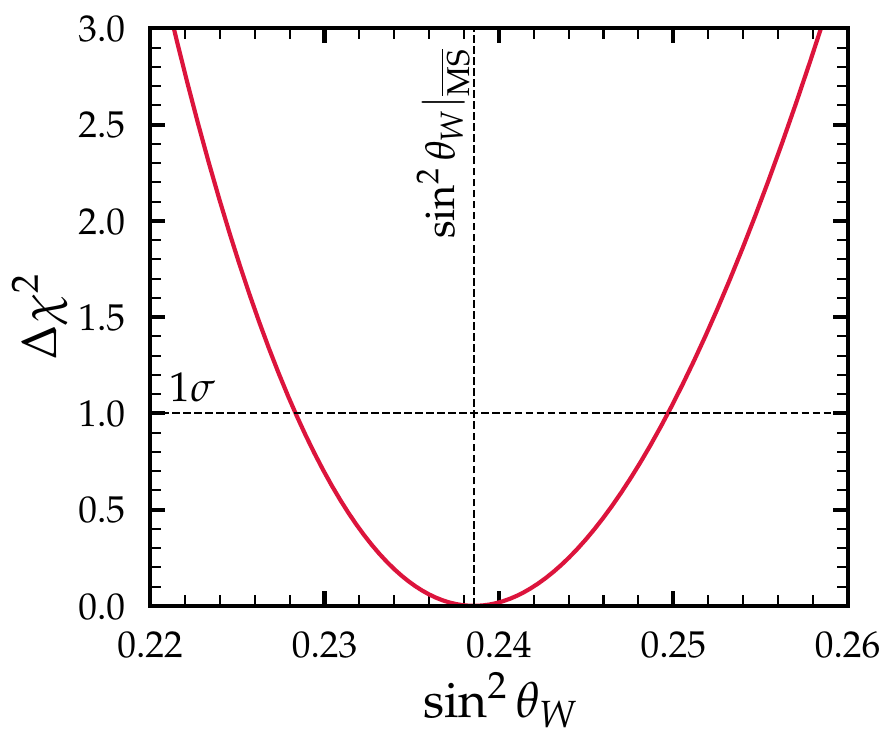}
    \caption{Projected sensitivity on the weak mixing angle from the combined analysis of the proposed ESS detectors.}
    \label{fig:sw2_Plot}
\end{figure}

\begin{figure}[ht!]
    \centering
    \includegraphics[width=0.8\linewidth]{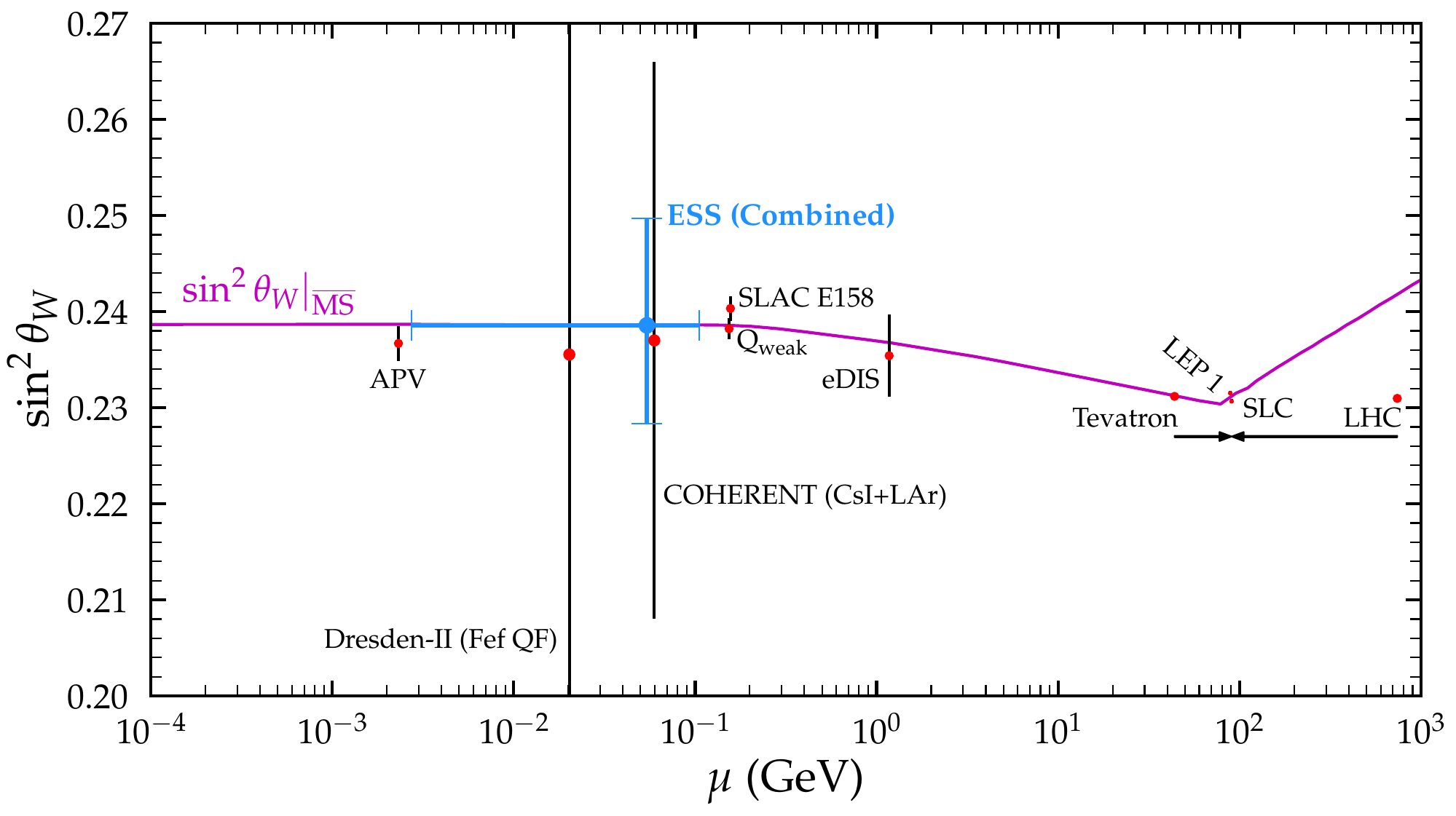}
    \caption{$\sin^2{\theta_W}$ evolution within the SM, depicted by the magenta solid line, under the $\overline{\mathrm{MS}}$ renormalization scheme as a function of the renormalization scale ($\mu$). The blue error bar denotes the $1\sigma$ determinations of $\sin^2{\theta_W}$ from the combined analysis of all detectors considered in this study. Other measurements from various experiments across different renormalization scales are also included for comparison. Notably for visual clarity, the Tevatron and LHC points are slightly shifted horizontally from the $Z$-pole mass scale.}
    \label{fig:sw2_Comparison_Plot}
\end{figure}

We begin by exploring the projected constraints on the SM weak mixing angle. With its high-intensity neutrino flux and potential for large statistical samples,  ESS is a favorable facility to constrain the weak mixing angle in the low-energy regime. The  $\Delta\chi^2$ profile of $\sin^2{\theta_W}$ from the combined analysis of all ESS detectors considered in this study is shown in Fig.~\ref{fig:sw2_Plot}, while the corresponding $1\sigma$ determination is found to be\footnote{The value of the RGE extrapolated weak mixing angle which coincides with our best fit point, has been rounded to match the number of significant figures.}

\begin{equation*}
        \sin^2{\theta_W} = 0.239{^{+0.011}_{-0.010}}\,.
\end{equation*}
The  individual  $1\sigma$ projections on $\sin^2{\theta_W}$ coming out from the analysis of  the different detectors are provided in Table~\ref{tab:sw2_values_for_different_detectors} in Appendix~\ref{Appendix_1}. As expected, the extracted best fit value matches exactly with the RGE-extrapolated value of  $\sin^2{\theta_W}$ in the low-energy regime. This is expected by recalling that in our performed statistical analysis when treating the weak mixing angle as a free parameter, we compare the corresponding predicted events with the expected SM events. Since the latter is evaluated by fixing the weak mixing angle to the value predicted by the low-energy RGE extrapolation,  we consistently obtain this to be the best fit point. In Fig.~\ref{fig:sw2_Comparison_Plot}, we compare our results with determinations from other probes~\cite{Majumdar:2022nby, DeRomeri:2022twg, MammenAbraham:2023psg, ParticleDataGroup:2024cfk} across a wide range of energies. While the $1\sigma$ uncertainty  of $\sin^2{\theta_W}$ in the low-energy regime will not be able to compete with other precision experiments such as atomic parity violation (APV),  the complementarity with other \cevns measurements is particularly noteworthy. Indeed, due to the high statistics  anticipated from the ESS facility, its sensitivity on $\sin^2{\theta_W}$ is expected to surpass that of current CE$\nu$NS-based measurements. For instance, our present results imply that the uncertainty will be reduced by $\sim 60$\%  with respect to the COHERENT limit~\cite{DeRomeri:2022twg} and $\sim 80\%$  compared to Dresden-II~\cite{AtzoriCorona:2022qrf, Majumdar:2022nby}. Further improvements can be achieved by combining the future ESS sensitivities presented here with APV data, as recently done in Refs.~\cite{Cadeddu:2021ijh,AtzoriCorona:2023ktl} for the case of COHERENT.

The  nuclear rms radius is another crucial SM parameter of the \cevns cross section which enters through the nuclear form factor (see Eqs.~\eqref{equn:CEvNS_SM_Lagrangian},~\eqref{Eq:Helm_FF} and~\eqref{eq:exact_Qw} for reference). From the theoretical perspective, the estimation of this parameter is heavily dependent on the nuclear structure model~\cite{Papoulias:2018uzy, Hoferichter:2020osn, Sahu:2020kwh, Kosmas:2021zve}. While the nuclear proton rms radius $(R_p)$ has been measured experimentally with high precision~\cite{DeVries:1987atn}, the corresponding nuclear neutron rms radius $(R_n)$ is yet poorly constrained. \cevns being a purely neutral-current process offers a valuable probe for a precise determination of $R_n$~\cite{Patton:2012jr,Papoulias:2015vxa,Papoulias:2019lfi,Sierra:2023pnf}.  In this study, different rms radii are considered for protons and neutrons, i.e., we take $\mathcal{F}_p(\mathfrak{q}^2)\neq \mathcal{F}_n(\mathfrak{q}^2)$ and rely on Eq.~\eqref{eq:exact_Qw}. For each detector, the theoretical event rate in Eq.~\eqref{Eq.:P_chi_2_Func} is calculated by fixing $R_p$ according to Table~\ref{Tab:Detectors_Specs}, while $R_n$ is treated as a free parameter. Instead, the expected rate is calculated by fixing $R_p$ as explained previously and by also taking $R_n=1.05R_p$ as done in Ref.~\cite{Coloma:2020nhf}.

\begin{figure}[t!]
    \centering
    \includegraphics[width=0.32\linewidth]{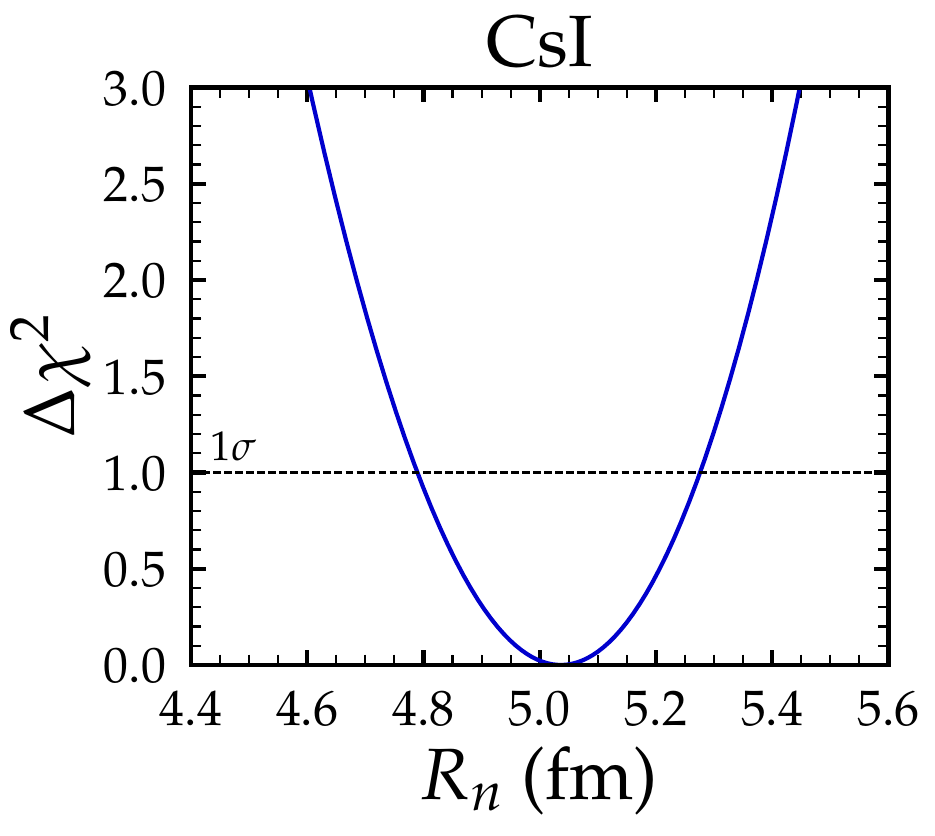}
    \includegraphics[width=0.32\linewidth]{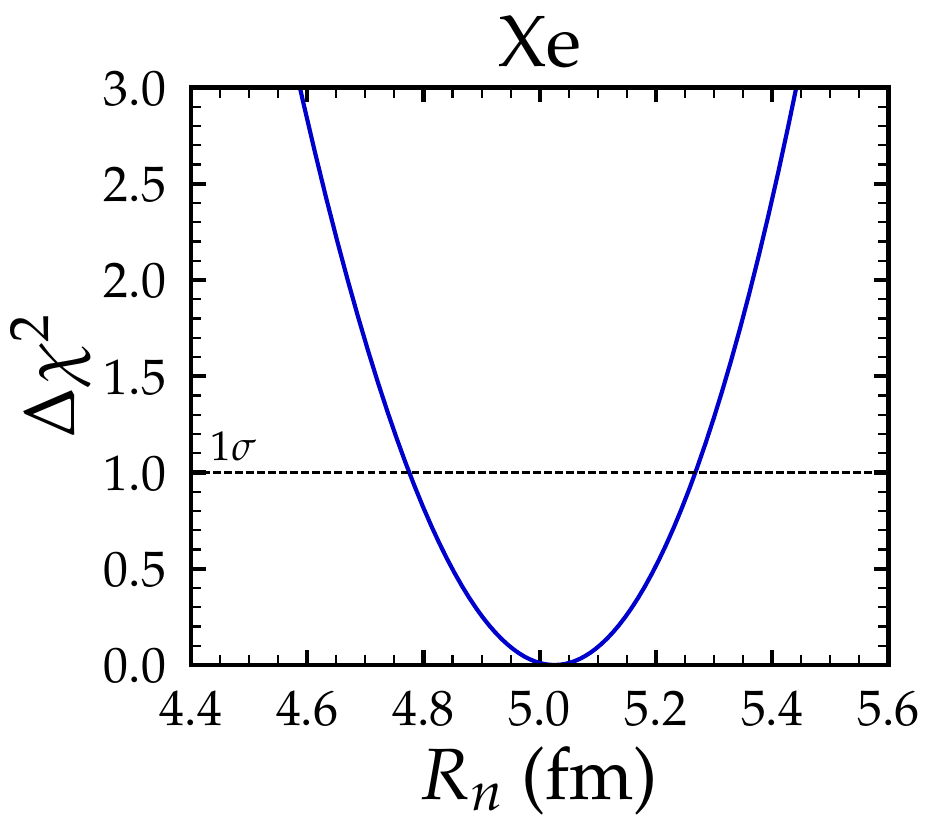}
    \includegraphics[width=0.32\linewidth]{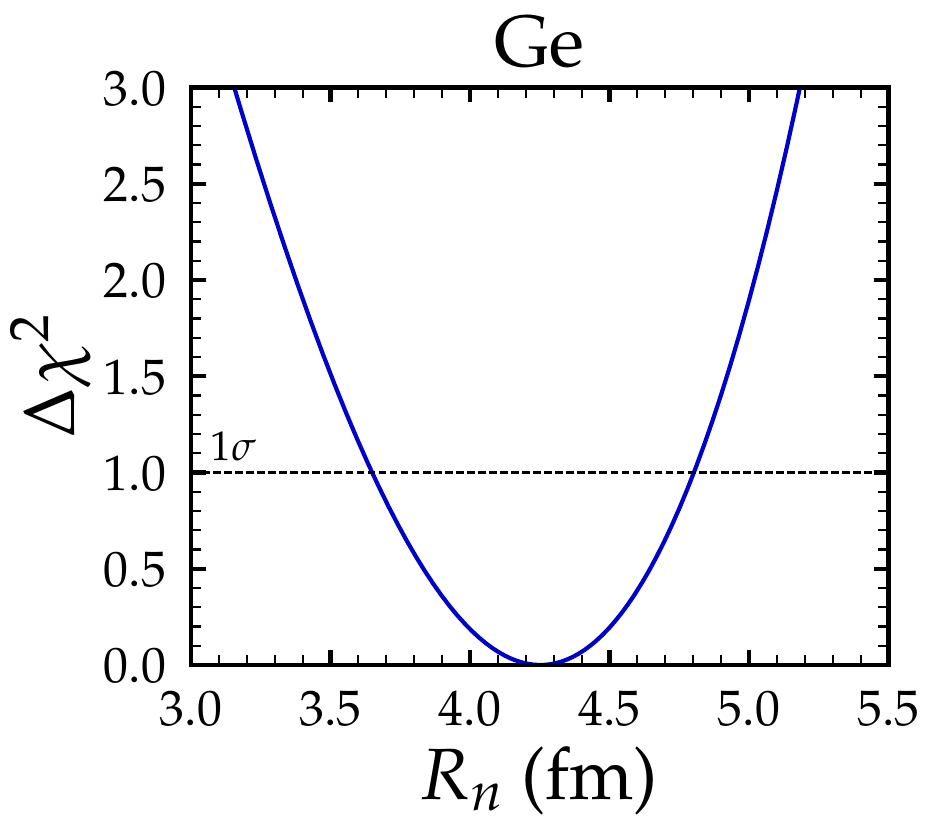}
    \includegraphics[width=0.32\linewidth]{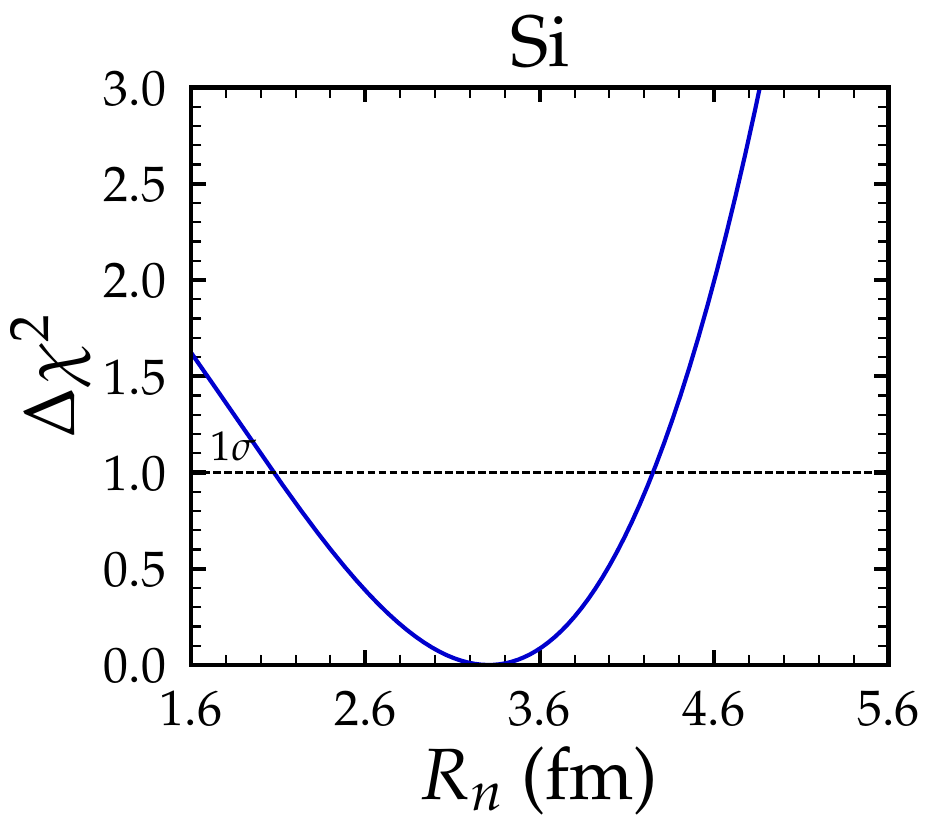}
    \includegraphics[width=0.32\linewidth]{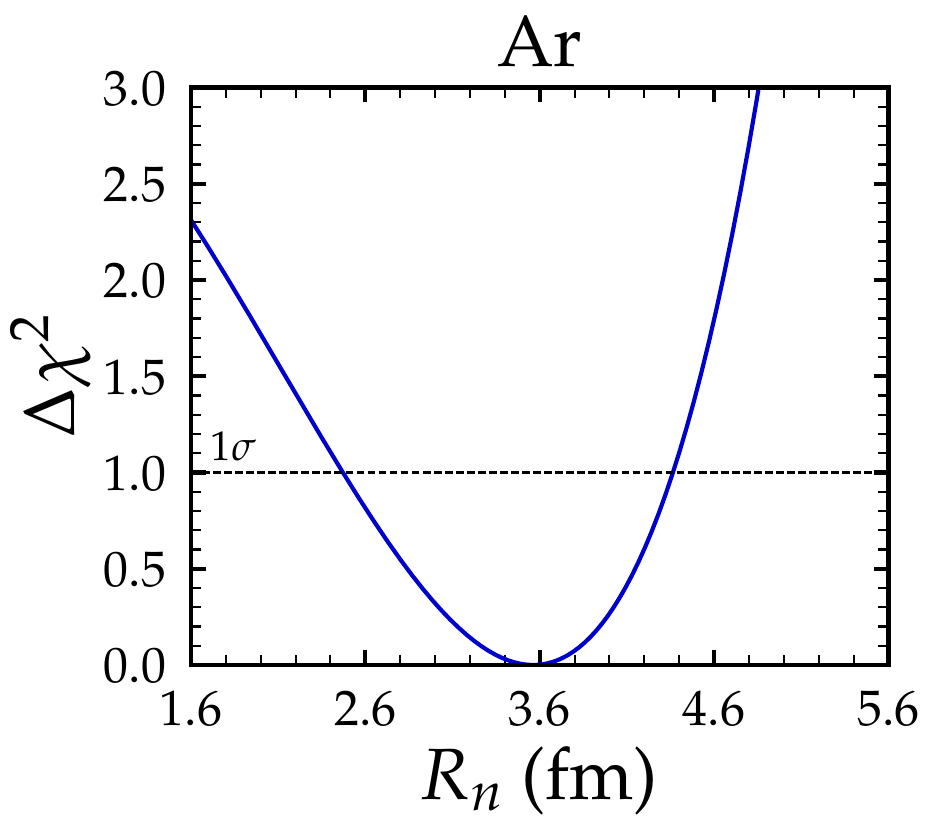}
    \includegraphics[width=0.32\linewidth]{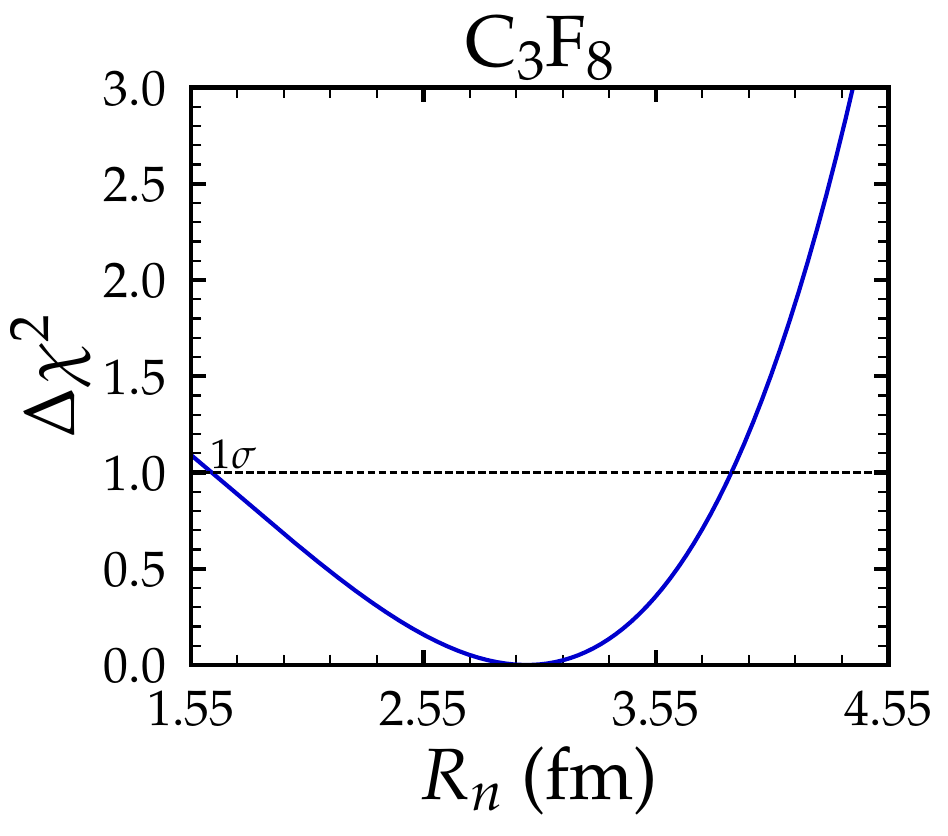}
    \caption{Projected sensitivity on the neutron rms radii of different detectors at the ESS.}
    \label{Fig:Rn_Sensitivity}
\end{figure}

\begin{table}[t!]
\centering
\begin{tabular}{|c|c|c|c|c|c|c|}
\hline
\hspace{0.5cm}\textbf{Detector}\hspace{0.5cm} & \hspace{0.7cm}CsI\hspace{0.7cm} & \hspace{0.7cm}Xe\hspace{0.7cm} & \hspace{0.7cm}Ge\hspace{0.7cm} & \hspace{0.7cm}Si\hspace{0.7cm} & \hspace{0.7cm}Ar\hspace{0.7cm} & \hspace{0.7cm}$\mathrm{C_3F_8}$\hspace{0.7cm} \\ 
\hline
 & & & & & & \\
$R_n$ (fm) & [4.79, 5.28] & [4.78, 5.27] & [3.65, 4.80] & [2.08, 4.25] & [2.47, 4.36] & [1.64, 3.88] \\
 & & & & & & \\
\hline
\end{tabular}
\caption{1$\sigma$ range of neutron rms radius $R_n$ for different detectors.}
\label{tab:rn_ranges}
\end{table}

Since a combined analysis has no meaning for $R_n$, in Fig.~\ref{Fig:Rn_Sensitivity} we show the projected sensitivities for the different proposed target nuclei at the future ESS experiment.  Let us note that for the CsI and $\mathrm{C_3F_8}$ nuclei, $R_n$ refers to the average rms radius. The corresponding  $1\sigma$ ranges  are summarized in Table~\ref{tab:rn_ranges}. Comparing with bounds placed from the analysis of available \cevns data, we conclude that a significant improvement is expected to be reached by the ESS. In particular,  for CsI  (Ar) we find an improvement of $\sim 40\%$   compared to an existing analysis of COHERENT-CsI~\cite{DeRomeri:2022twg}. Notice that the analysis of COHERENT-LAr data~\cite{Miranda:2020tif, DeRomeri:2022twg} yielded an upper limit only, thus the future ESS measurement will provide valuable new information. For the case of germanium, the first COHERENT data~\cite{Adamski:2024yqt} have low statistics, while existing data from the Dresden-II and CONUS+ reactor experiments ---for which the momentum transfer is rather low--- are less sensitive to nuclear physics effects. Finally, for Xe target nuclei utilized by the direct dark matter detection experiments, while the momentum transfer is not as low as in the reactor experiments, the statistics is still poor to provide a strong constraint. 

\begin{figure}[t!]
    \centering
    \includegraphics[width=0.32\linewidth]{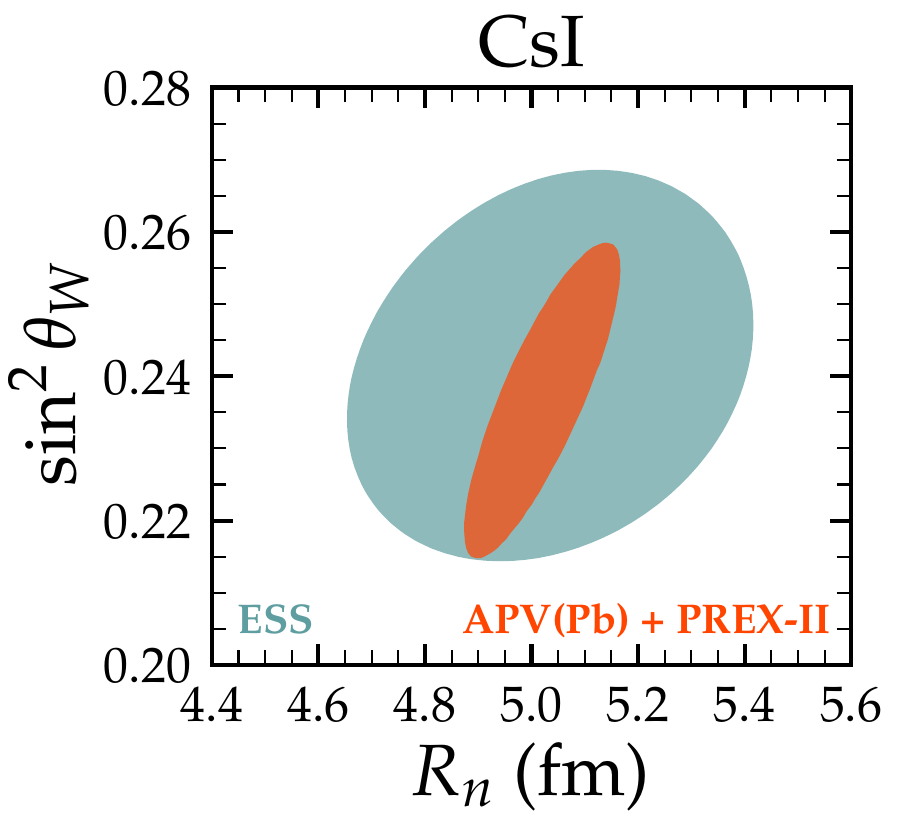}
    \includegraphics[width=0.32\linewidth]{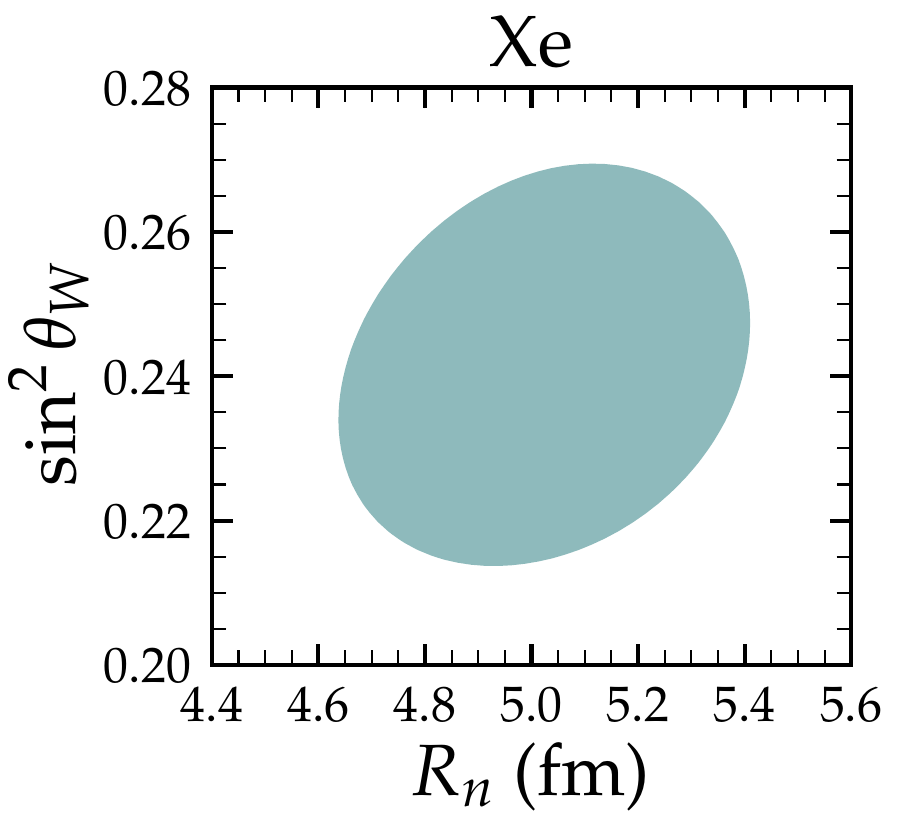}
    \includegraphics[width=0.32\linewidth]{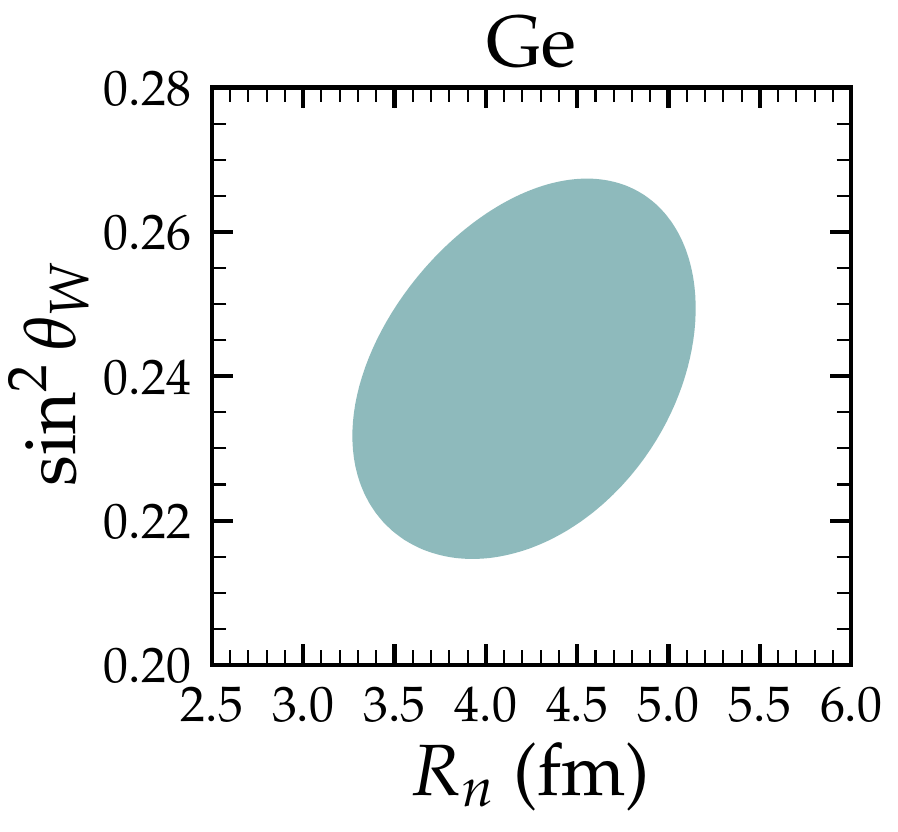}
    \includegraphics[width=0.32\linewidth]{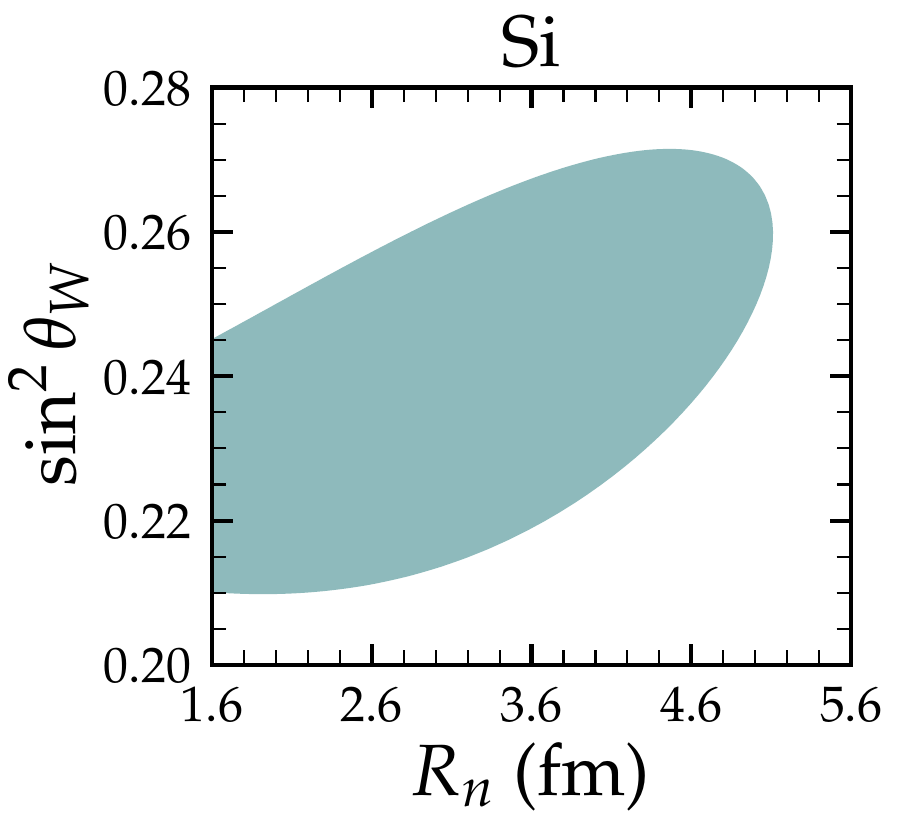}
    \includegraphics[width=0.32\linewidth]{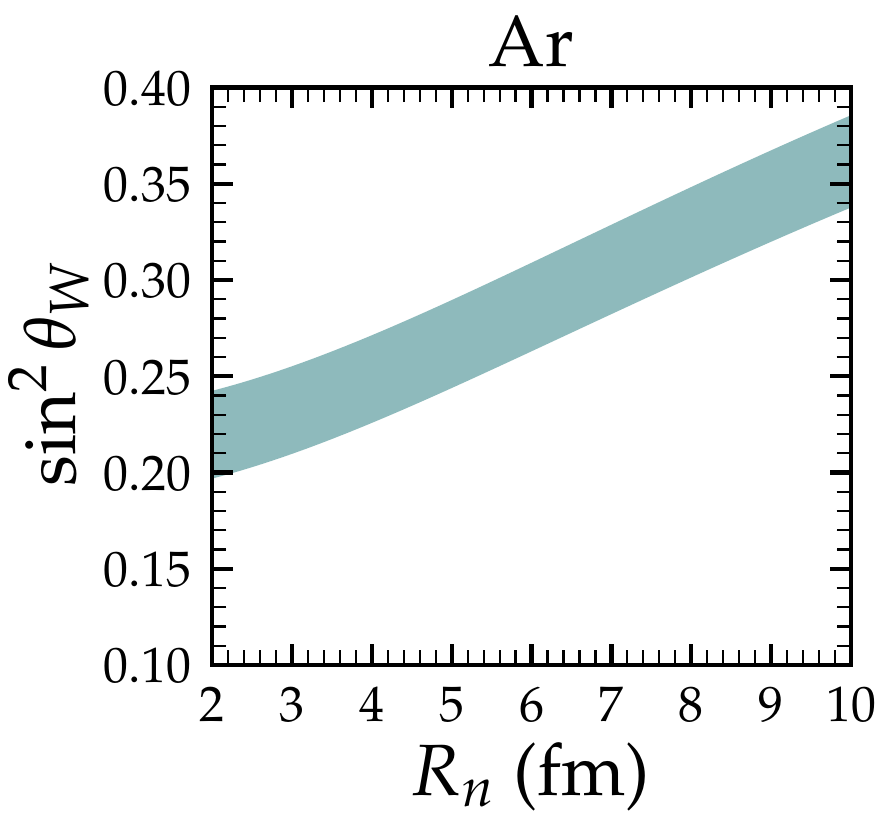}
    \includegraphics[width=0.32\linewidth]{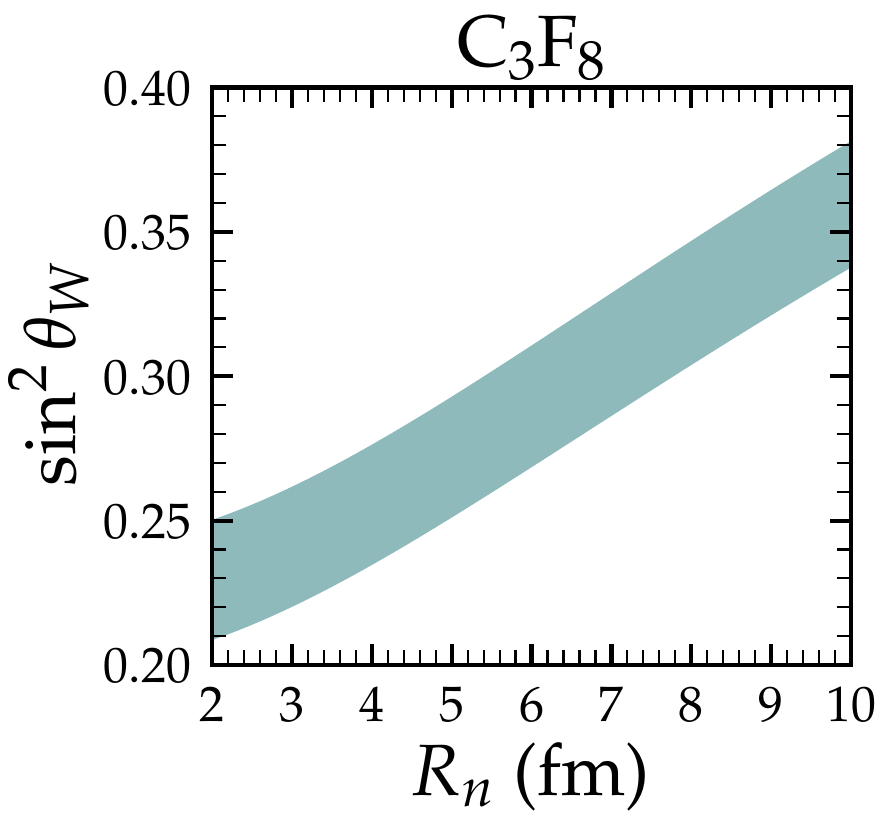}
    \caption{Projected sensitivity in the $(R_n,~\sin^2{\theta_W})$ plane for the various proposed ESS detectors. The shaded teal region represents the allowed $1\sigma$ region. The orange area corresponding to the APV(Pb) + PREX-II analysis of Ref.~\cite{AtzoriCorona:2024vhj} is also shown for comparison.}
    \label{Fig:Rn_vs_sW2_contour}
\end{figure}

Finally, we perform a simultaneous fit allowing both $\sin^2{\theta_W}$ and $R_n$ to vary. Figure~\ref{Fig:Rn_vs_sW2_contour} displays the projected $1\sigma$ sensitivity in the $(R_n,~\sin^2{\theta_W})$ plane for different detectors. The different panels provide a comprehensive understanding of the interplay between the weak mixing angle and the  neutron nuclear rms radius, highlighting the capability of ESS to simultaneously constrain these parameters.  From the various plots one can see that the use of heavier target nuclei leads to an enhanced sensitivity on the nuclear neutron rms radius. This  is expected since for the latter cases the effect of the nuclear form factor becomes more pronounced at lower recoil energies, leading to spectral features. Before closing this discussion, let us also mention that the band-shaped regions found for Ar and $\mathrm{C_3F_8}$ detectors are due to the fact that a single-bin analysis is done for these cases. For the case of CsI, the ESS results will improve previous limits placed by the analysis of COHERENT-only data~\cite{Cadeddu:2021ijh,AtzoriCorona:2023ktl}. However, we must note that in a recent work~\cite{AtzoriCorona:2024vhj} a global fit of all available electroweak data was performed, combining \cevns with APV (on Cs and Pb) and PREX-II data. This led to a dramatic improvement in the simultaneous determination of both $\sin^2 \theta_W$ and $R_n$, as shown in the upper-left panel of Fig.~\ref{Fig:Rn_vs_sW2_contour}. A similar analysis, including future \cevns data from the ESS will offer further improvement. 

\begin{figure}[t!]
    \centering
    \includegraphics[width=0.49\linewidth]{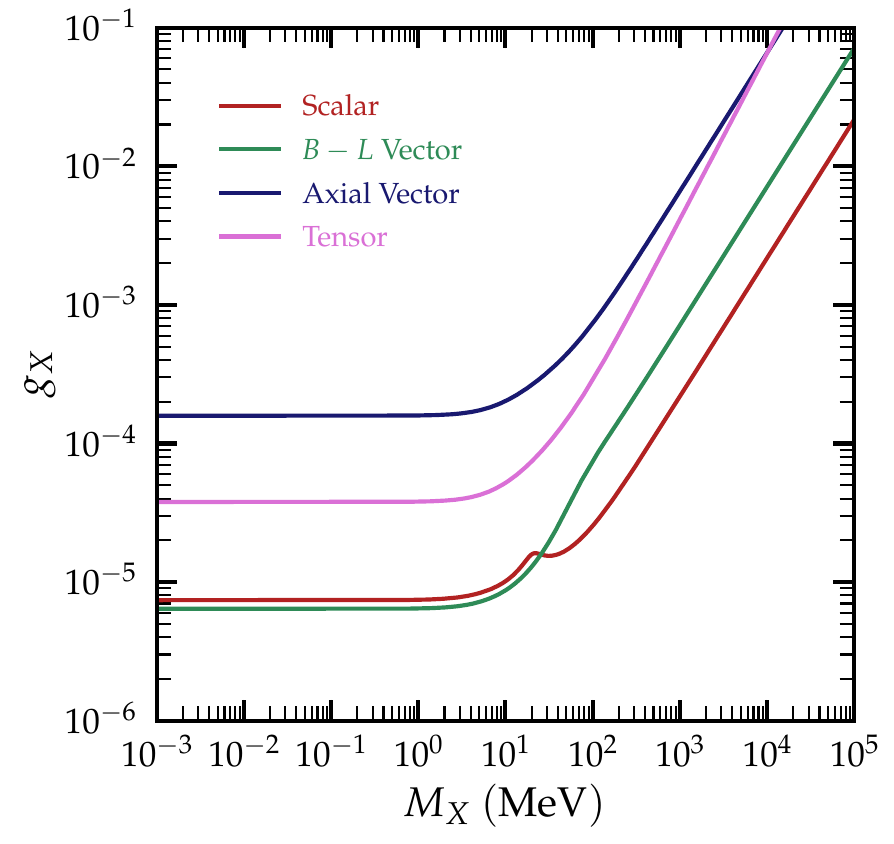}
    \caption{Projected sensitivities at 90\% C.L. for the various $X = \{S, V, A, T\}$ interactions derived from a combined analysis of all proposed detectors at the ESS. }
    \label{Fig:NGI_combined_contour}
\end{figure}

We now explore the prospect of constraining several BSM physics scenarios at the ESS via the \cevns channel. We begin our discussion by focusing on NGIs. In Fig.~\ref{Fig:NGI_combined_contour}, we present the 90\% C.L. projected sensitivities in the $(M_X, \textsl{g}_X)$ parameter plane, obtained from a combined analysis of all proposed detectors. Our purpose here is to highlight the relative strength of the constraints for the different interaction channels. As can be seen, the scalar and vector interactions are expected to yield the most stringent limits, while the nuclear spin-suppressed axial vector interaction will  be the least constrained. The projections concerning the various NGIs $X = \{S, V, A, T\}$ derived for each ESS detector individually, are demonstrated in Appendix~\ref{Appendix_1}.

\begin{figure}[t!]
    \centering
    \includegraphics[width=0.49\linewidth]{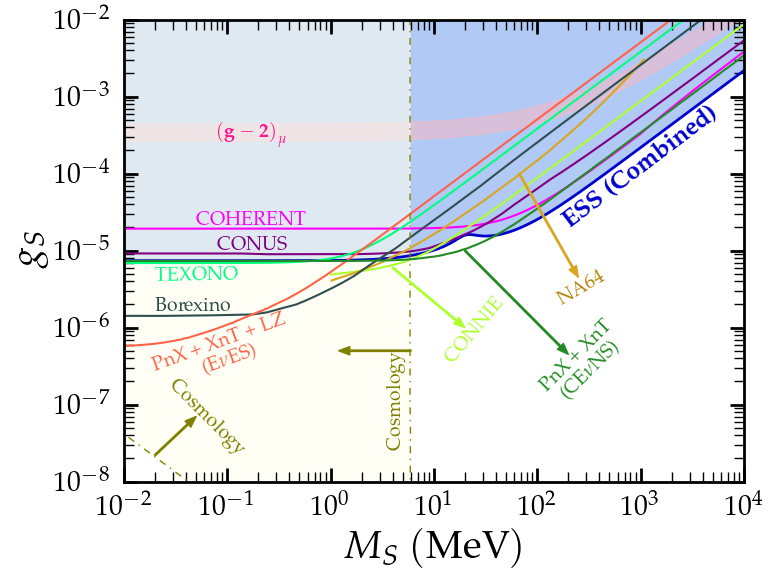}
    \includegraphics[width=0.49\linewidth]{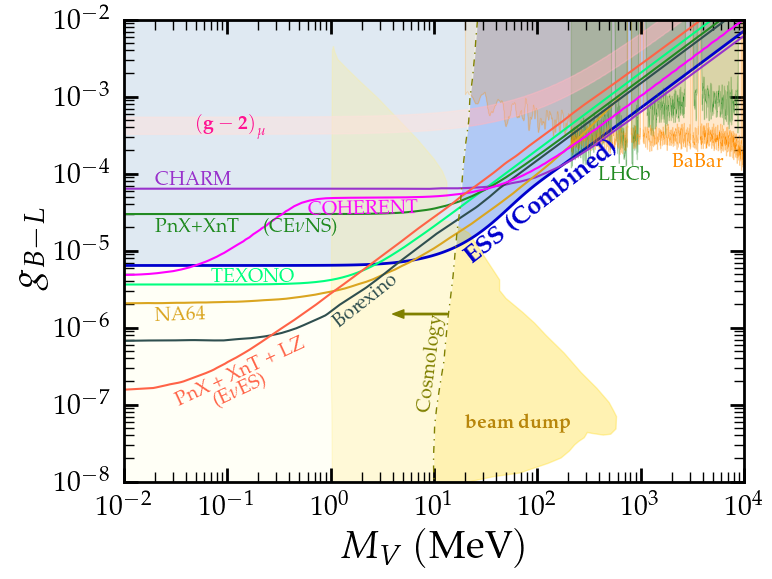}
    \includegraphics[width=0.49\linewidth]{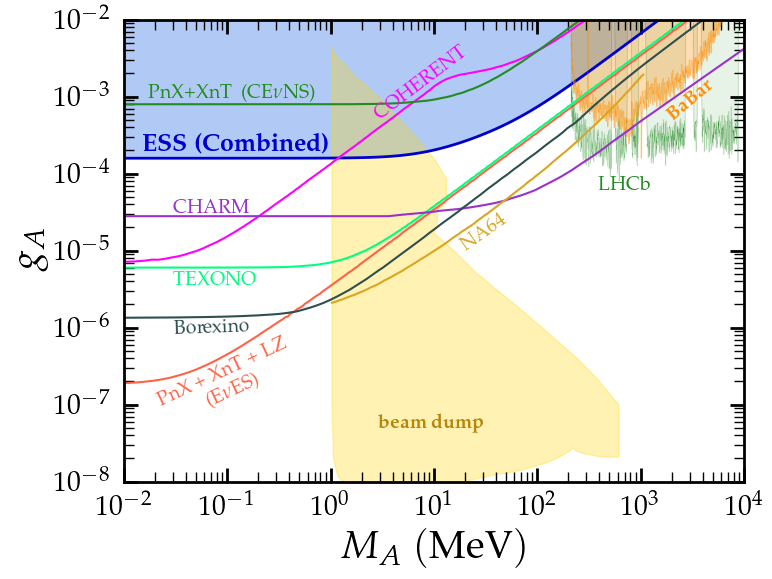}
    \includegraphics[width=0.49\linewidth]{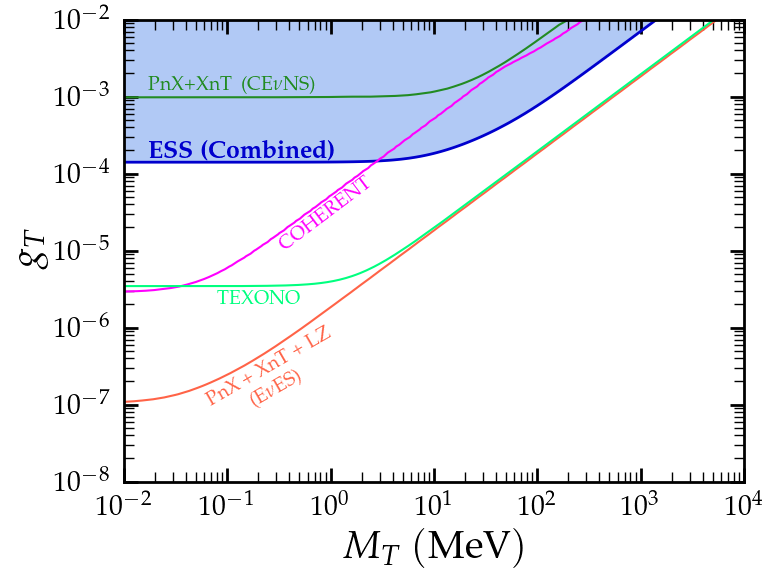}    
    \caption{Projected sensitivities at 90\% C.L. in the $(M_X, \textsl{g}_X)$ parameter space for scalar, vector $B-L$, axial vector, and tensor NGIs, derived from the combined analysis of all the proposed ESS detectors (blue contours). Existing constraints from other experimental and astrophysical probes are included for comparison (see main text for more details).}
    \label{Fig:NGI_limit_comparison}
\end{figure}

It is interesting to examine the  complementarity of the projected ESS limits obtained in the present work, with existing constraints in the literature from various experimental and astrophysical sources. Figure~\ref{Fig:NGI_limit_comparison} overlays the ESS sensitivities (blue contours) with constraints derived from available \cevns data, in particular, from a combined analysis of COHERENT CsI and LAr data\footnote{The  constraints incorporate both \cevns and \eves events for COHERENT-CsI data, whereas only \cevns events are considered for COHERENT-LAr data analysis. For further details see Ref.~\cite{DeRomeri:2022twg}.}~\cite{DeRomeri:2022twg, AtzoriCorona:2022moj, Majumdar:2024dms} as well as from a combined analysis of the recently measured solar $^8\mathrm{B}$ neutrino-induced \cevns signals of PandaX-4T and XENONnT~\cite{DeRomeri:2024iaw, Blanco-Mas:2024ale}. We also include constraints from E$\nu$ES-based analyses of data available from Borexino~\cite{Coloma:2022avw}, CHARM-II~\cite{CHARM-II:1994wkf},  TEXONO~\cite{Bilmis:2015lja, Majumdar:2024dms}, and from a combined analysis of PandaX-4T, XENONnT, and LZ  data~\cite{A:2022acy, DeRomeri:2024dbv, Majumdar:2024dms}. Limits from the invisible decay of novel bosons at NA64~\cite{NA64:2021xzo, NA64:2022yly} are also included. We furthermore depict limits from dark photon searches at fixed target electron beam-dump experiments (such as CHARM~\cite{CHARM:1985anb,Gninenko:2012eq}, NA64~\cite{NA64:2016oww,NA64:2019auh, NA64:2023wbi}, NOMAD~\cite{NOMAD:2001eyx}, E141~\cite{Riordan:1987aw,Bjorken:2009mm}, E137~\cite{Bjorken:1988as,Andreas:2012mt}, E774~\cite{Bross:1989mp}, KEK~\cite{Konaka:1986cb}, Orsay~\cite{Andreas:2012mt}, U70/$\nu$-CAL~I~\cite{Blumlein:2011mv,Blumlein:2013cua}, APEX~\cite{APEX:2011dww}, etc.) and high-energy collider experiments (e.g., BaBar~\cite{BaBar:2014zli,BaBar:2017tiz} and LHCb~\cite{LHCb:2019vmc}). These dark photon search limits have been recast into the relevant parameters using the \href{https://gitlab.com/philten/darkcast}{\texttt{darkcast}} software package, as detailed in Refs.~\cite{Ilten:2018crw, Baruch:2022esd}. Additionally, constraints from astrophysical observations, such as cosmological parameters like $N_\mathrm{eff}$~\cite{Esseili:2023ldf, Li:2023puz, Ghosh:2024cxi}, which also incorporate Big Bang Nucleosynthesis (BBN) limits~\cite{Blinov:2019gcj,Suliga:2020jfa}, are superimposed. 
 
From the present analysis we conclude that future CE$\nu$NS measurements at the ESS will substantially advance the sensitivity to NGIs, complementing and extending existing bounds from diverse experimental and observational datasets. Specifically, for the case of scalar and vector $B-L$ interactions, the ESS will dominate the constraints for $M_S >40$~MeV and $25<M_V<200$~MeV, respectively. Notably these regions are not in conflict with bounds from cosmology. On the other hand, for the spin-dependent axial vector and tensor interactions, the ESS will not be able to compete with existing constraints obtained from E$\nu$ES analyses. However, this comparison is valid only under the assumption of universal couplings between the novel mediators with the quarks and leptons. Focusing on constraints solely from CE$\nu$NS, the ESS has the prospect to  surpass all the CE$\nu$NS-based constraints for all the interactions\footnote{Let us remind that the COHERENT constraints reported in Refs.~\cite{DeRomeri:2022twg, DeRomeri:2024iaw} includes also \eves data, which drive the depicted constraints to lower couplings for low mediator masses.}.

\begin{figure}[t!]
    \centering
    \includegraphics[width=0.49\linewidth]{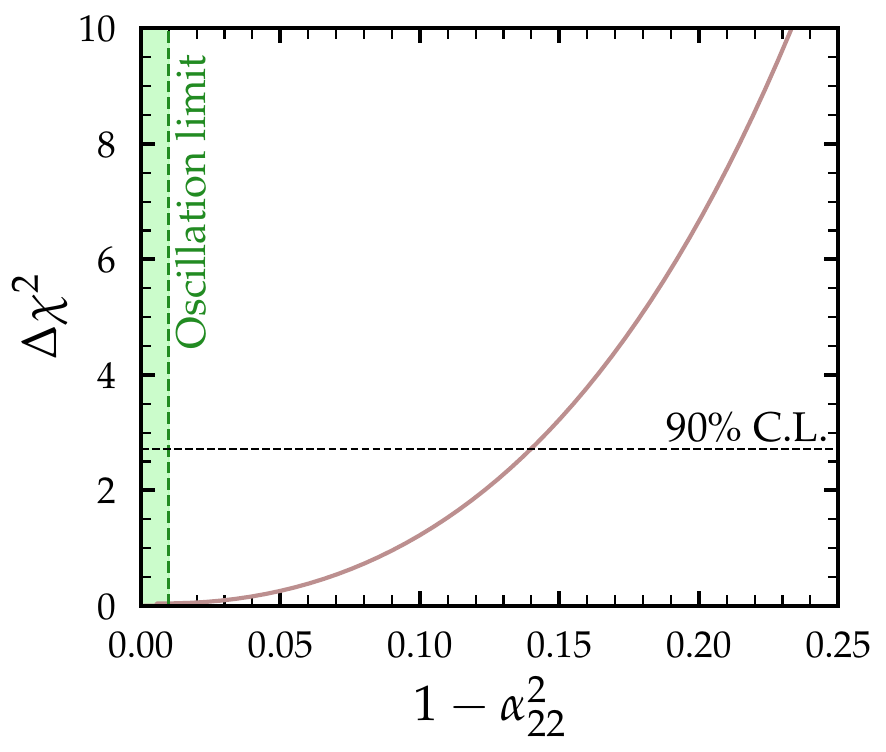}
    \caption{Projected sensitivity on the NU parameter $\alpha_{22}$ from the combined analysis of all the proposed ESS detectors considered in this study. For comparison, the corresponding 90\% C.L. constraint derived from neutrino oscillation global fits is also shown.}
    \label{Fig:NU_Sensitivity}
\end{figure}

At this point we  consider the potential to probe unitarity violation effects in the neutrino mixing matrix at ESS. As discussed in Sec.~\ref{SubSec:Violation_of_lepton_unitarity}, for the case of ESS the relevant NU parameters are $\alpha_{11}$ and $\alpha_{22}$. However, since the different components of SM events spectra, $R_{\nu_\mu \mathcal{N}}^\mathrm{SM}$, $R_{\bar{\nu}_\mu \mathcal{N}}^\mathrm{SM}$, and $R_{\nu_e \mathcal{N}}^\mathrm{SM}$, are approximately equal, using Eq.~\eqref{equn:NU_Oscillation_Probability} the ratio of the total number of events can be expressed as $\sum_\alpha R_{\nu_\alpha \mathcal{N}}^\mathrm{NU} / \sum_\alpha R_{\nu_\alpha \mathcal{N}}^\mathrm{SM} \approx 3\alpha_{22}^2$.  As a result, the total number of events is predominantly sensitive to $\alpha_{22}$, rendering ESS incapable of severely constraining $\alpha_{11}$. 
In Fig.~\ref{Fig:NU_Sensitivity}, we present the $\Delta\chi^2$ profile of $1-\alpha_{22}^2 $ from the combined analysis of all detectors considered in this study. As for the previously studied cases, the individual detector sensitivities are shown in Appendix~\ref{Appendix_1}. From the combined analysis, the projected $90$\% C.L. sensitivity on $1-\alpha_{22}^2 $ is determined to be
\begin{equation*}
    1-\alpha_{22}^2 < 0.14\,.
\end{equation*}
For comparison, Fig.~\ref{Fig:NU_Sensitivity} also includes the constraints on $\alpha_{22}$ derived from a global fit of neutrino oscillation data~\cite{Forero:2021azc}. It becomes evident that given the considered experimental setup, the sensitivities projected for the ESS are not expected to be competitive with those of large-scale oscillation experiments.

\begin{figure}[t!]
    \centering
    \includegraphics[width=0.49\linewidth]{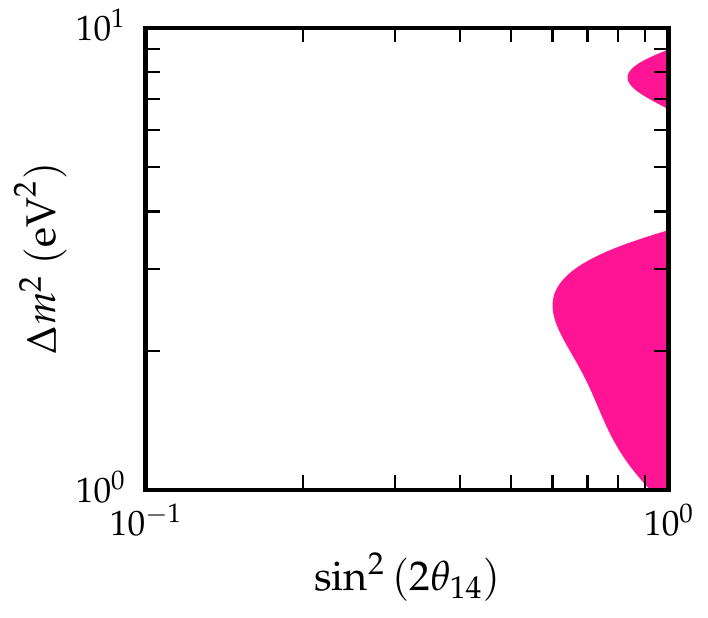}
    \includegraphics[width=0.49\linewidth]{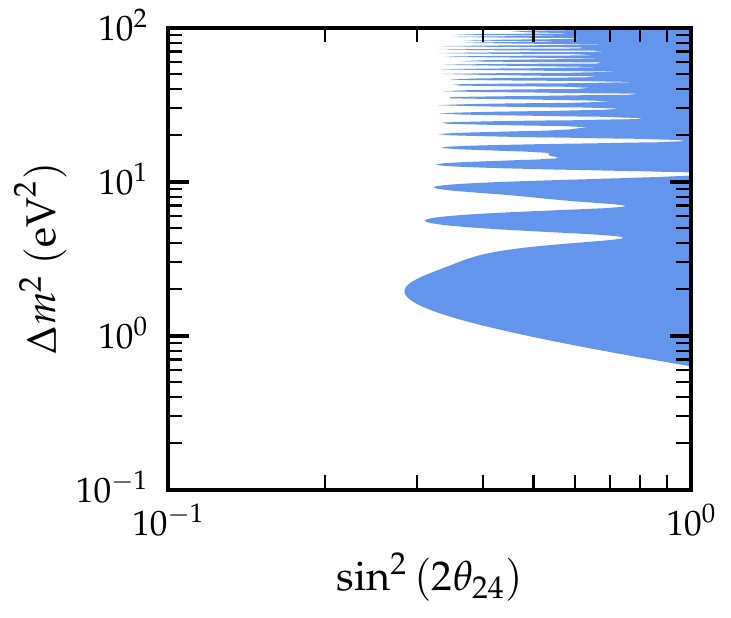}
    \caption{Projected 90\% C.L. sensitivity regions for sterile neutrino oscillations in the $(\sin^2{2\theta_{14}}, \Delta m^2)$ (left, magenta contour) and $(\sin^2{2\theta_{24}}, \Delta m^2)$ (right, blue contour) parameter spaces, obtained from the combined analysis of all the proposed ESS detectors considered in this work. The analysis assumes $\Delta m^2_{41} \approx \Delta m^2_{42} = \Delta m^2$.}
    \label{Fig:Sterile_Neutrino_Oscillation_Sensitivity}
\end{figure}

Now we turn our attention on exploring the potential of probing sterile neutrino oscillations via neutral current \cevns measurements at the ESS. The left panel of Fig.~\ref{Fig:Sterile_Neutrino_Oscillation_Sensitivity} shows the 90\% C.L. projected sensitivity in the $(\sin^2{2\theta_{14}}, \Delta m^2)$ plane, corresponding to oscillations of electron neutrinos into light sterile states. The right panel displays the corresponding constraints in the $(\sin^2{2\theta_{24}}, \Delta m^2)$ plane for oscillations involving muon neutrinos. Since the ESS will exploit $\pi$-DAR neutrinos, the sensitivity to $(\sin^2{2\theta_{14}}, \Delta m^2)$ is notably weaker compared to the $(\sin^2{2\theta_{24}}, \Delta m^2)$ case. It becomes evident that the ESS is not expected to provide stringent constraints in comparison to short baseline neutrino experiments, see e.g., 
 Ref.~\cite{Machado:2019oxb,Giunti:2022btk}.

\begin{figure}[ht!]
    \centering
    \includegraphics[width=0.49\linewidth]{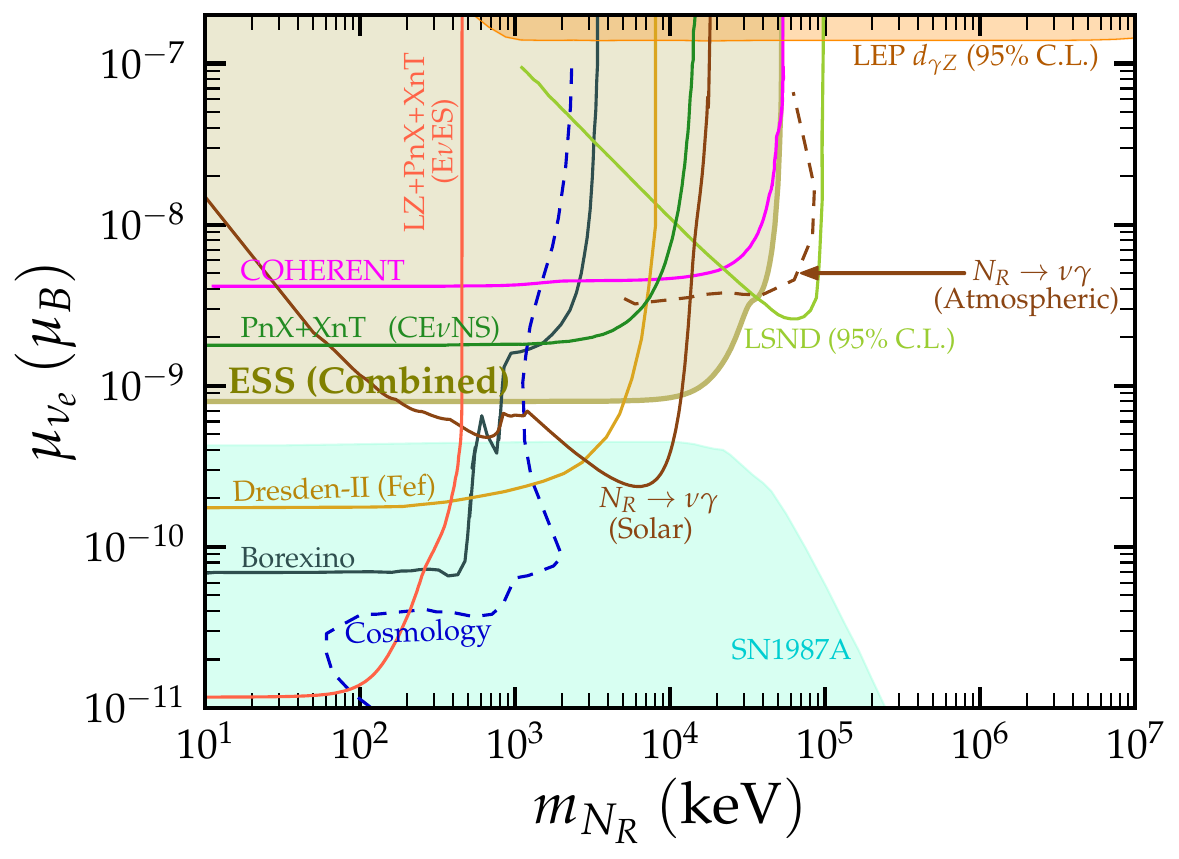}
    \includegraphics[width=0.49\linewidth]{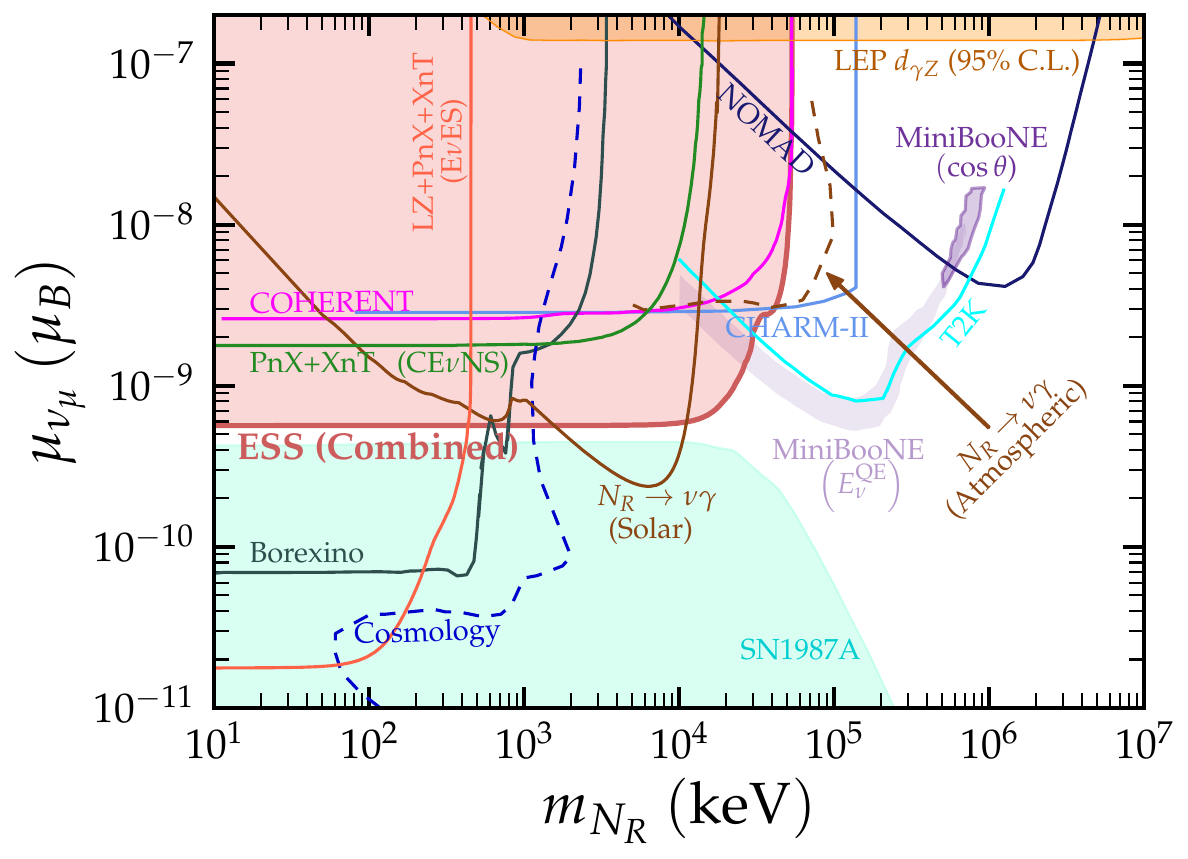}
    \caption{Projected 90\% C.L. sensitivity on the sterile dipole portal scenario. The left and right panels show the contours in the $(m_{N_R}, \mu_{\nu_e})$ and $(m_{N_R}, \mu_{\nu_\mu})$  planes, respectively. The results are obtained from the combined analysis of all detectors proposed for the ESS \cevns experiment. For comparison, constraints from other experimental and astrophysical probes have been superimposed  (see main text for further details).}
    \label{fig:dipole-combined}
\end{figure}

Next we discuss the future ESS sensitivity on active-sterile neutrino transitions in the presence of  nonzero TMMs. Figure~\ref{fig:dipole-combined} presents the projected 90\% C.L. exclusion regions for $\mu_{\nu_e}$ (left panel) and $\mu_{\nu_\mu}$ (right panel) as a function of the sterile neutral lepton mass $m_{N_R}$. These projections are derived from the combined analysis of all detectors considered in this work. Due to the specific nuclear recoil ranges of the various ESS detectors and the energy range of $\pi$-DAR neutrinos produced at the ESS, the generated SNL mass ($m_{N_R}$) is kinematically limited to approximately $50$ MeV, as implied by Eq.~\eqref{Eq:Sterile_kinematics_limit_on_m4}. 
For a comparison with our present ESS sensitivities we include existing constraints from \cevns analyses of COHERENT~\cite{Miranda:2021kre,DeRomeri:2022twg},
Dresden-II~\cite{AristizabalSierra:2022axl}\footnote{This analysis is based on  the iron-filter (Fef) quenching factor model.}, and from the combined analysis of PandaX-4T and XENONnT data~\cite{DeRomeri:2024hvc}. Also included are limits coming from \eves analyses using  Borexino~\cite{Brdar:2020quo,Plestid:2020vqf} and
CHARM-II~\cite{Coloma:2017ppo} data, as well as  
from the recent combined analysis of XENONnT, PandaX-4T and LUX-ZEPLIN data performed in Ref.~\cite{DeRomeri:2024hvc}. Additional bounds from LSND~\cite{Magill:2018jla}, LEP~\cite{Magill:2018jla}, NOMAD~\cite{NOMAD:1997pcg,Gninenko:1998nn},  MiniBooNE~\cite{Kamp:2022bpt}, 
T2K~\cite{T2K:2019jwa,Liu:2024cdi}, and from  $N_R \to \nu \gamma$ decay~\cite{Plestid:2020vqf,Plestidlumsolnu} using solar (Borexino and Super-Kamiokande) and atmospheric (Super-Kamiokande)~\cite{Gustafson:2022rsz} data, are also shown\footnote{We account for a factor 2 difference in the Lagrangian defined in Refs.~\cite{Magill:2018jla,Plestid:2020vqf}. For the case of MinibooNE, the depicted constraints are relevant for muon neutrinos only, for which the preferred regions are shown from the analysis of energy ($E_\nu^\mathrm{QE}$) and angular (cos$\theta$) data.}. We finally superimpose bounds from astrophysical and cosmological data such as
SN1987A~\cite{Magill:2018jla,Chauhan:2024nfa}, BBN~\cite{Magill:2018jla,Brdar:2020quo}, and CMB constraints on $\Delta N_\mathrm{eff}$~\cite{Brdar:2020quo}.

For SNL masses below $10$ MeV, the ESS analysis performed in this work results in exclusion limits as low as $\mu_{\nu_e} \sim 8 \times 10^{-10} \, \mu_B$ and $\mu_{\nu_\mu} \sim 6 \times 10^{-10} \, \mu_B$. These projections demonstrate a significant improvement over existing limits. For instance, the ESS is expected to improve the COHERENT bound by a factor 5 for both electron and muon neutrinos. It will furthermore improve the combined PandaX-4T and XENONnT \cevns result by factor 2 (3) for the case of electron (muon) neutrinos. While the projected exclusion for $\mu_{\nu_e}$ is slightly weaker than the limits inferred from Dresden-II reactor data~\cite{AristizabalSierra:2022axl}, the ESS experiment offers a broader reach in SNL masses. In particular, the ESS facility using the \cevns channel has the prospect to investigate a completely unexplored part of the parameter space as it will dominate the constraints in the mass range $10 \lesssim m_{N_R} \lesssim 40$~MeV ($10 \lesssim m_{N_R} \lesssim 40$~MeV) for electron (muon) neutrinos. Before closing this discussion we must warn that the limits depicted in the figure are not always directly comparable. This is due to the fact that the effective magnetic moments corresponding to the various experiments depend on different combinations of TMMs, CP phases and oscillation parameters  (for a discussion see Ref.~\cite{Miranda:2021kre}). Since both COHERENT and ESS exploit $\pi$-DAR neutrinos, a direct comparison is possible for these two experiments only, where the more intense ESS neutrino beam will offer  a significant improvement.

\begin{figure}[t!]
    \centering
    \includegraphics[width=0.49\linewidth]{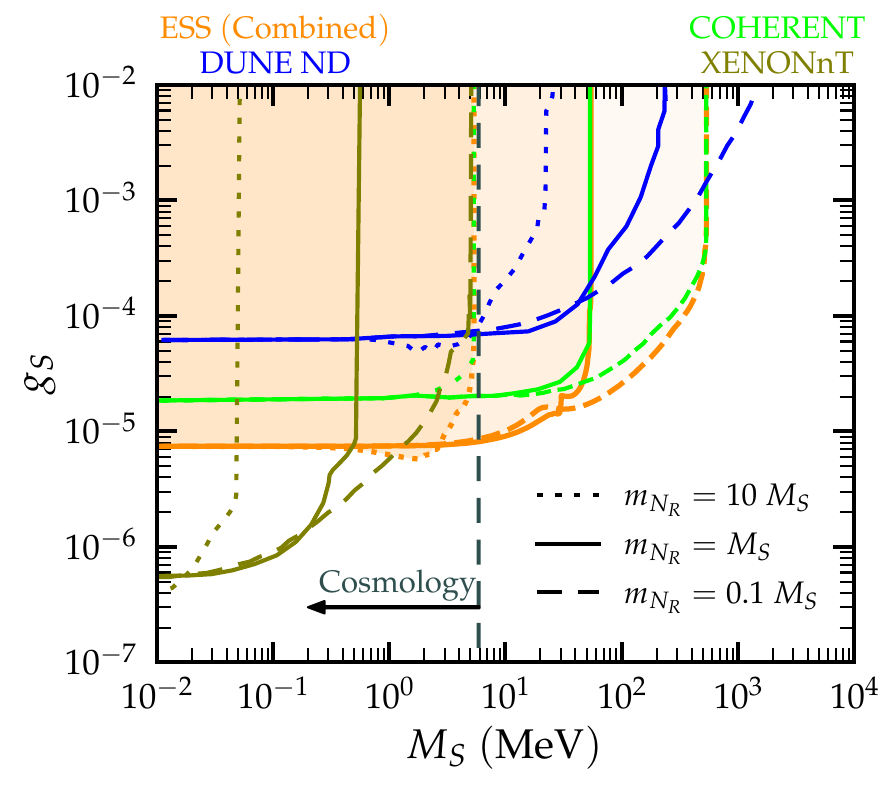}
    \includegraphics[width=0.49\linewidth]{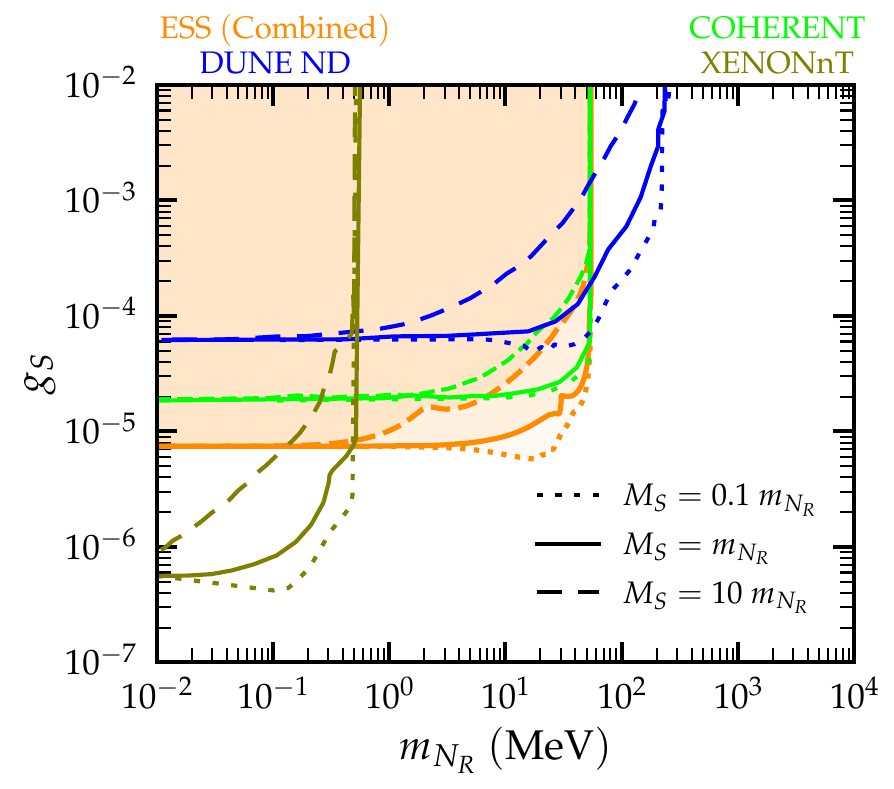}
    \caption{Projected sensitivity to scalar-mediated SNL production at the future ESS experiment, derived from the combined analysis of all detectors. The left (right) panels display the 90\% C.L. exclusion contours for ESS with orange lines, projecting on the mediator (SNL) mass, considering three benchmark scenarios $m_{N_R} = \{0.1, 1, 10\} \times M_S$. For comparison, the ESS projections are shown together with constraints from current XENONnT and COHERENT data, as well as with sensitivity projections from DUNE Near Detector measurements.}
    \label{fig:Upscattering_Scalar_Combined}
\end{figure}

\begin{figure}[ht!]
    \centering
    \includegraphics[width=0.49\linewidth]{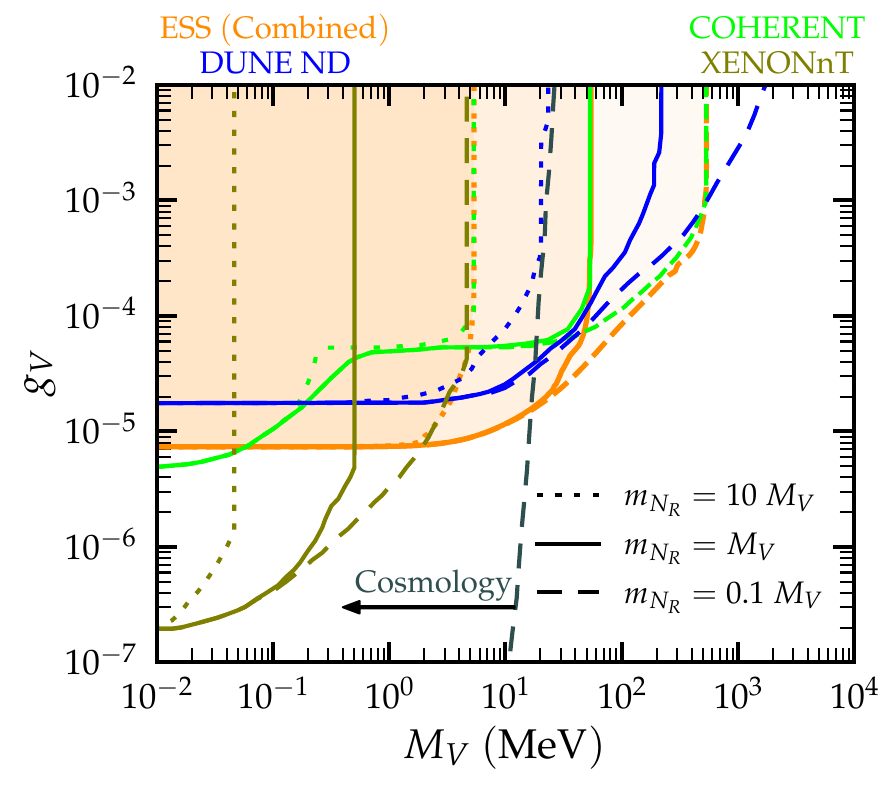}
    \includegraphics[width=0.49\linewidth]{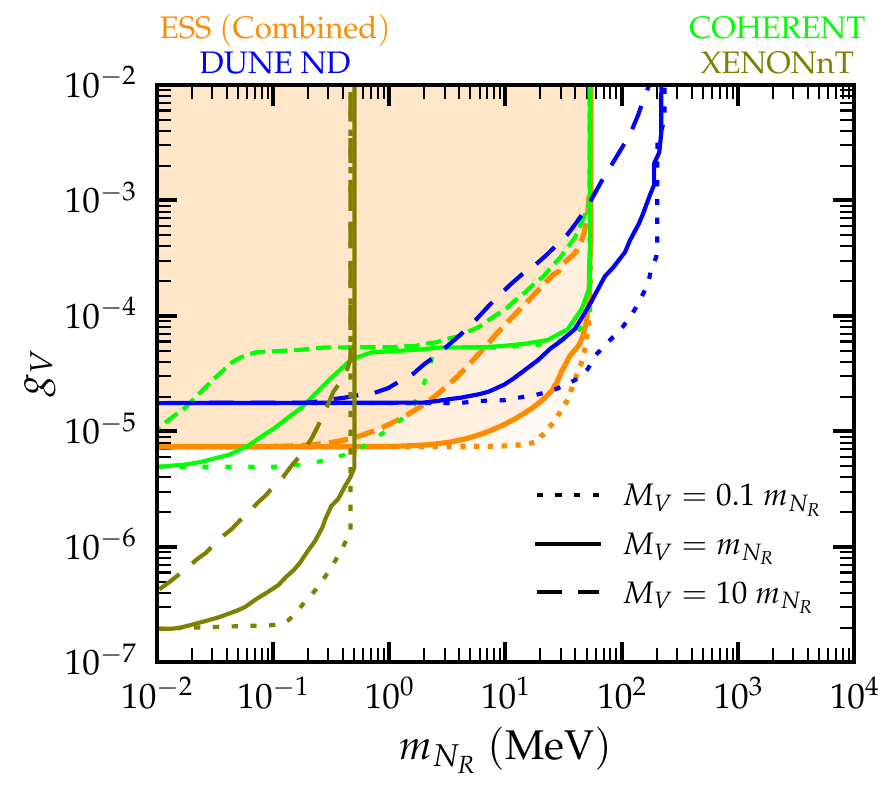}
    \caption{Same as Fig.~\ref{fig:Upscattering_Scalar_Combined}, but for vector-mediated SNL production.}
    \label{fig:Upscattering_Vector_Combined}
\end{figure}

\begin{figure}[ht!]
    \centering
    \includegraphics[width=0.49\linewidth]{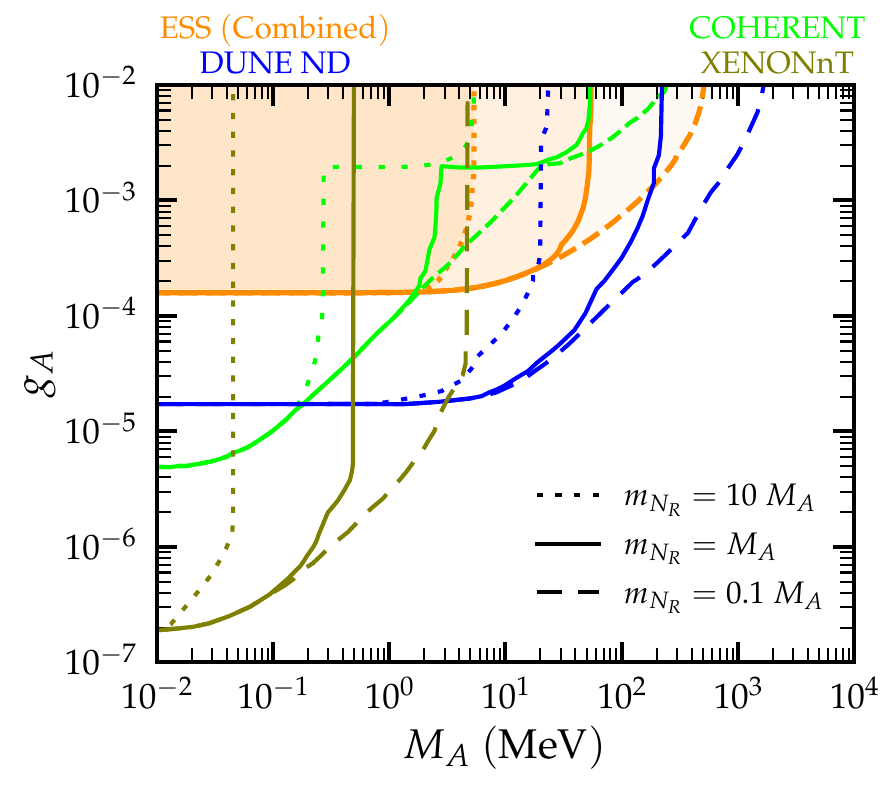}
    \includegraphics[width=0.49\linewidth]{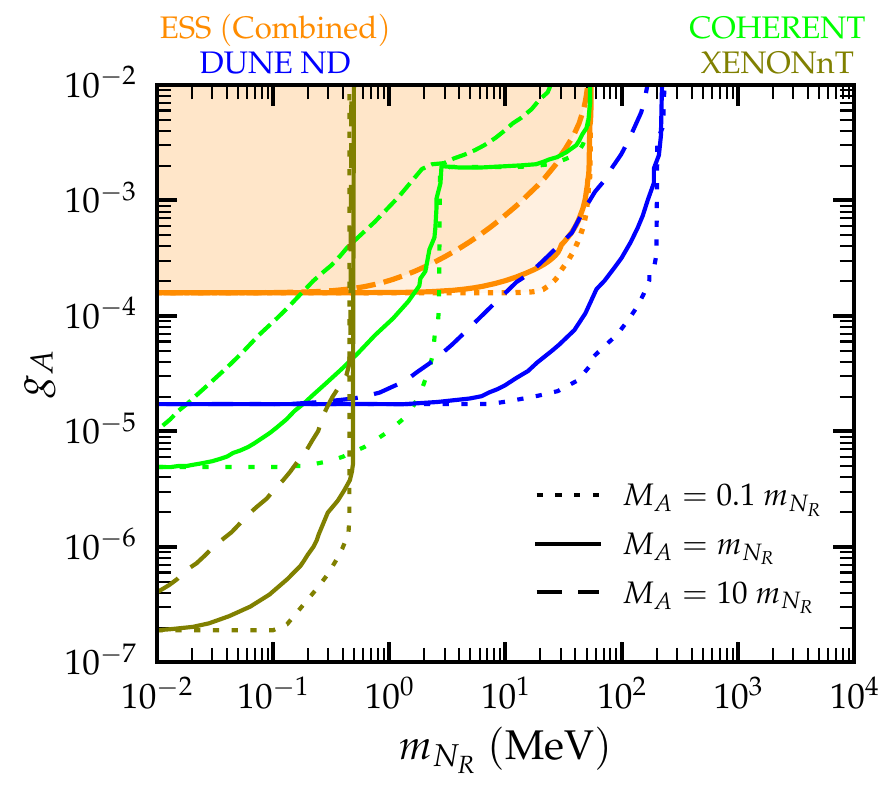}
    \caption{Same as Fig.~\ref{fig:Upscattering_Scalar_Combined}, but for axial vector-mediated SNL production.}
    \label{fig:Upscattering_Axial_Vector_Combined}
\end{figure}

\begin{figure}[ht!]
    \centering
    \includegraphics[width=0.49\linewidth]{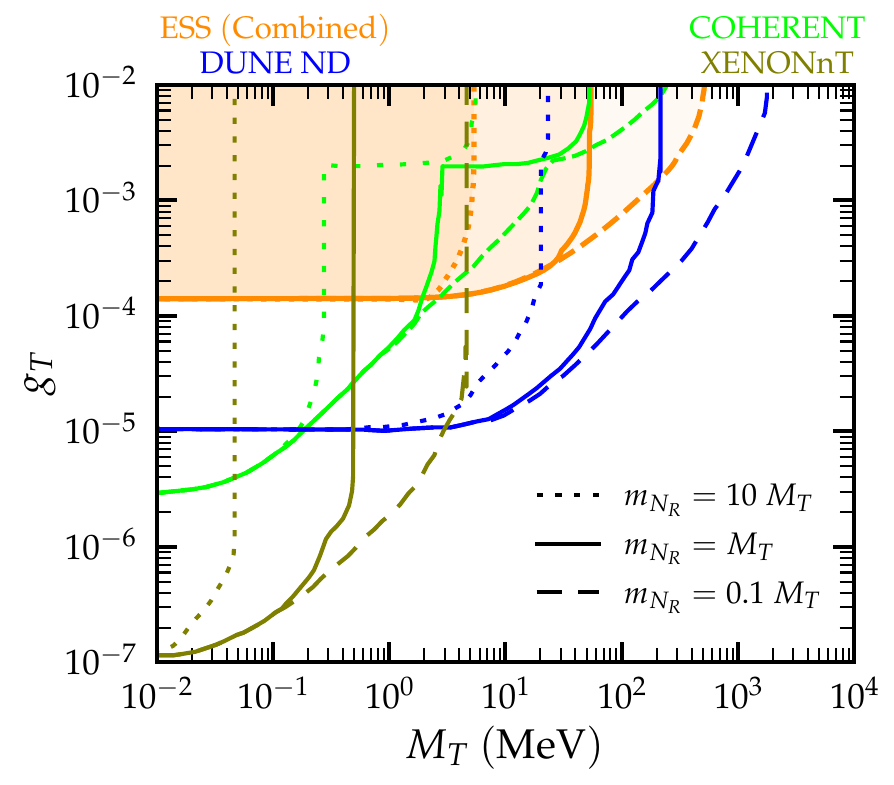}
    \includegraphics[width=0.49\linewidth]{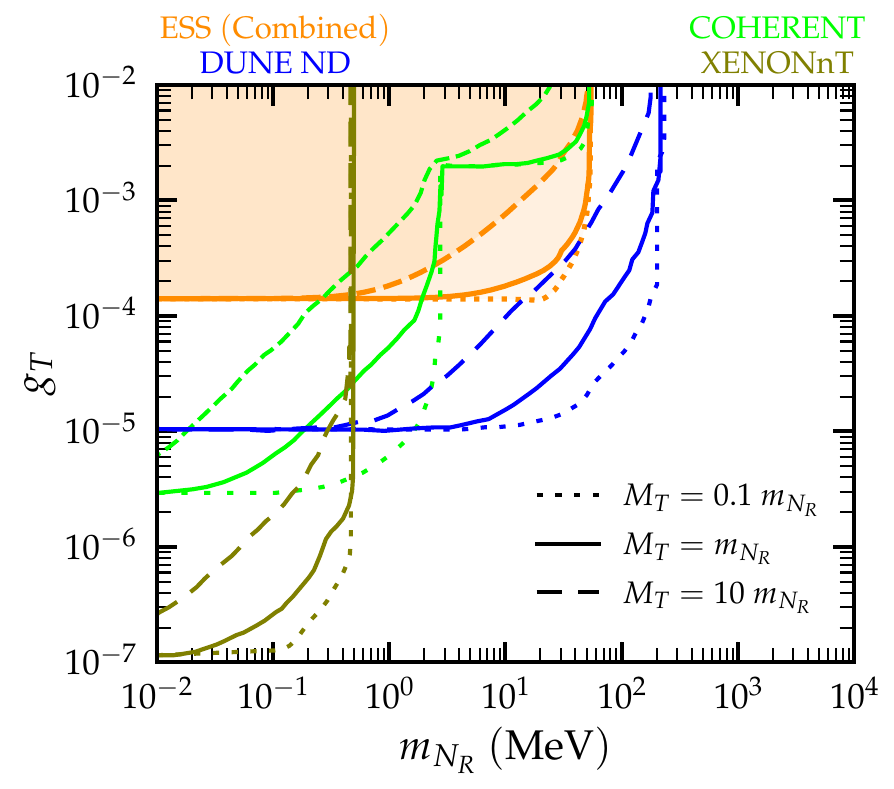}
    \caption{Same as Fig.~\ref{fig:Upscattering_Scalar_Combined}, but for tensor-mediated SNL production.}
    \label{fig:Upscattering_Tensor_Combined}
\end{figure}

We now present the projected sensitivities for upscattering production of SNL in the presence of NGIs at the future ESS experiment. Figures~\ref{fig:Upscattering_Scalar_Combined}, \ref{fig:Upscattering_Vector_Combined}, \ref{fig:Upscattering_Axial_Vector_Combined}, and \ref{fig:Upscattering_Tensor_Combined} illustrate the 90\% C.L. projected exclusion limits for scalar, vector, axial vector, and tensor-mediated channels respectively, denoted by $X = \{S, V, A, T\}$. As previously, the obtained results come out from the combined analysis of all ESS detectors, while the results corresponding to the individual detectors are given in Appendix~\ref{Appendix_1}. The figures demonstrate three representative benchmark scenarios with $m_{N_R} = \{0.1, 1, 10\} \times M_X$. In each case, the left panels illustrate the limits in the coupling-mediator mass plane, $\textsl{g}_X$ vs $M_X$, while the right panels show the contours in the coupling-SNL mass plane, $\textsl{g}_X$ vs $m_{N_R}$. At small mediator ($M_X$) or SNL masses ($m_{N_R}$), the sensitivity contours plateau, reflecting the saturation of constraints, and become essentially identical to the NGI cases shown in Fig.~\ref{Fig:NGI_limit_comparison}.  Conversely, for larger $M_X$ or $m_{N_R}$, the behavior diverges. In the left panels the exclusion limits diminish at different mediator masses $M_X$ based on the different $m_{N_R}/M_X$ ratios, while in the right panels the sensitivity loss occurs at a fixed SNL mass, as dictated by the kinematic constraint of Eq.~\eqref{Eq:Sterile_kinematics_limit_on_m4}.

To assess the complementarity of the ESS projections, the figures also overlay existing constraints from XENONnT and COHERENT CsI, along with projected limits from the DUNE Near Detector, as reported in Ref.~\cite{Candela:2024ljb}. It becomes evident that different experiments dominate in different regions of the parameter space. For instance, XENONnT using the \eves channel provides the most stringent constraints for very low $M_X$ or $m_{N_R}$, while for $m_{N_R} \geq 0.5~\text{MeV}$, the ESS projections are expected to surpass  XENONnT sensitivities. Notably, the ESS sensitivity is expected to exceed current limits from COHERENT CsI-2021 across most of the parameter space, except at low $m_{N_R}$ for some cases\footnote{Notice that the authors in Ref.~\cite{Candela:2024ljb} incorporated both E$\nu$ES and \cevns signals in their analysis of COHERENT CsI-2021 data; this dual-channel strategy enhances the sensitivity for very low $m_{N_R}$, i.e., in a region where E$\nu$ES events dominate over the \cevns ones.}. For scalar and vector interactions, ESS constraints outperform those anticipated from DUNE Near Detector measurements until they reach the kinematical constraint of about 50~MeV. Instead for the spin-dependent axial vector and tensor interactions, the DUNE Near Detector offers enhanced sensitivity since the corresponding \cevns cross sections are suppressed by the nuclear spin, which is not the case for the E$\nu$ES-based DUNE constraints. Additionally, the high neutrino beam accessible at DUNE enables a  broader reach of SNL masses compared to the ESS across all channels. As a result, the ESS experiment and the DUNE Near Detector provide complementary results, with the two experiments dominating in different mass regimes. For completeness, whenever available we superimpose cosmological  constraints  (see the discussion of Fig.~\ref{Fig:NGI_limit_comparison}).

\section{Conclusions}
\label{Sec:Conclusions}

  The highly intense ESS neutrino beam will produce a wealth of new \cevns data, opening a new avenue for probing interesting physics phenomena within and beyond the SM. The unprecedented statistics characterizing the  new era of \cevns measurements at the ESS, will offer improved  sensitivities by up to one order of magnitude  ---or more depending on the physics scenario in question--- in comparison to existing ones from \cevns measurements exploiting $\pi$-DAR (COHERENT), reactor (Dresden-II) and solar  (PandaX-4T and XENONnT) data. In this work, a comprehensive exploration of the ESS potential is carried out focusing on the various detectors, highlighting the promising potential of the ESS to explore both fundamental and exotic neutrino physics. For the various physics scenarios,  by performing a combined analysis of all the proposed ESS detectors, the projected sensitivities are quantified. The attainable sensitivities resulted by analyzing each detector individually are given in  Appendix~\ref{Appendix_1}, where their relative performance is also discussed.
 
 The explored physics scenarios focus on important SM parameters such as the weak mixing angle and the nuclear size through the neutron rms radius. For the former, we find that the ESS will improve the precision reached by COHERENT (Dresden-II) by  $\sim 60\%$ ($\sim 80\%$), with Ar and $\mathrm{C_3F_8}$ expected to set the most stringent limits. Concerning  the nuclear neutron rms radius, the multitarget strategy of the ESS \cevns experiment will constrain  the Si and $\mathrm{C_3F_8}$ rms radii for the first time.  In the case of Ar and CsI where constraints already exist from e.g. COHERENT, an improvement of $\sim 40\%$ is anticipated for the case of CsI, while for the case of Ar only an upper limit exists. Moreover,  while there exist \cevns measurements on Xe and Ge targets from the Dresden-II and the dark matter direct detection experiments (PandaX-4T and XENONnT), it should be stressed that the corresponding sensitivities are weak. This is because reactor experiments are not ideal to probe nuclear physics effects as they are sensitive to the (almost fully) coherent regime, while the statistics collected by the dark matter direct detection experiments is yet poor. Turning to new physics scenarios, we explored NGIs and presented the corresponding constraints for the Lorentz-invariant scalar, vector, axial vector and tensor interactions. We found that in general the ESS will drastically improve previous CE$\nu$NS-based constraints in all cases, with the different proposed detector performing equally well (for a detailed discussion see the Appendix). Moreover, for the scalar and vector interactions, the projected constraints will not only complement collider and astrophysical limits but are expected to dominate in certain regions of the parameter space e.g. for $M_S >40$~MeV and $25<M_V<200$~MeV, respectively. On the other hand, as expected the spin-dependent axial vector and tensor ESS sensitivities will not be able to compete with existing bounds resulting from \eves analyses. Our present analysis implies that future \cevns measurements at the ESS will not be able to provide competitive sensitivities for the case of active-sterile neutrino oscillations and the violation of lepton unitarity. However, by focusing on upscattering channels we have verified that the ESS will serve as a valuable probe of sterile neutral lepton phenomenology. To this aim, we explored two interesting scenarios for producing massive sterile particle states, e.g., the so-called sterile dipole portal in the presence of neutrino magnetic moments and via NGIs. For the former case, the ESS will surpass previous \cevns sensitivities and will furthermore probe a previously unexplored region in the parameter space i.e., for $10 \lesssim m_{N_R} \lesssim 40$~MeV  and $10 \lesssim m_{N_R} \lesssim 40$~MeV for electron and muon neutrinos, respectively. Finally, we investigated the possible production of final state sterile neutral leptons within the NGI framework and discussed the complementarity with existing $\pi$-DAR-induced  \cevns at COHERENT,  solar neutrino-induced \eves  searches at dark matter direct detection experiments, as well as with prospects from the future DUNE Near detector.

%%%%%%%%%%%%%%%%%%%%%%%%%%%%%%%%%%%
\acknowledgments
%%%%%%%%%%%%%%%%%%%%%%%%%%%%%%%%%%%%

The authors are grateful to Gonzalo Sanchez Garcia for fruitful discussions.
AM expresses sincere thanks for the financial support provided through the Prime Minister Research Fellowship (PMRF), funded by the Government of India (PMRF ID: 0401970).
The work of DKP is supported by CNS2023-144124 (MCIN/AEI/10.13039/ 501100011033 and ``Next Generation E''/PRTR), and partially by the Spanish grants PID2023-147306NB-I00 and CEX2023-001292-S (MCIU/AEI/10.13039/501100011033), as well as CIDEXG/2022/20 (Generalitat Valenciana).

\appendix
\section{Projected limits from individual detectors}
\label{Appendix_1}

Sec.~\ref{Sec:Results} presents the combined ESS limits on various physics scenarios derived by combining the results from all the proposed detectors considered in this study. For completeness, in this Appendix we present the projected limits obtained from the analysis of  the individual detectors.

\begin{figure}[ht!]
    \centering
    \includegraphics[width=0.6\linewidth]{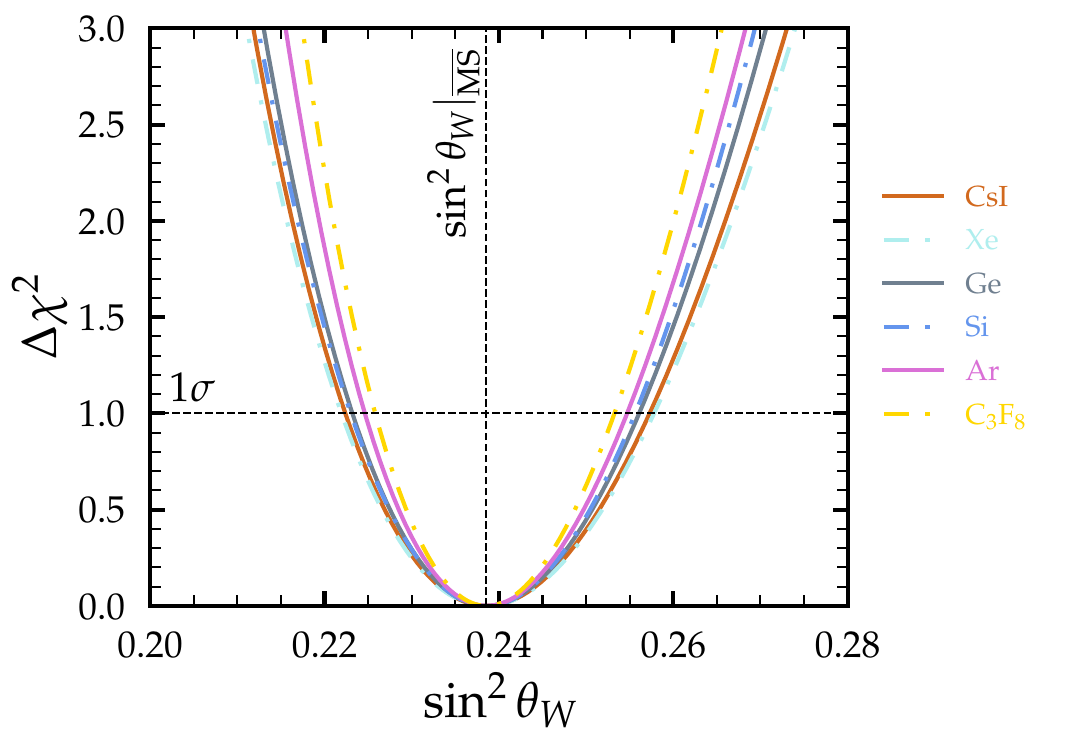}
    \caption{Projected sensitivity on the weak mixing angle obtained exploiting different ESS detectors.}
    \label{fig:sw2_Plot_for_different_detectors}
\end{figure}

\begin{table}[ht!]
    \centering
    \begin{tabular}{|c|c|c|c|}
        \hline
        \hspace{0.5cm}Detector\hspace{0.5cm} & \hspace{2cm}$\sin^2{\theta_W}$\hspace{2cm} & \hspace{0.5cm}Detector\hspace{0.5cm} & \hspace{2cm}$\sin^2{\theta_W}$\hspace{2cm} \\
        \hline
        \hline
         & & & \\
        CsI & $0.239{^{+0.019}_{-0.016}}$ & Si & $0.239{^{+0.017}_{-0.016}}$ \\
         & & & \\
        \hline
         & & & \\
        Xe & $0.239{^{+0.019}_{-0.017}}$ & Ar & $0.239{^{+0.016}_{-0.014}}$ \\
         & & & \\
        \hline
         & & & \\
        Ge & $0.239{^{+0.018}_{-0.015}}$ & $\mathrm{C_3F_8}$ & $0.239{^{+0.015}_{-0.013}}$ \\
         & & & \\
        \hline
        \hline
        & & & \\
        & & \textbf{ESS Combined} & $\mathbf{0.239{^{\mathbf{+0.011}}_{\mathbf{-0.010}}}}$ \\
         & & & \\
         \hline
    \end{tabular}
    \caption{Best fit values and $1\sigma$ uncertainties of $\sin^2{\theta_W}$ obtained for the various  ESS detectors.} 
    \label{tab:sw2_values_for_different_detectors}
\end{table}

The $\Delta\chi^2$ profiles of $\sin^2{\theta_W}$ for the different detectors are shown in Fig.~\ref{fig:sw2_Plot_for_different_detectors}, while the projected $1\sigma$ determinations  are listed in Table~\ref{tab:sw2_values_for_different_detectors}. These results indicate that, within the specified experimental setup, the most precise limits on $\sin^2{\theta_W}$ can be expected from the $\mathrm{C_3F_8}$ detector, while the least precise limits are expected from the Xe detector. Our results are in good agreement with those extracted in Ref.~\cite{Baxter:2019mcx}. We should further note, that while the most stringent constraints are expected for the heavier target nuclei, since in these cases the statistics is higher, the different backgrounds corresponding to each detector together with the nuclear physics suppression and the different thresholds are forcing the limits to be stronger for the $\mathrm{C_3F_8}$ detector.

\begin{figure}[ht!]
    \centering
    \includegraphics[width=0.32\linewidth]{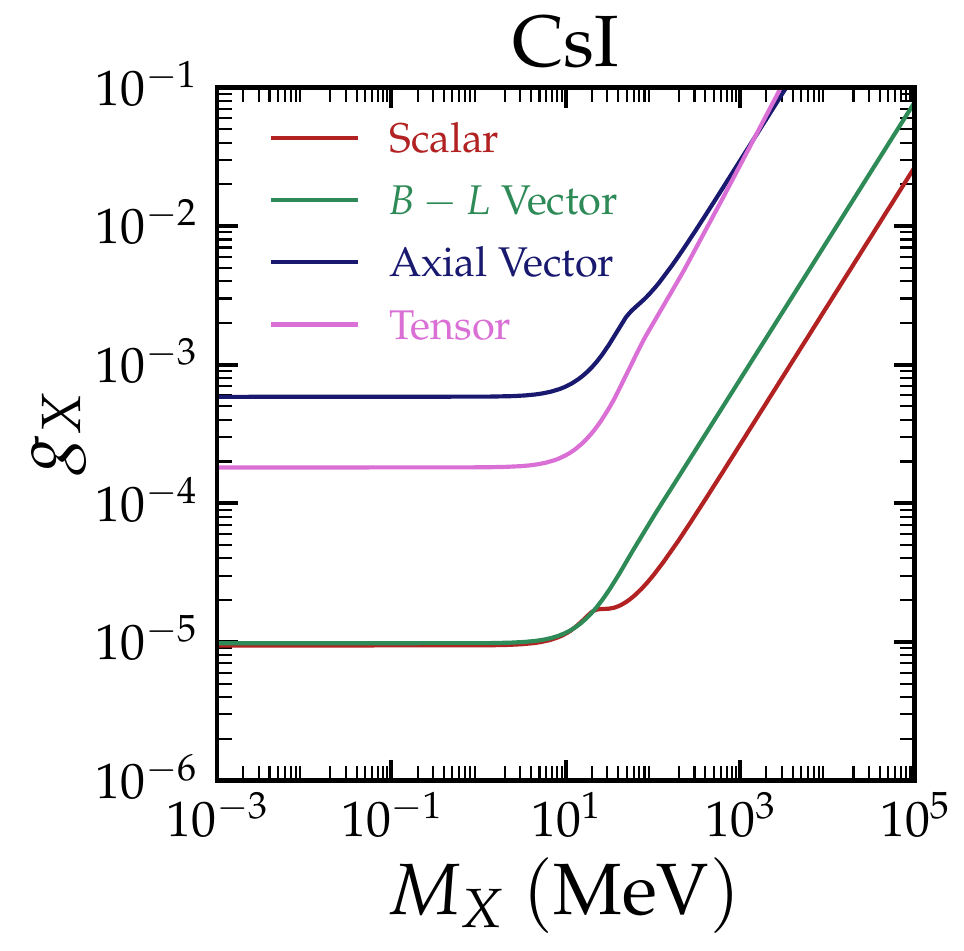}
    \includegraphics[width=0.32\linewidth]{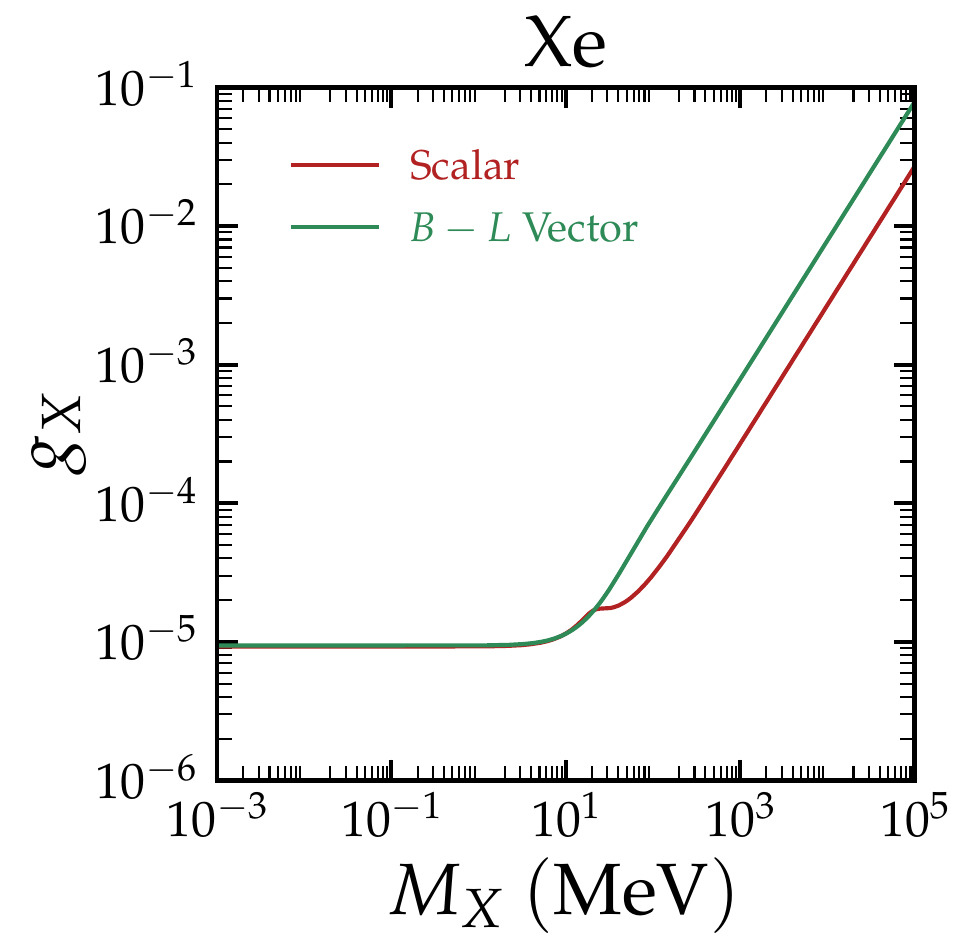}
    \includegraphics[width=0.32\linewidth]{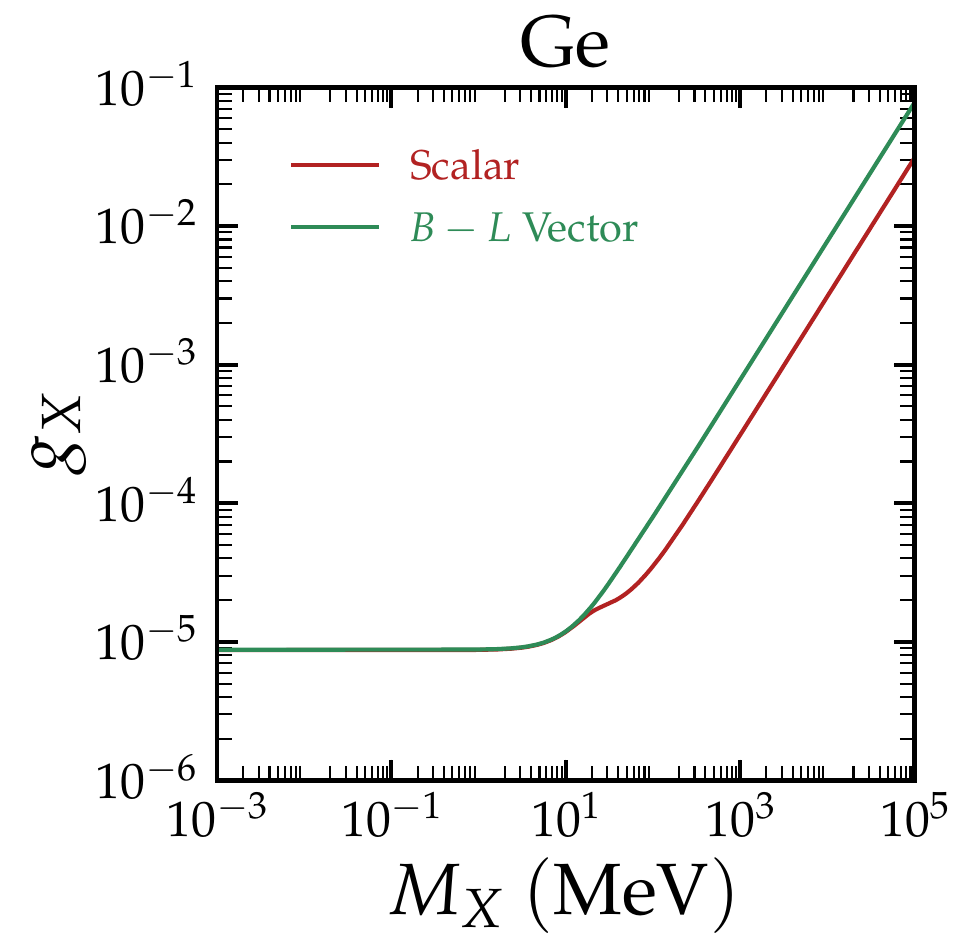}
    \includegraphics[width=0.32\linewidth]{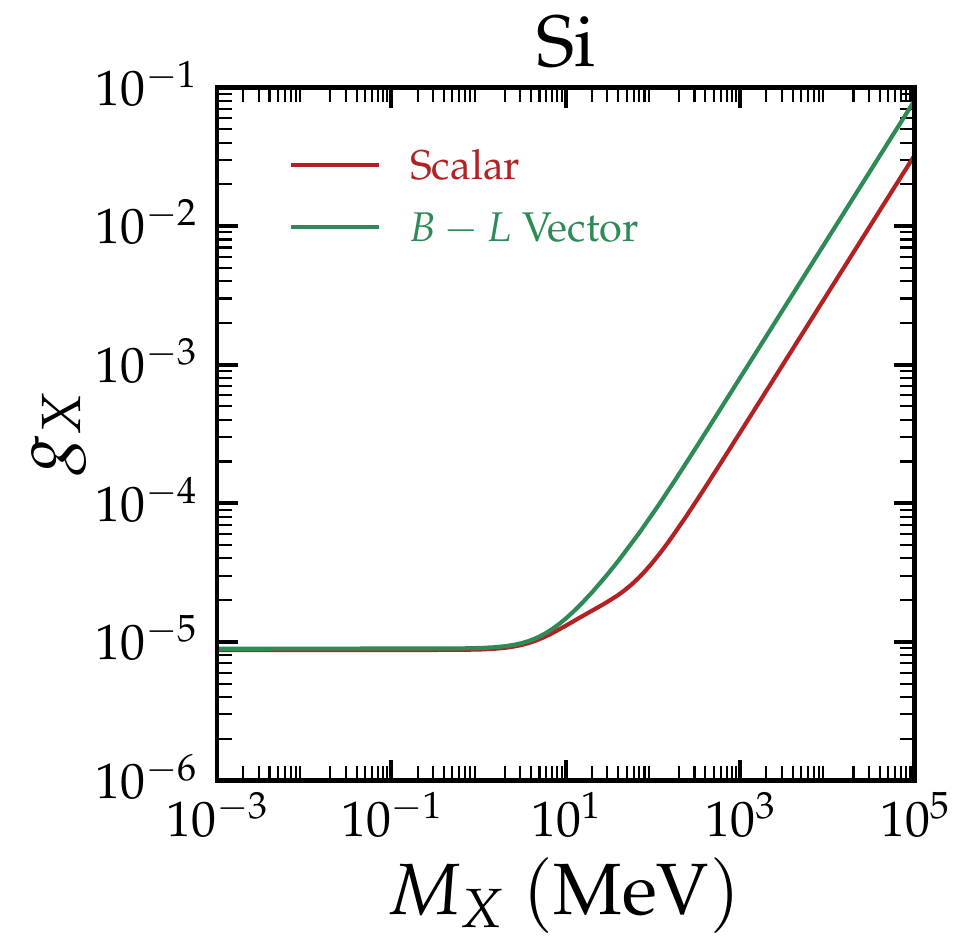}
    \includegraphics[width=0.32\linewidth]{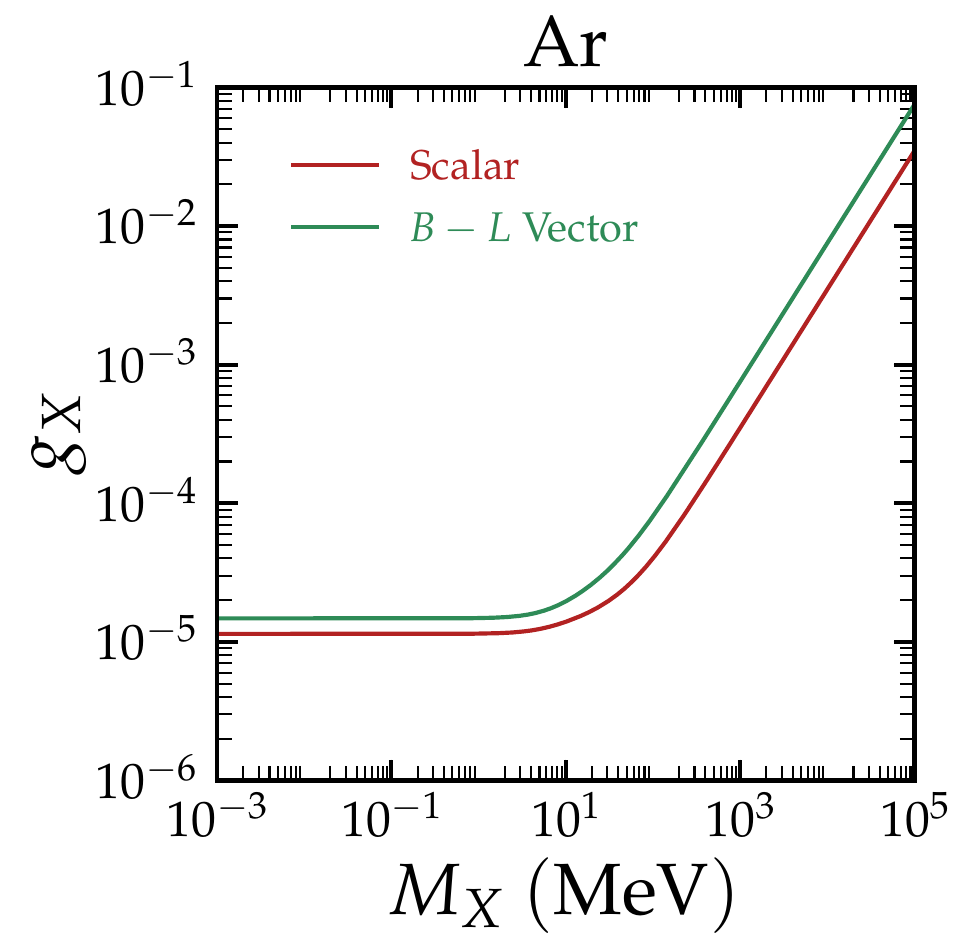}
    \includegraphics[width=0.32\linewidth]{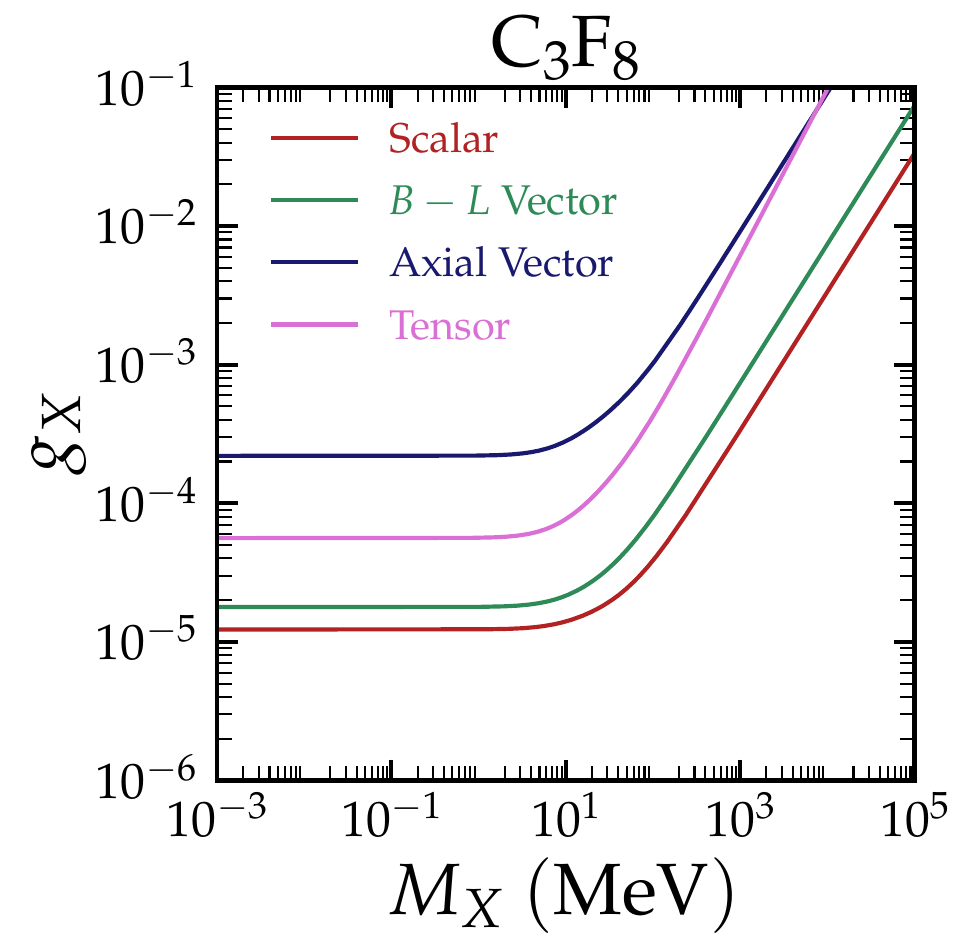}
    \caption{Projected sensitivities at 90\% C.L. in the $(M_X, \textsl{g}_X)$ parameter space for scalar, vector $B-L$,  axial vector and tensor interactions, derived for the individual ESS detectors.}
    \label{Fig:NGI_contour_different_detector}
\end{figure}

Figure~\ref{Fig:NGI_contour_different_detector} depicts the 90\% C.L. projected NGI sensitivities at the ESS in the $(M_X, \textsl{g}_X)$ parameter space for scalar, vector $B-L$, axial vector, and tensor interactions.  Let us remind that only the CsI and $\mathrm{C_3F_8}$ detectors exhibit sensitivity to nuclear spin-dependent axial vector and tensor interactions, while the rest detectors being spin-zero nuclei are not sensitive to these interactions (see also Table~\ref{Tab:Detectors_Specs}). Among these, the $\mathrm{C_3F_8}$ detector is projected to impose more stringent constraints on spin-dependent interactions compared to the CsI detector. This is because the relative axial vector contribution compared to the dominant SM vector component is more significant for light nuclei, since the latter are composed by fewer nucleons and hence their vector part is less enhanced. For the case of spin-independent scalar and vector $B-L$ interactions, all the detectors provide similar constraints, except the Ar and $\mathrm{C_3F_8}$ cases which perform slightly worse. Further, in the case of scalar interactions, it is worth noting that, unlike the CsI, Xe, Ge and Si detectors, the little kink in the contour around $M_S \approx 10$ MeV is absent for the Ar and $\mathrm{C_3F_8}$ cases. This feature and the performance difference arise due to the adoption of a single-bin analysis strategy for these detectors. 

\begin{figure}[ht!]
    \centering
    \includegraphics[width=0.49\linewidth]{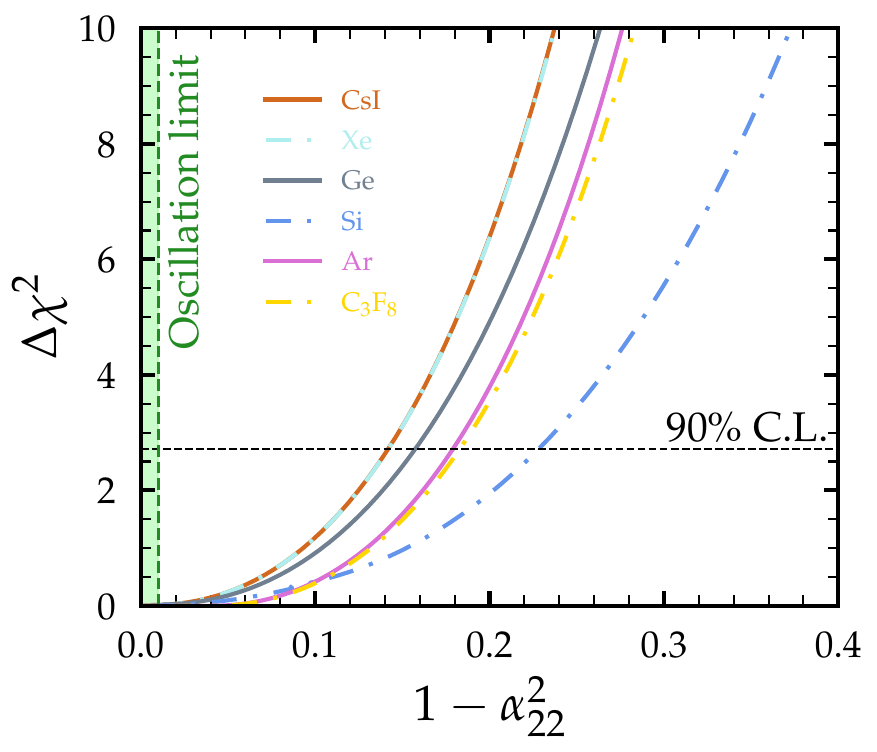}
    \caption{$\Delta\chi^2$ profile of $1-\alpha_{22}^2$ for the different ESS detectors. The green dashed line represent the corresponding 90\% C.L. upper limit derived from neutrino oscillation global fits.}
    \label{fig:NU_sensitivity_Plot_for_different_detectors}
\end{figure}

\begin{table}[ht!]
    \centering
    \begin{tabular}{|c|c|c|c|}
        \hline
        \hspace{1cm}Detector\hspace{1cm} & \hspace{1cm}$1-\alpha_{22}^2$\hspace{1cm} & \hspace{1cm}Detector\hspace{1cm} & \hspace{1cm}$1-\alpha_{22}^2$\hspace{1cm} \\
        \hline
        \hline
         & & & \\
        CsI & $<0.14$ & Si & $<0.22$ \\
         & & & \\
        \hline
         & & & \\
        Xe & $<0.14$ & Ar & $<0.18$ \\
         & & & \\
        \hline
         & & & \\
        Ge & $<0.16$ & $\mathrm{C_3F_8}$ & $<0.18$ \\
         & & & \\
         \hline
         \hline
          & & & \\
         \textbf{ESS Combined} & $\mathbf{<0.14}$ & \textbf{Oscillation limit} & $\mathbf{<0.01}$~\cite{Forero:2021azc}\\
          & & & \\
         \hline
         \end{tabular}
    \caption{Projected 90\% C.L. upper limits on $1-\alpha_{22}^2$ for different ESS detectors considered in this study. For comparative purposes, the 90\% C.L. upper limit on $1-\alpha_{22}^2$ derived from global fits of neutrino oscillation data is also included.}
         \label{tab:alpha22_limits_for_different_detectors}
\end{table}

Turning to the NU scenario, we present the $\Delta\chi^2$ profile of $1-\alpha_{22}^2$ for the different ESS detectors in Fig.~\ref{fig:NU_sensitivity_Plot_for_different_detectors}. The projected 90\% C.L. upper limits on $1-\alpha_{22}^2$ for each detector are listed in Table~\ref{tab:alpha22_limits_for_different_detectors}. Under the specified experimental setup, the CsI and Xe detectors are projected to provide the most stringent constraints on $\alpha_{22}$, while the Si detector is expected to yield the least stringent limit. As explained in the main text, the NU sensitivity at the ESS using the \cevns channel will not be able to compete with oscillation experiments.

\begin{figure}[ht!]
    \centering
    % First row of six figures
    \begin{subfigure}{0.16\textwidth}
        \centering
        \includegraphics[width=1.08\textwidth]{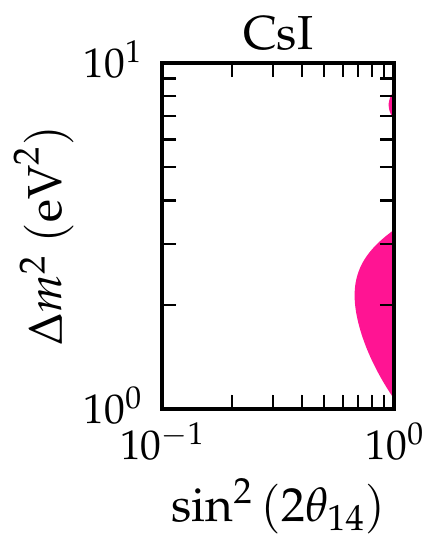}
    \end{subfigure}
    \begin{subfigure}{0.16\textwidth}
        \centering
        \includegraphics[width=1.08\textwidth]{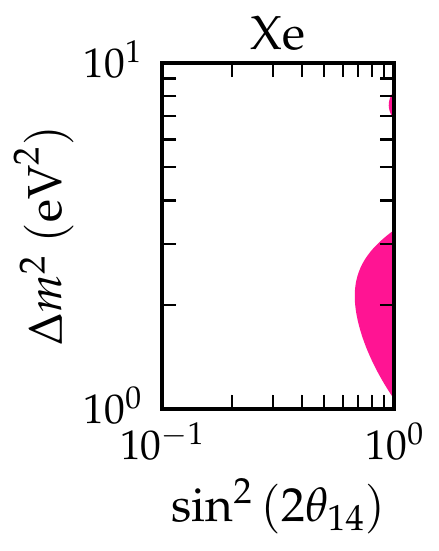}
    \end{subfigure}
    \begin{subfigure}{0.16\textwidth}
        \centering
        \includegraphics[width=1.08\textwidth]{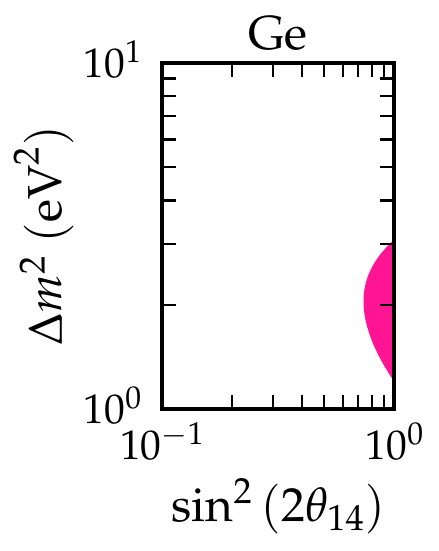}
    \end{subfigure}
    \begin{subfigure}{0.16\textwidth}
        \centering
        \includegraphics[width=1.08\textwidth]{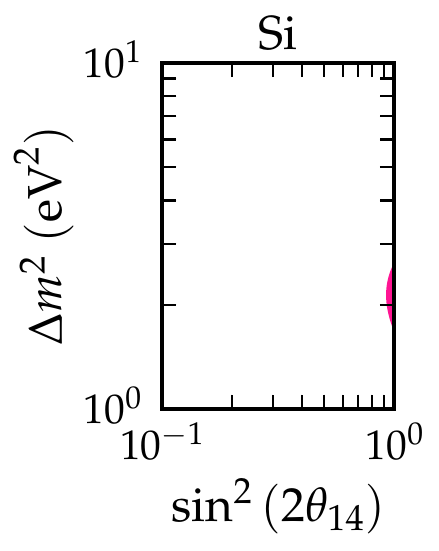}
    \end{subfigure}
    \begin{subfigure}{0.16\textwidth}
        \centering
        \includegraphics[width=1.08\textwidth]{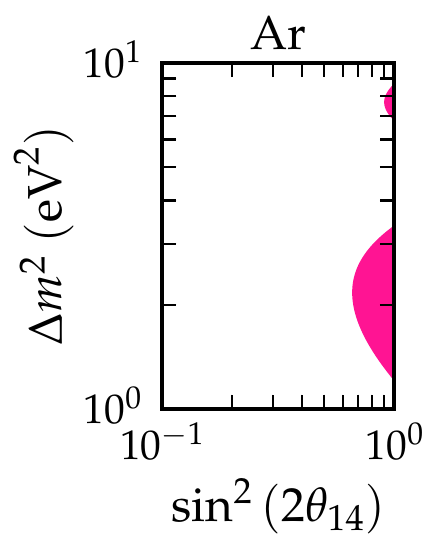}
    \end{subfigure}
    \begin{subfigure}{0.16\textwidth}
        \centering
        \includegraphics[width=1.08\textwidth]{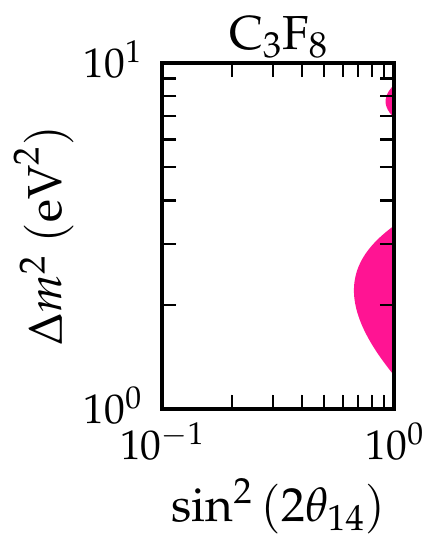}
    \end{subfigure}
    
    % Second row of six figures
    \begin{subfigure}{0.16\textwidth}
        \centering
        \includegraphics[width=1.08\textwidth]{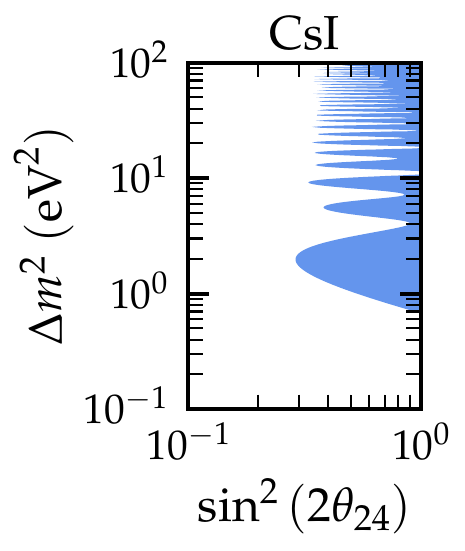}
    \end{subfigure}
    \begin{subfigure}{0.16\textwidth}
        \centering
        \includegraphics[width=1.08\textwidth]{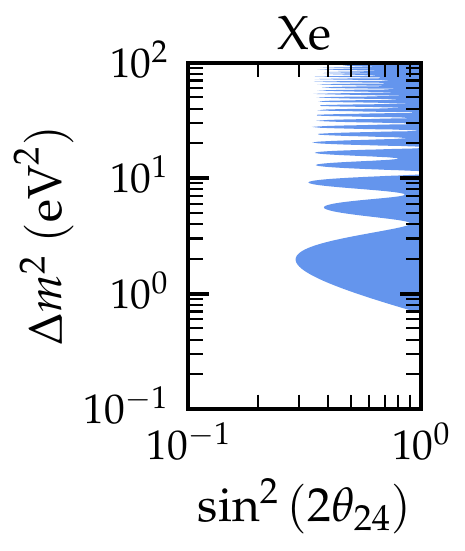}
    \end{subfigure}
    \begin{subfigure}{0.16\textwidth}
        \centering
        \includegraphics[width=1.08\textwidth]{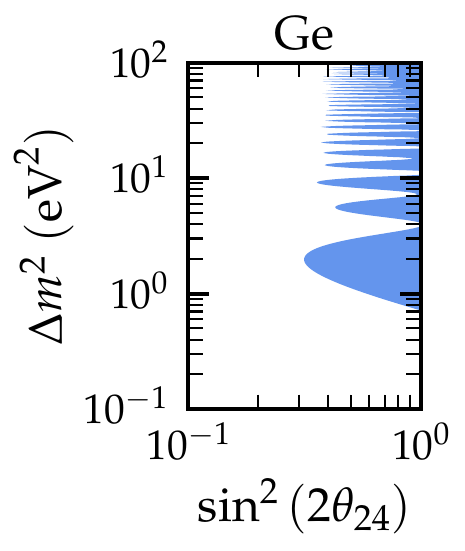}
    \end{subfigure}
    \begin{subfigure}{0.16\textwidth}
        \centering
        \includegraphics[width=1.08\textwidth]{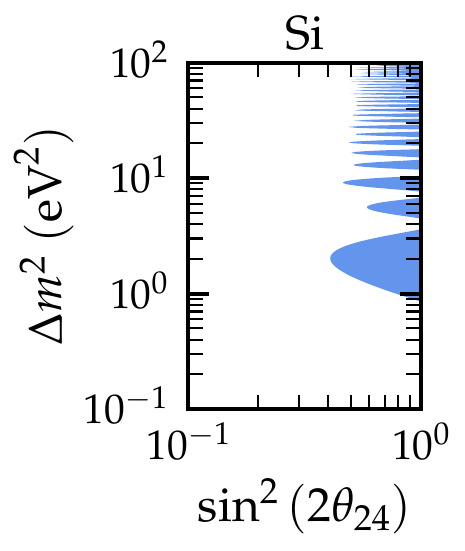}
    \end{subfigure}
    \begin{subfigure}{0.16\textwidth}
        \centering
        \includegraphics[width=1.08\textwidth]{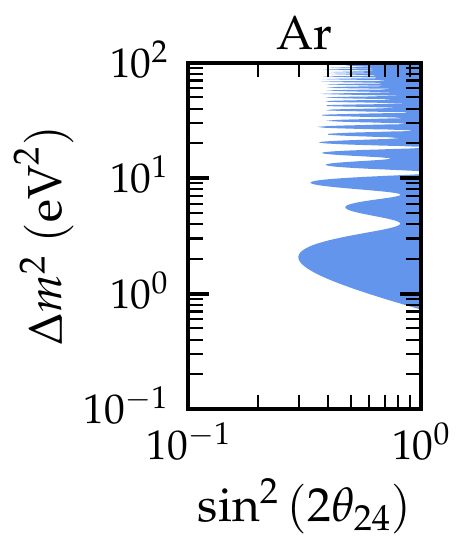}
    \end{subfigure}
    \begin{subfigure}{0.16\textwidth}
        \centering
        \includegraphics[width=1.08\textwidth]{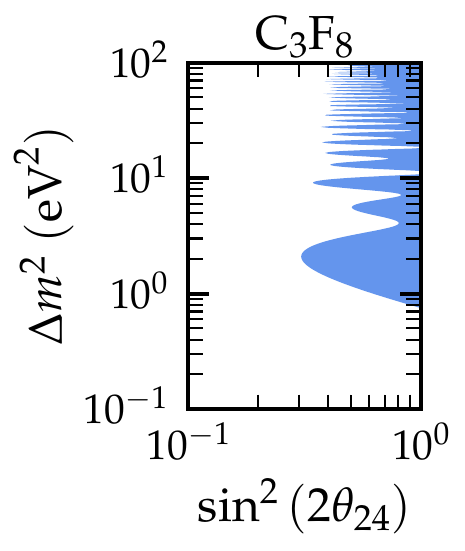}
    \end{subfigure}
\caption{Sensitivity projections at the 90\% C.L. in the $(\sin^2{2\theta_{14}}, \Delta m^2)$  and $(\sin^2{2\theta_{24}}, \Delta m^2)$ planes,  for the different ESS detectors. The  magenta (blue) contours represent the  exclusion regions for active-sterile oscillations involving electron (muon) neutrinos in the upper (lower) panel.}
\label{tab:Sterile_Neutrino_Oscillation_for_different_detectors}
\end{figure}

Focusing now on the active-sterile neutrino oscillation scenario,  the upper and lower panels of Fig.~\ref{tab:Sterile_Neutrino_Oscillation_for_different_detectors} illustrate the 90\% C.L. sensitivities in the $(\sin^2{2\theta_{14}}, \Delta m^2)$ and $(\sin^2{2\theta_{24}}, \Delta m^2)$ parameter space, respectively. As can be seen from the plot, similar sensitivities are found from the analyses of the individual ESS detectors. We conclude that, the ESS sensitivity on this scenario is rather weak and quite far from that of dedicated short baseline neutrino experiments~\cite{Machado:2019oxb}.

\begin{figure}[ht!]
    \centering
    \includegraphics[width=0.32\linewidth]{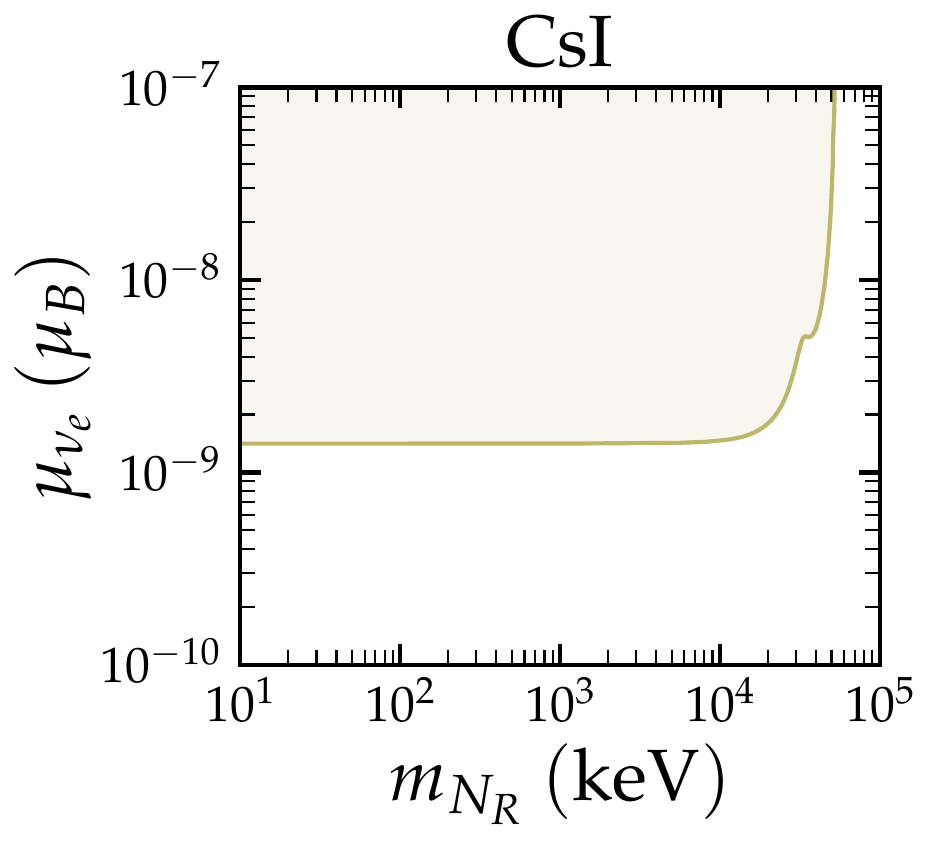}
    \includegraphics[width=0.32\linewidth]{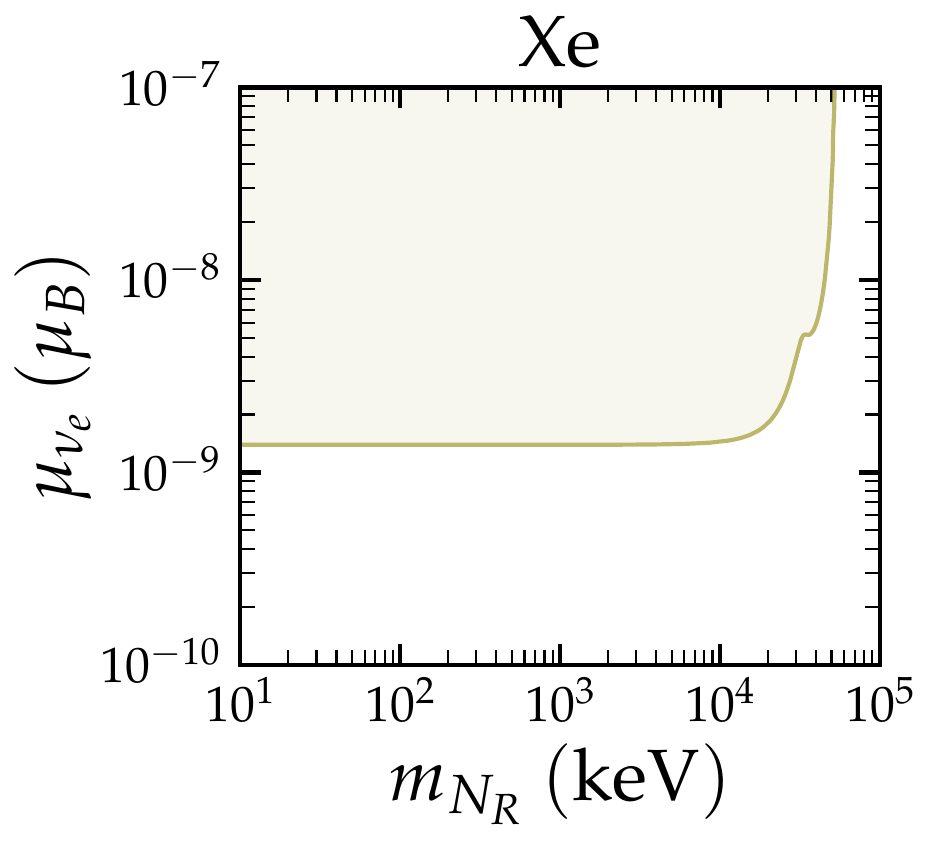}
    \includegraphics[width=0.32\linewidth]{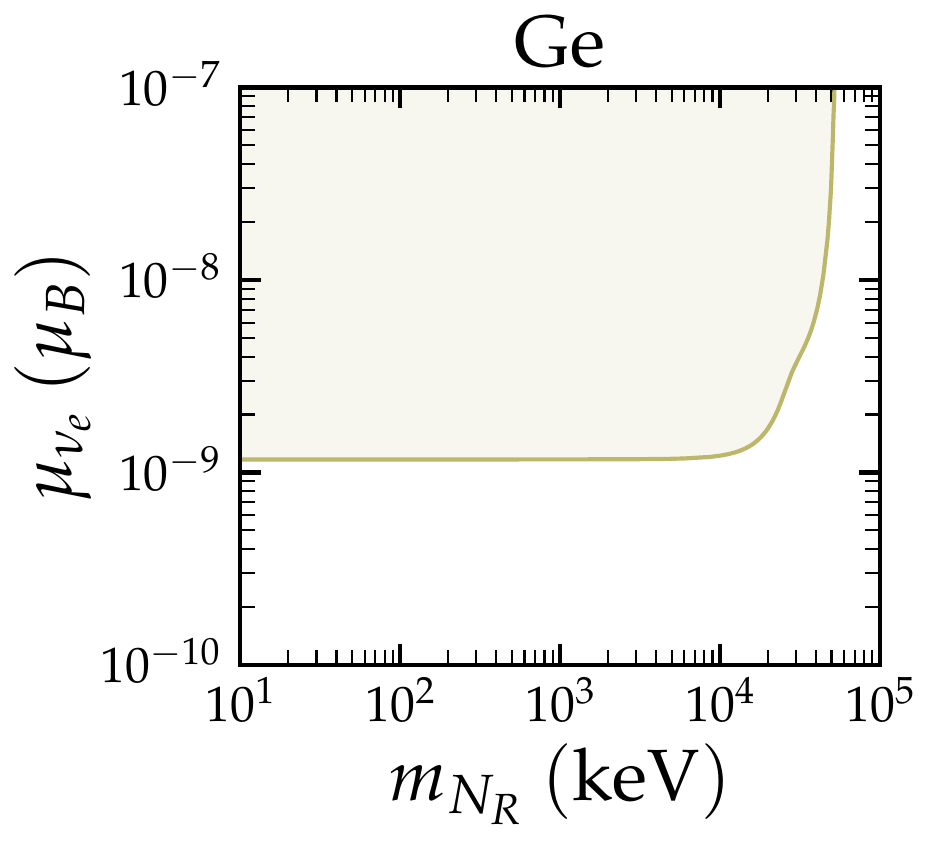}
    \includegraphics[width=0.32\linewidth]{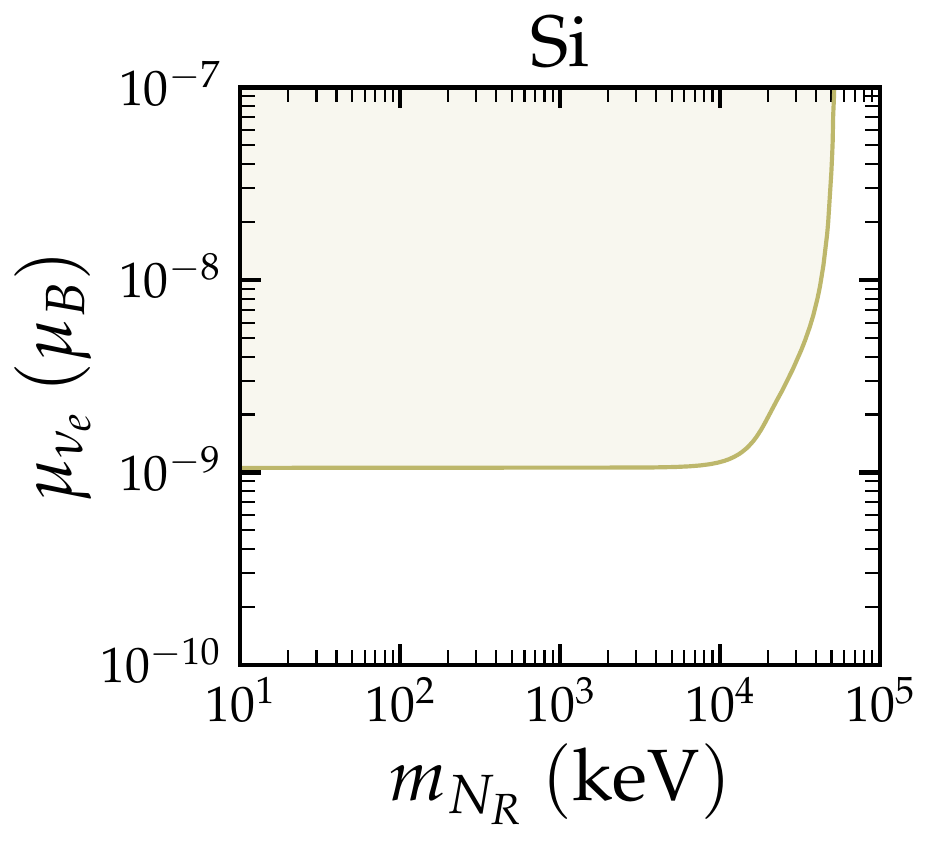}
    \includegraphics[width=0.32\linewidth]{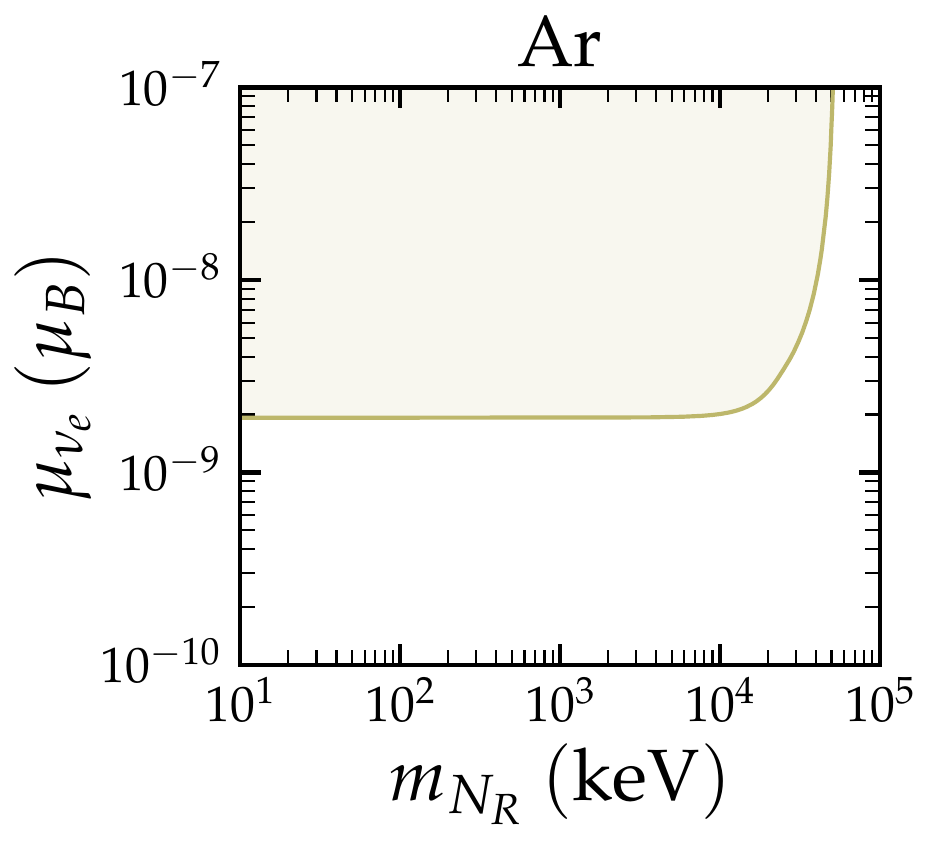}
    \includegraphics[width=0.32\linewidth]{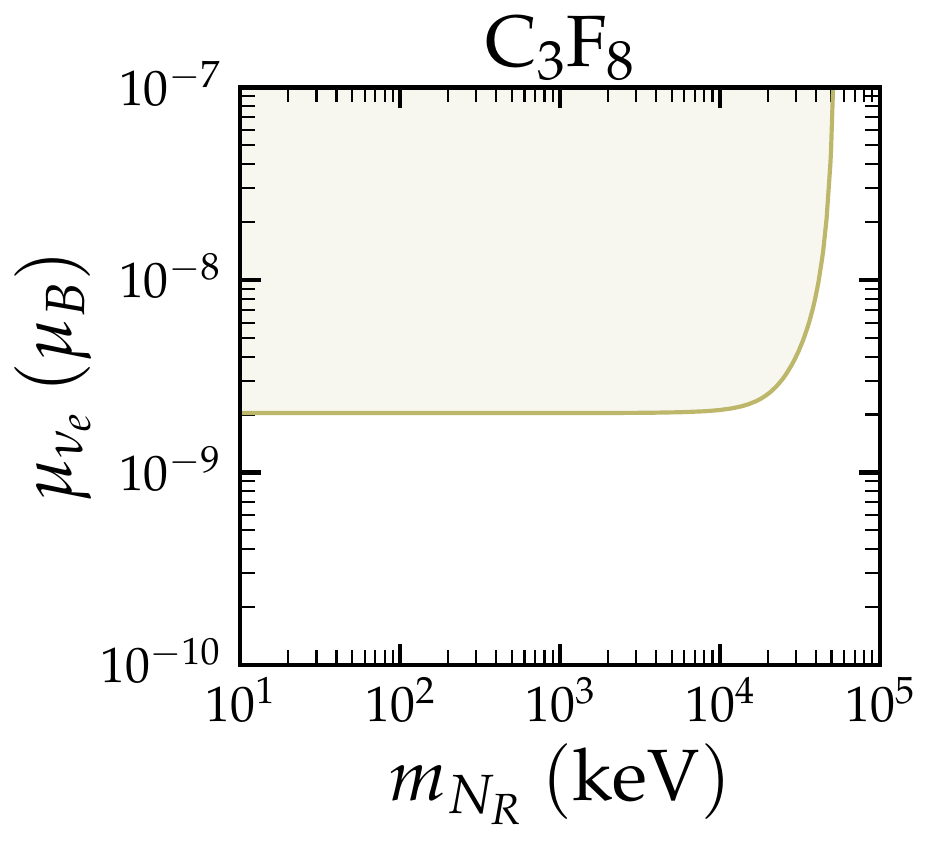}
    \includegraphics[width=0.32\linewidth]{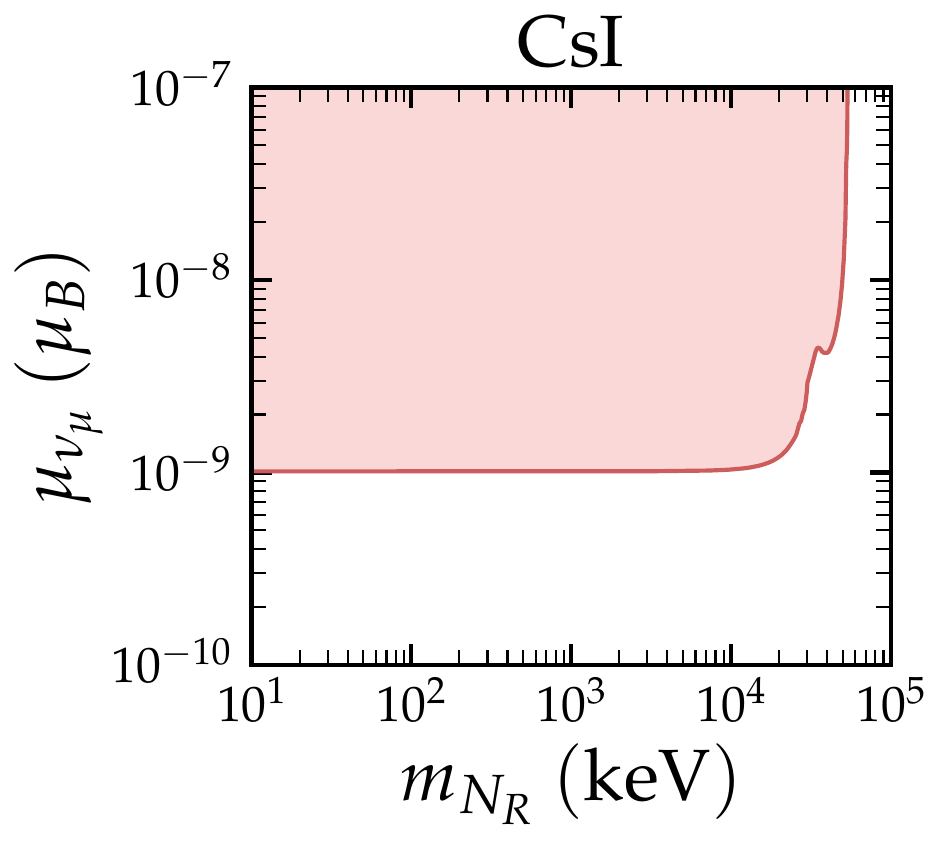}
    \includegraphics[width=0.32\linewidth]{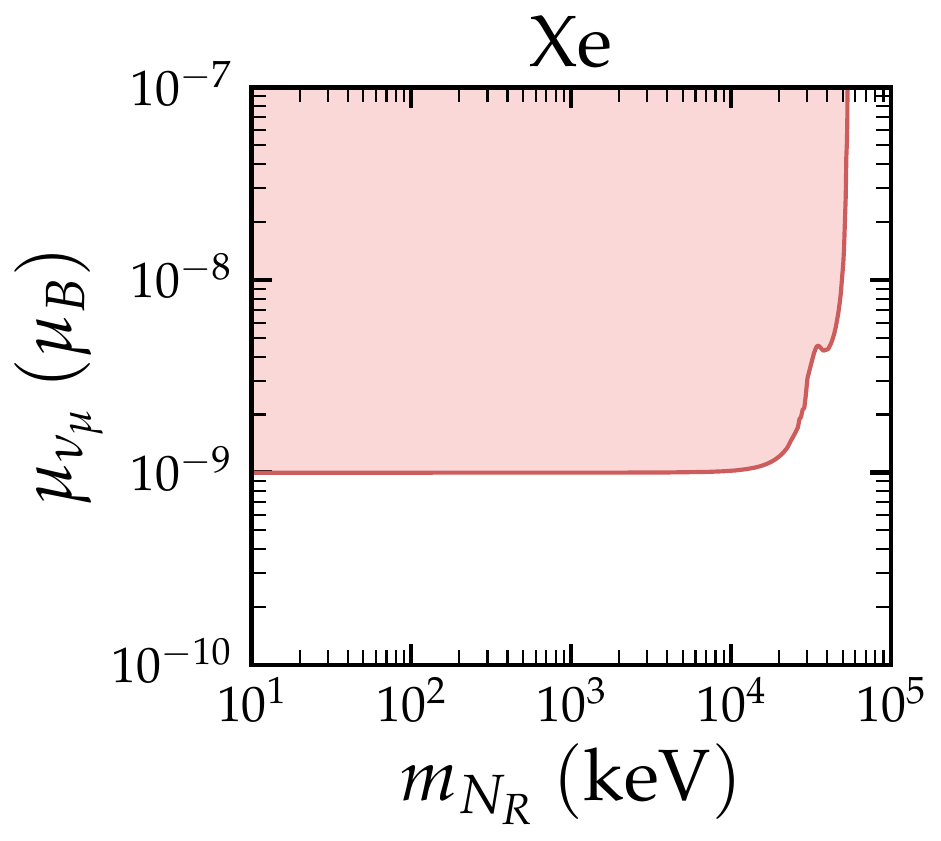}
    \includegraphics[width=0.32\linewidth]{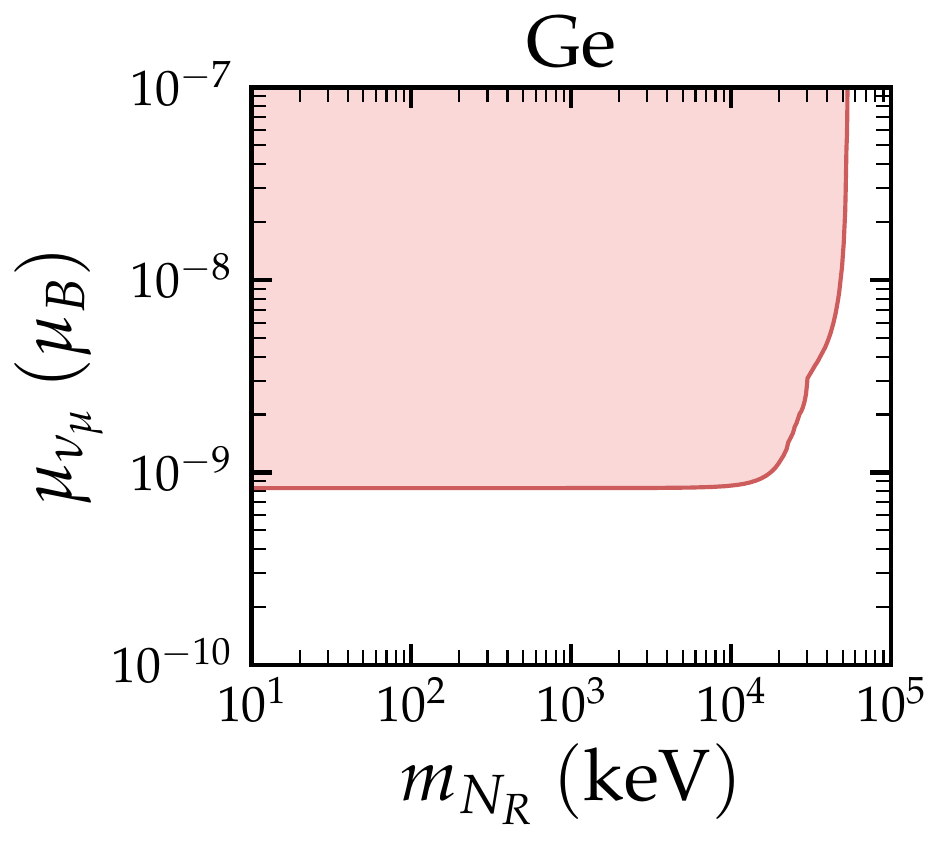}
    \includegraphics[width=0.32\linewidth]{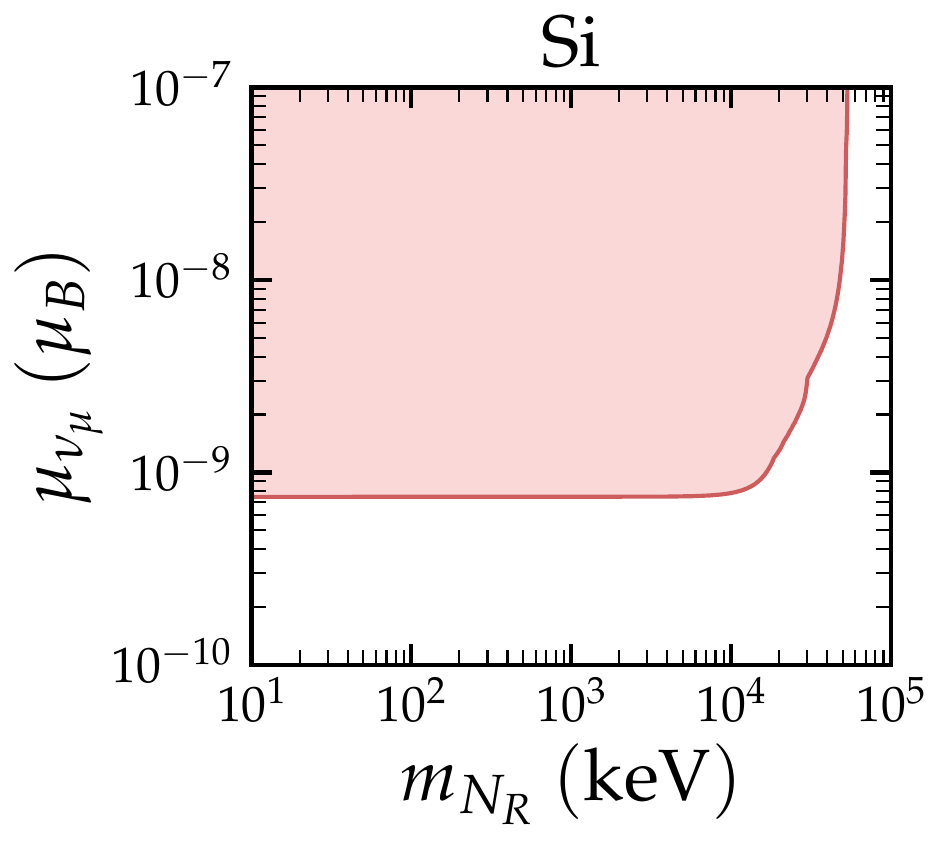}
    \includegraphics[width=0.32\linewidth]{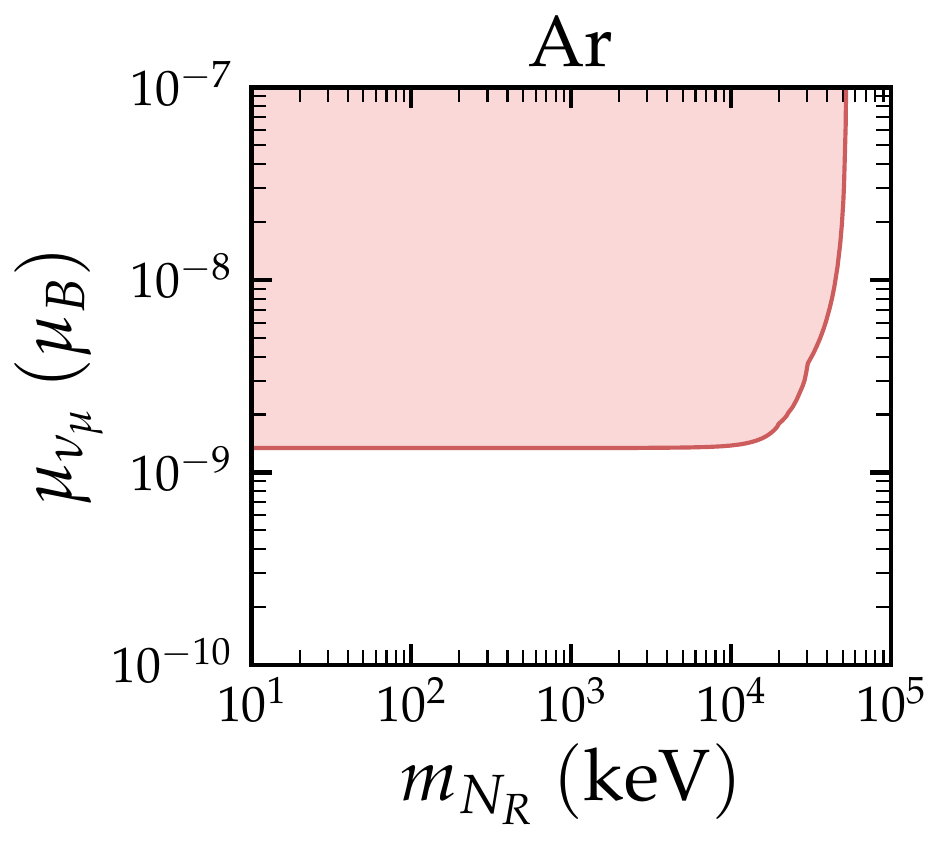}
    \includegraphics[width=0.32\linewidth]{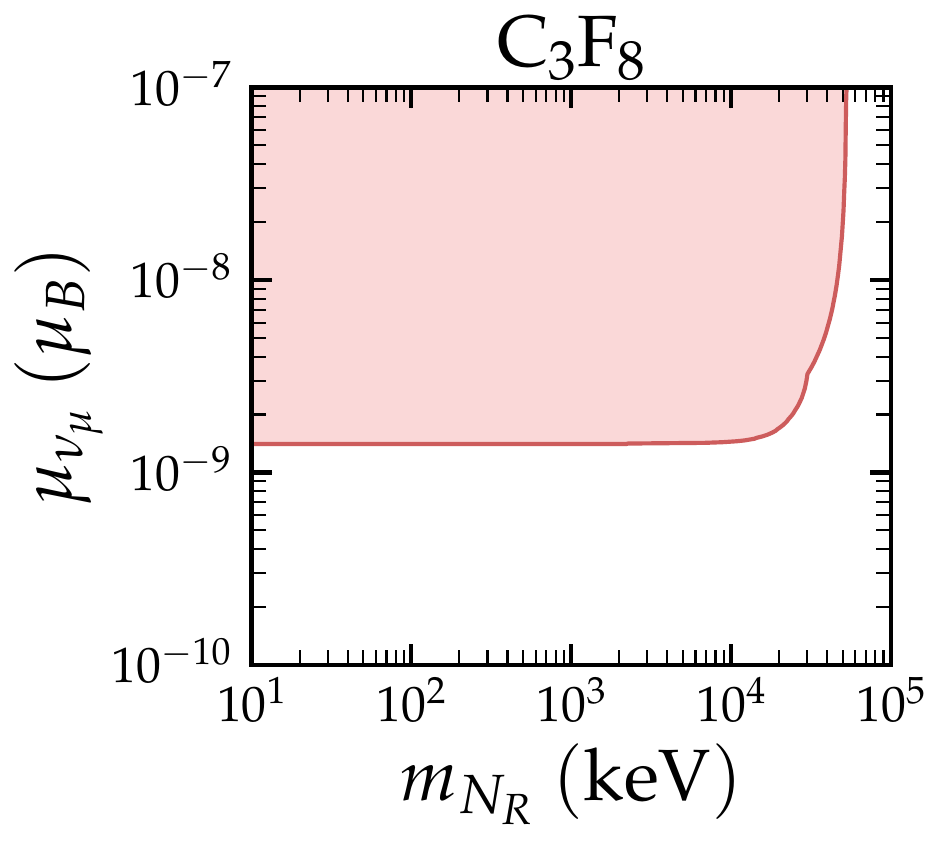}
    \caption{Projected 90\% C.L. exclusion regions in the sterile dipole portal for  the individual ESS detectors. The results are shown for electron (muon) neutrinos in the 1st and 2nd (3rd and 4th) rows.}
    \label{Fig:dipole_mu_nu_e_different_detector}
\end{figure}

We now focus on the active-sterile dipole portal scenario, i.e., we explore the possibility of generating a massive final state sterile neutral leptons via upscattering in the presence of nonzero TMMs. For the different ESS detectors, the upper two rows of Fig.~\ref{Fig:dipole_mu_nu_e_different_detector} present the projected exclusion limits at 90\% C.L. in the  $(m_{N_R}, \mu_{\nu_e})$ plane, while the lower two rows show the corresponding result in the  $(m_{N_R}, \mu_{\nu_\mu})$ planes. As can be seen, the Si and Ge detectors are anticipated to provide the most stringent constraints, whereas the Ar and $\mathrm{C_3F_8}$ detectors ---for which a single-bin analysis is performed--- are expected to yield the least stringent constraints. 

\begin{figure}
    \centering
    \includegraphics[width=0.328\linewidth]{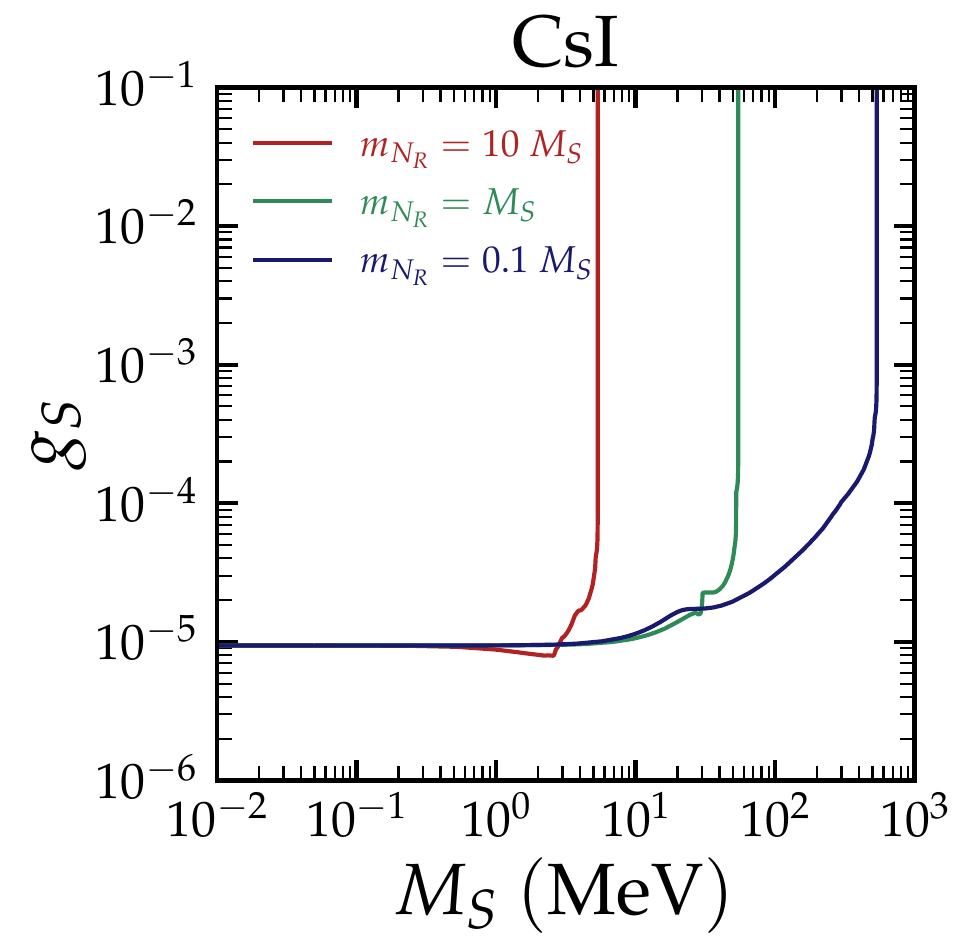}
    \includegraphics[width=0.328\linewidth]{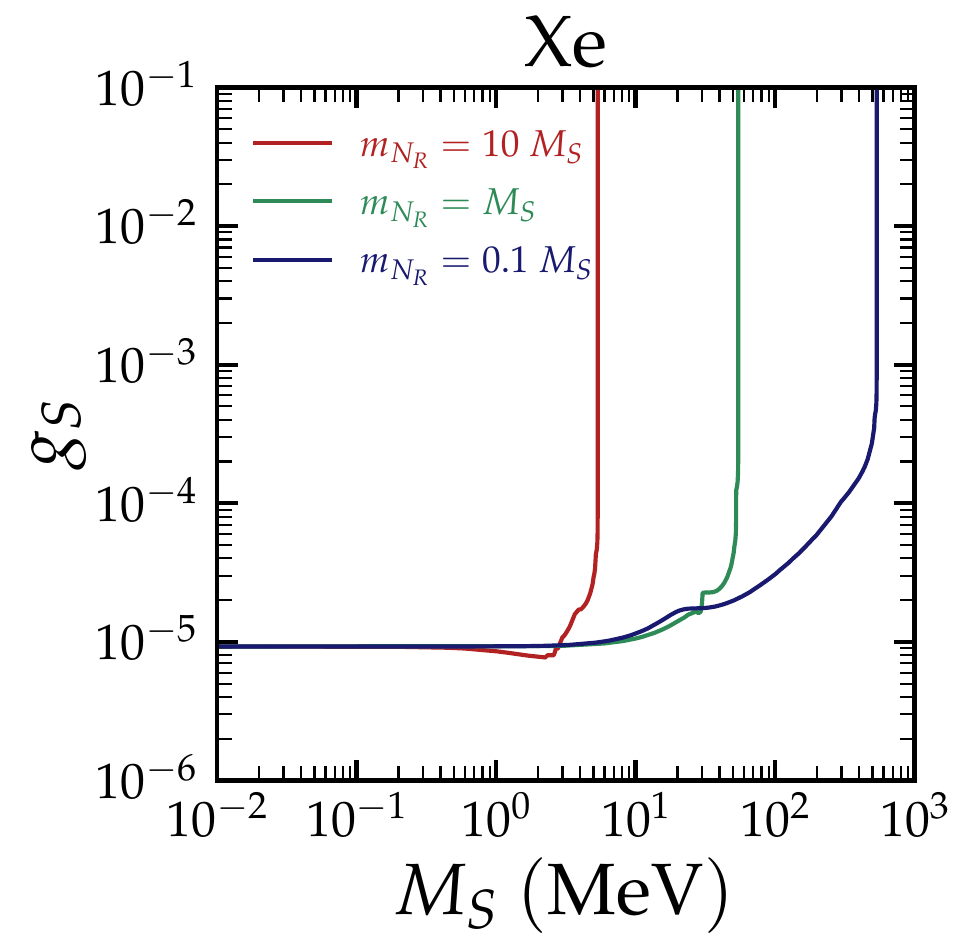}
    \includegraphics[width=0.328\linewidth]{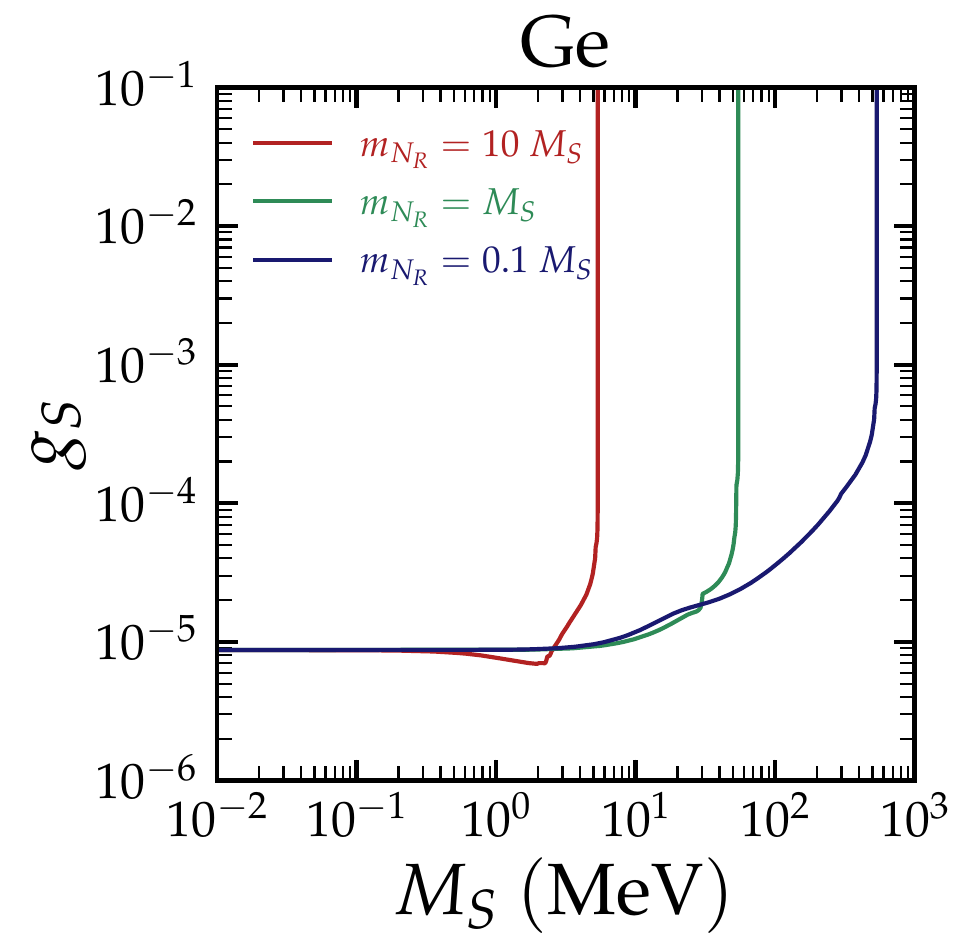}
    \includegraphics[width=0.328\linewidth]{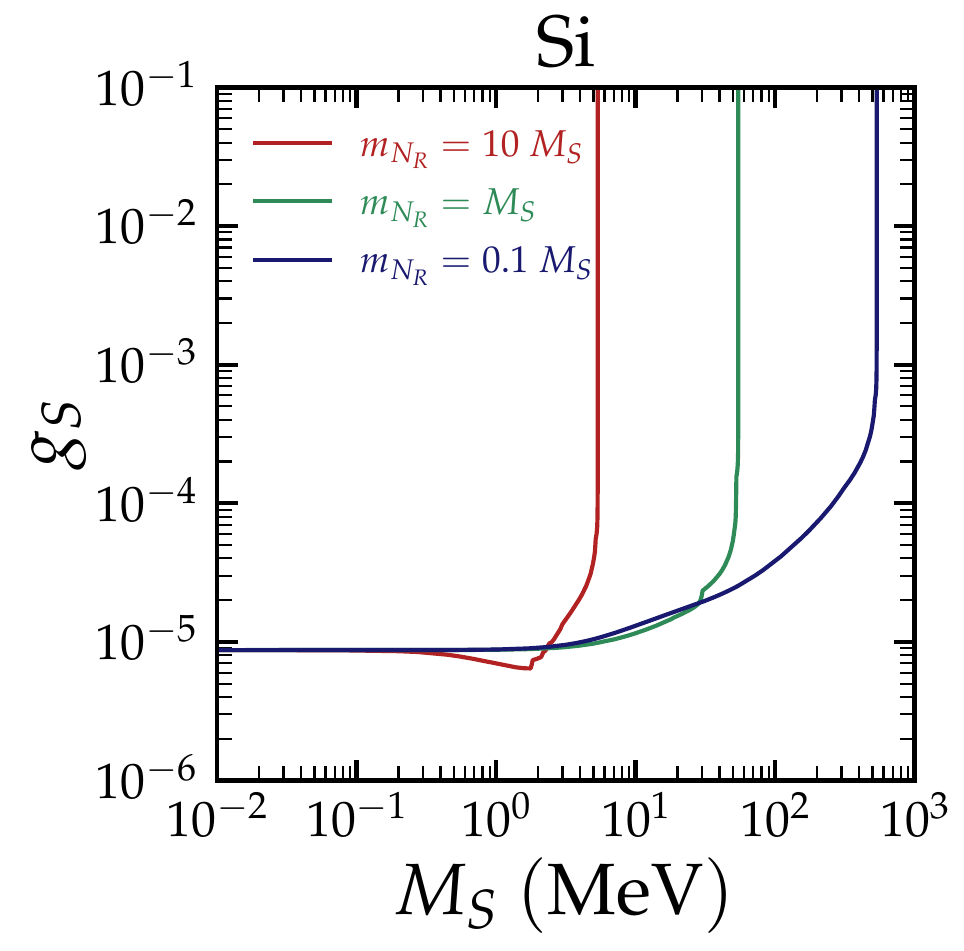}
    \includegraphics[width=0.328\linewidth]{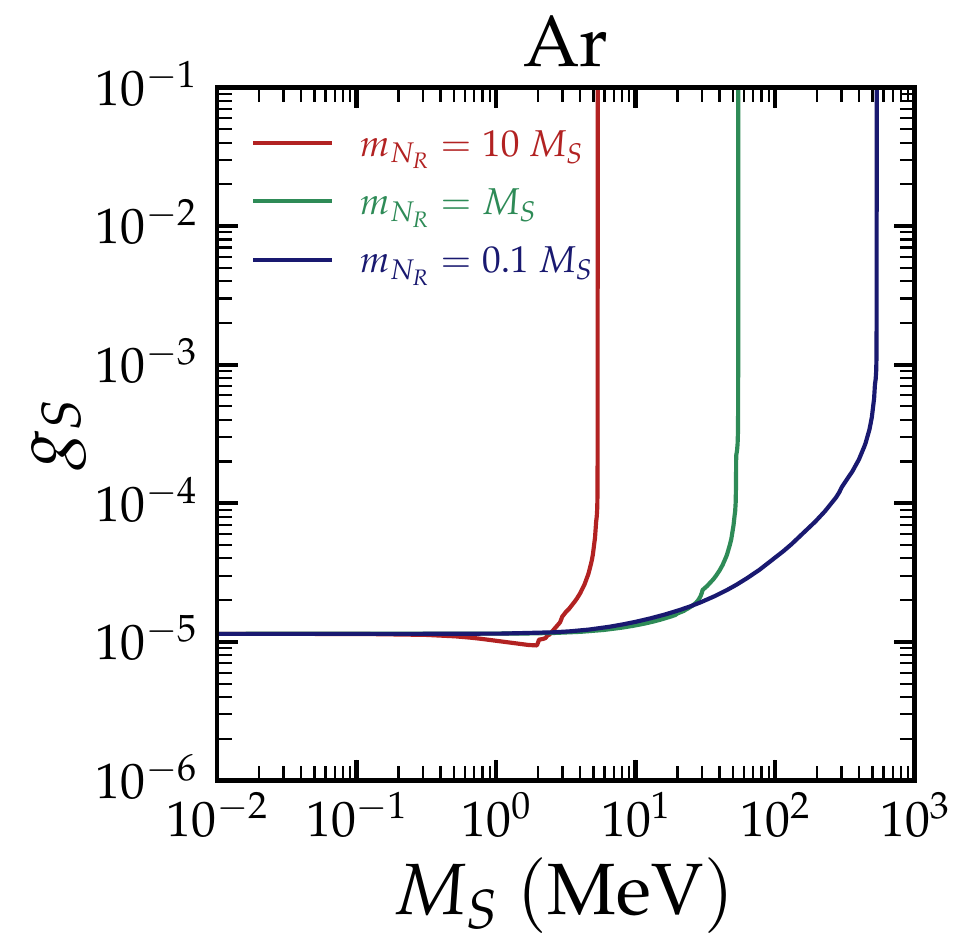}
    \includegraphics[width=0.328\linewidth]{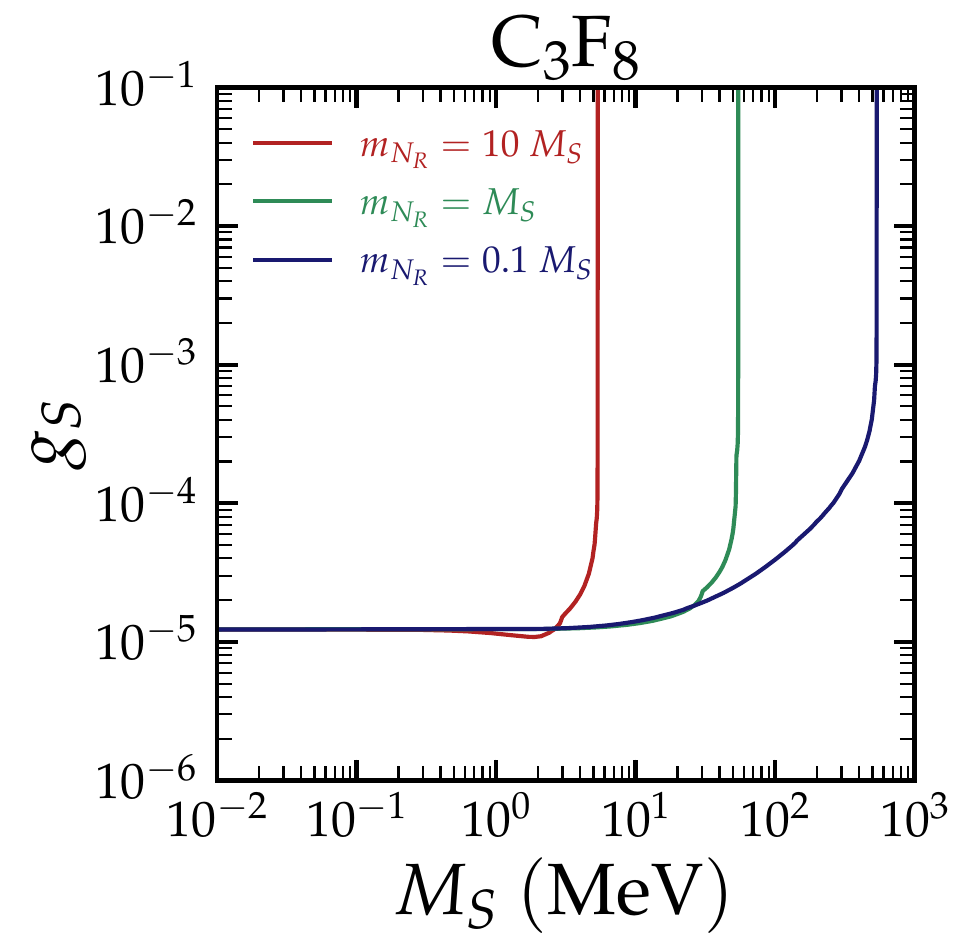}
    \includegraphics[width=0.328\linewidth]{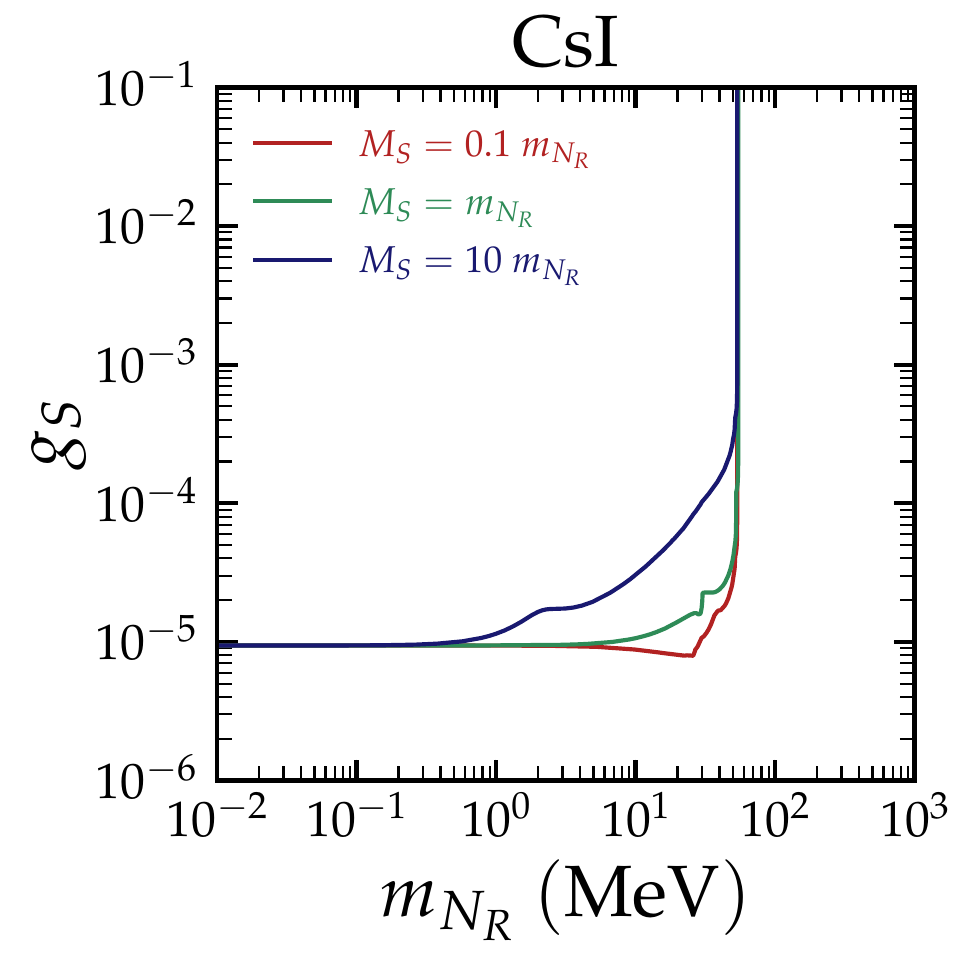}
    \includegraphics[width=0.328\linewidth]{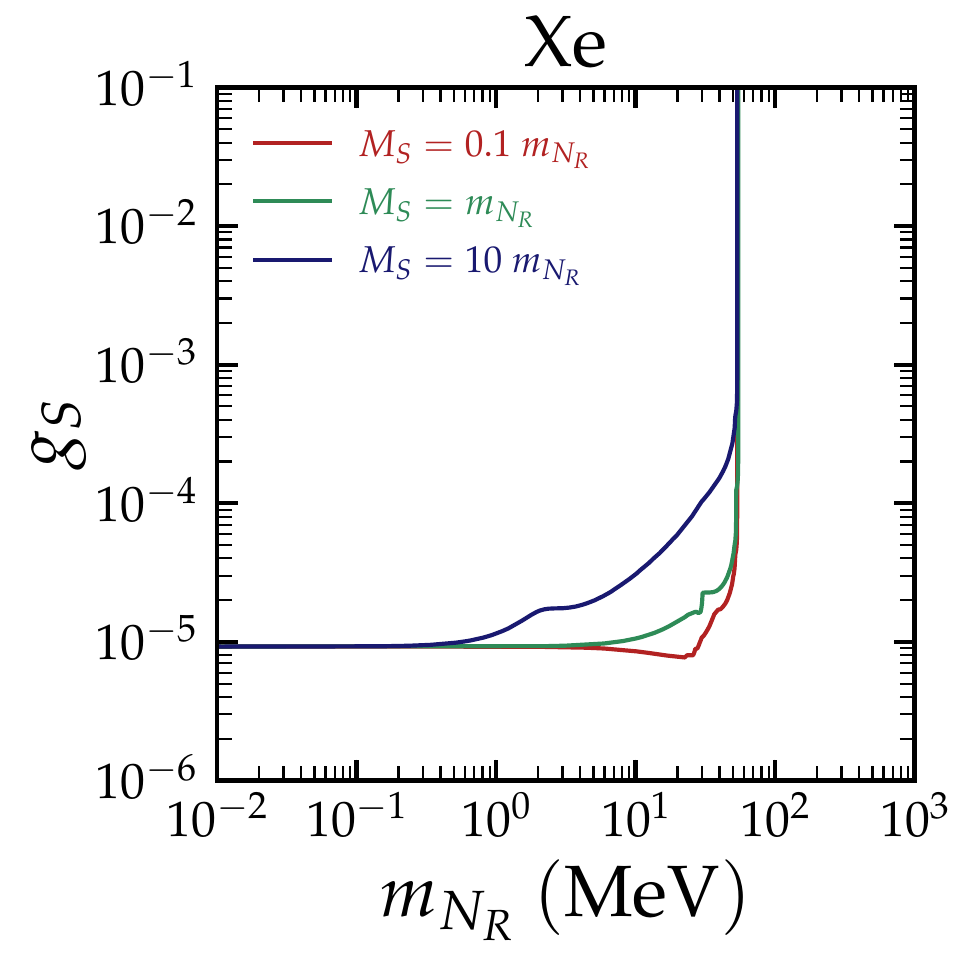}
    \includegraphics[width=0.328\linewidth]{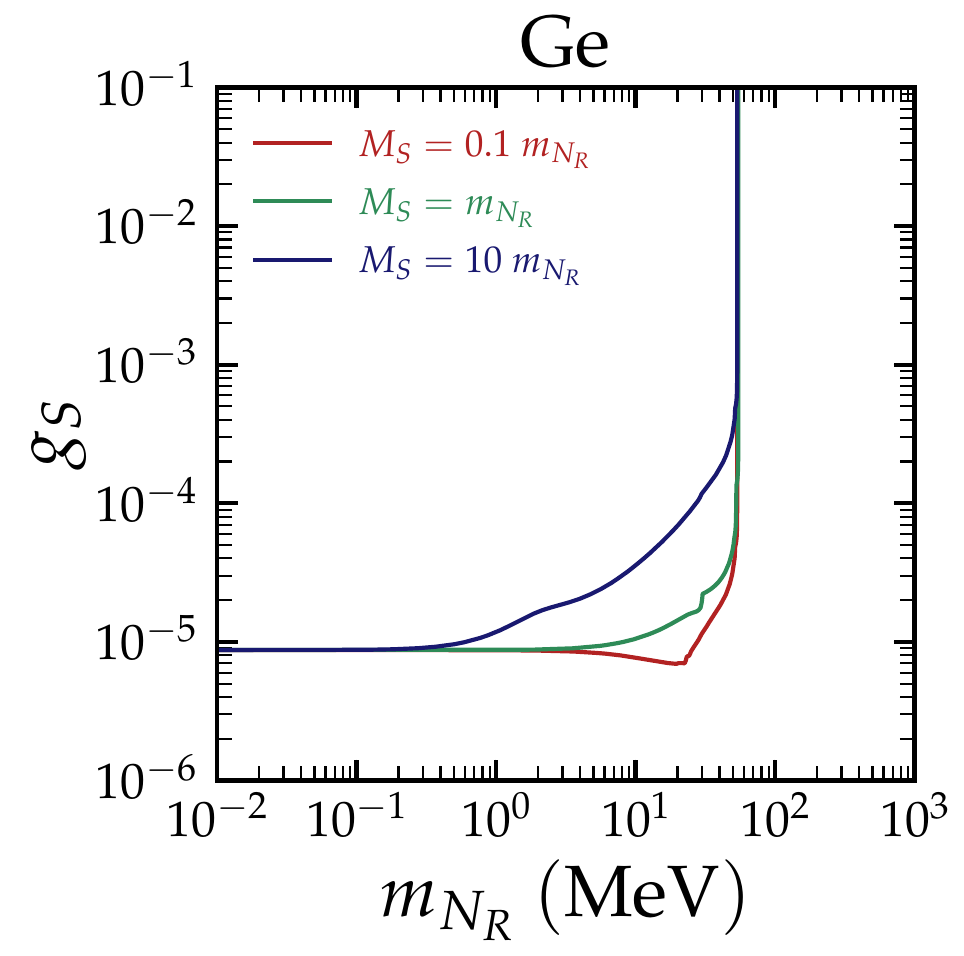}
    \includegraphics[width=0.328\linewidth]{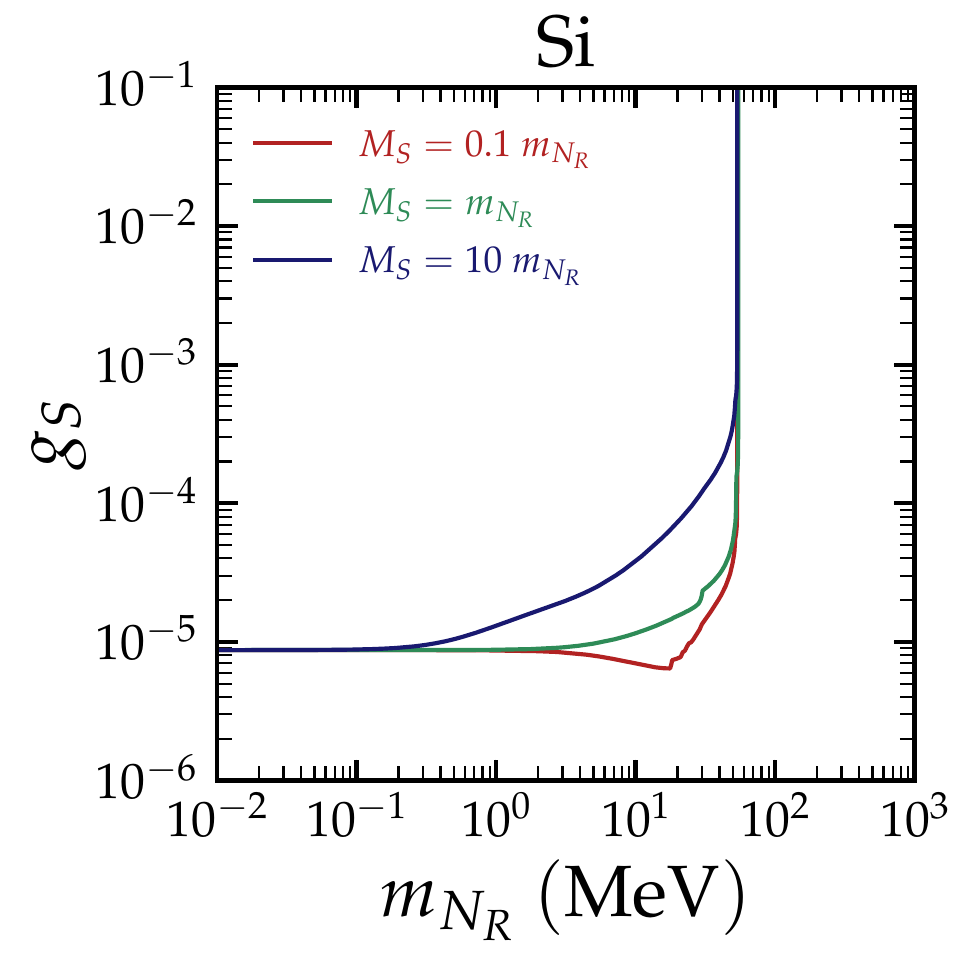}
    \includegraphics[width=0.328\linewidth]{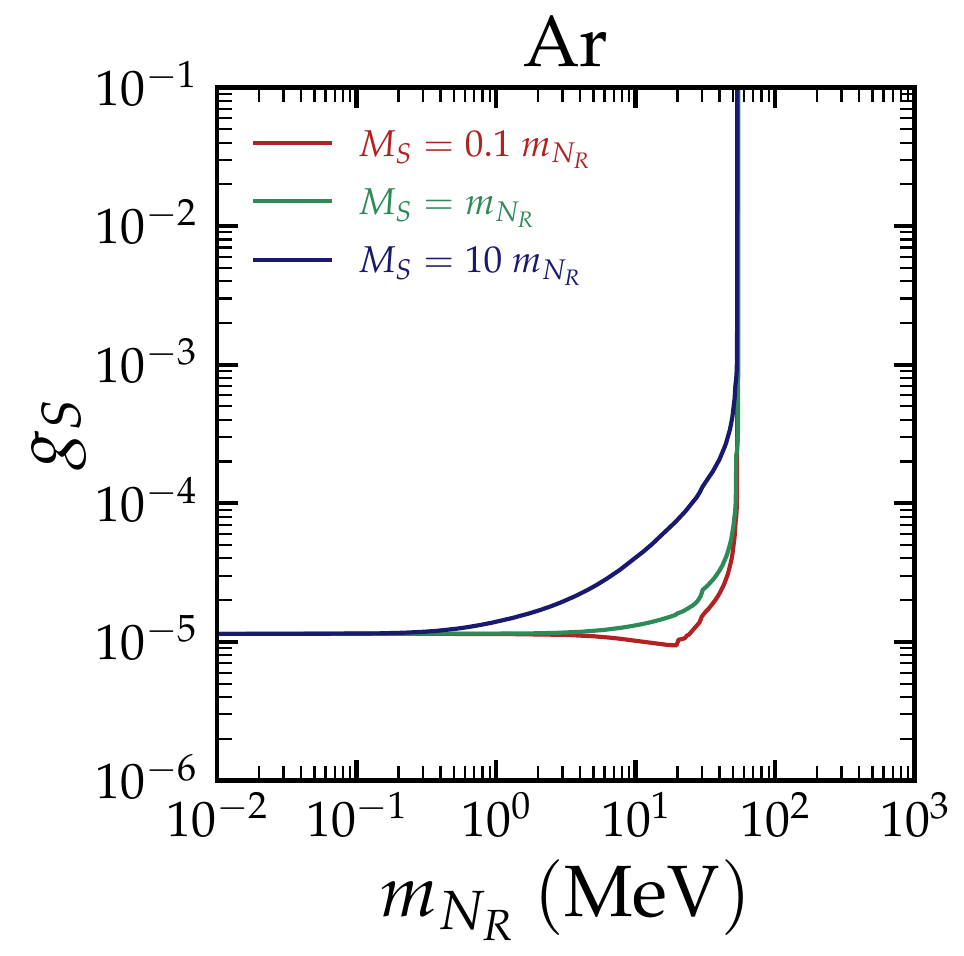}
    \includegraphics[width=0.328\linewidth]{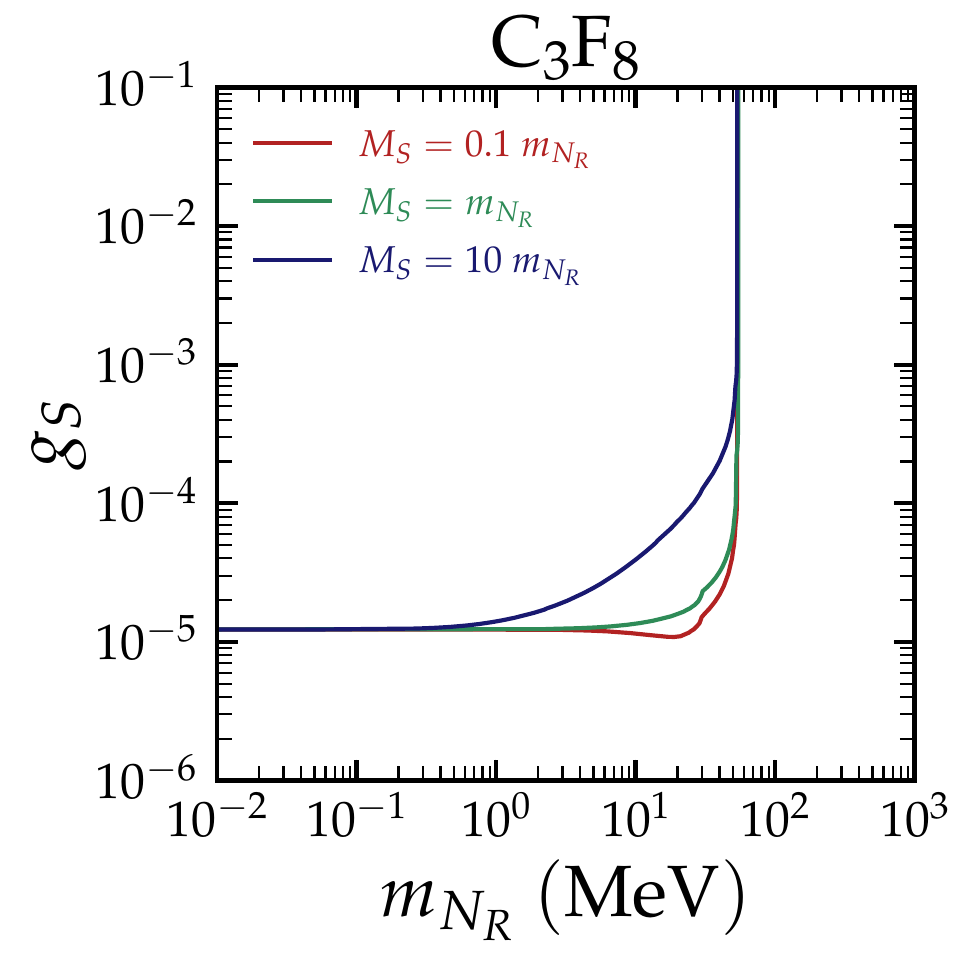}
    \caption{Projected 90\% C.L. limits for the case of the scalar-mediated scenario for SNL  production via upscattering, assuming  three   benchmark scenarios: $m_{N_R} = \{0.1, 1, 10\} \times M_S$. The results are projected in the $(M_S, g_S)$ plane (1st and 2nd rows) and in the $(m_{N_R}, g_S)$  plane (3rd and 4th rows), for the individual detectors at ESS.}
    \label{fig:Upscattering_Scalar_different_detector_1}
\end{figure}

\begin{figure}
    \centering
    \includegraphics[width=0.328\linewidth]{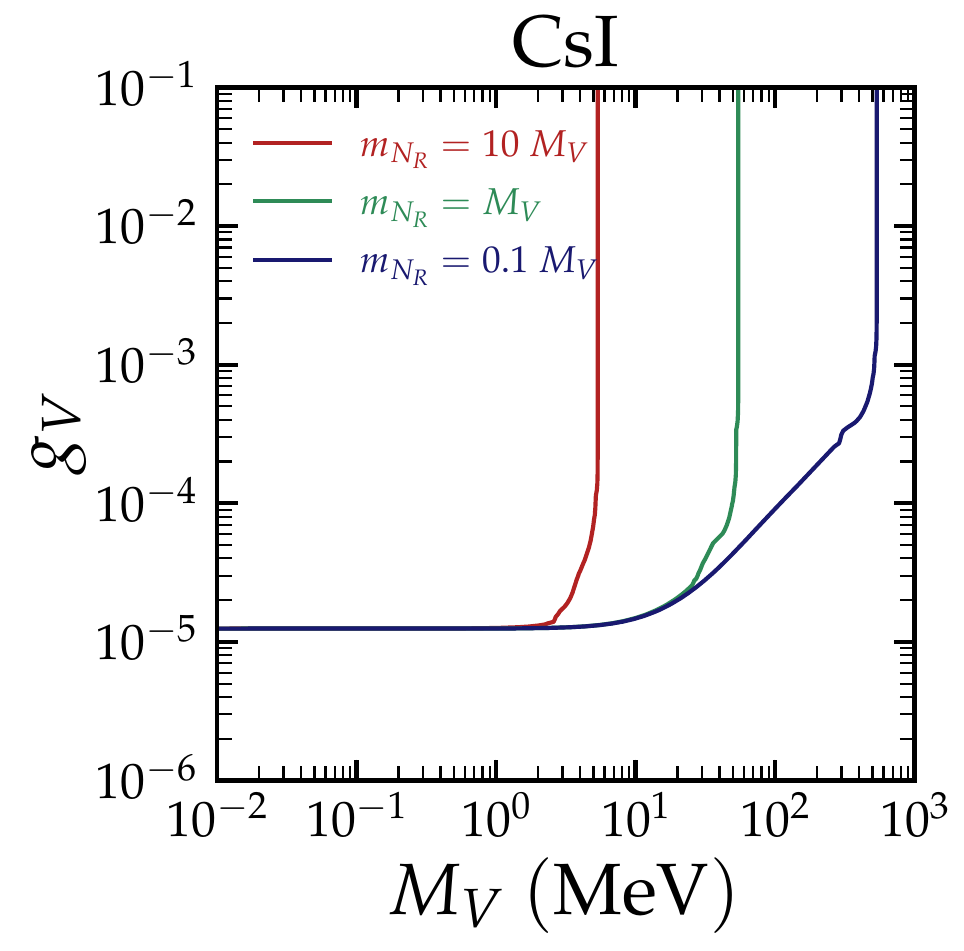}
    \includegraphics[width=0.328\linewidth]{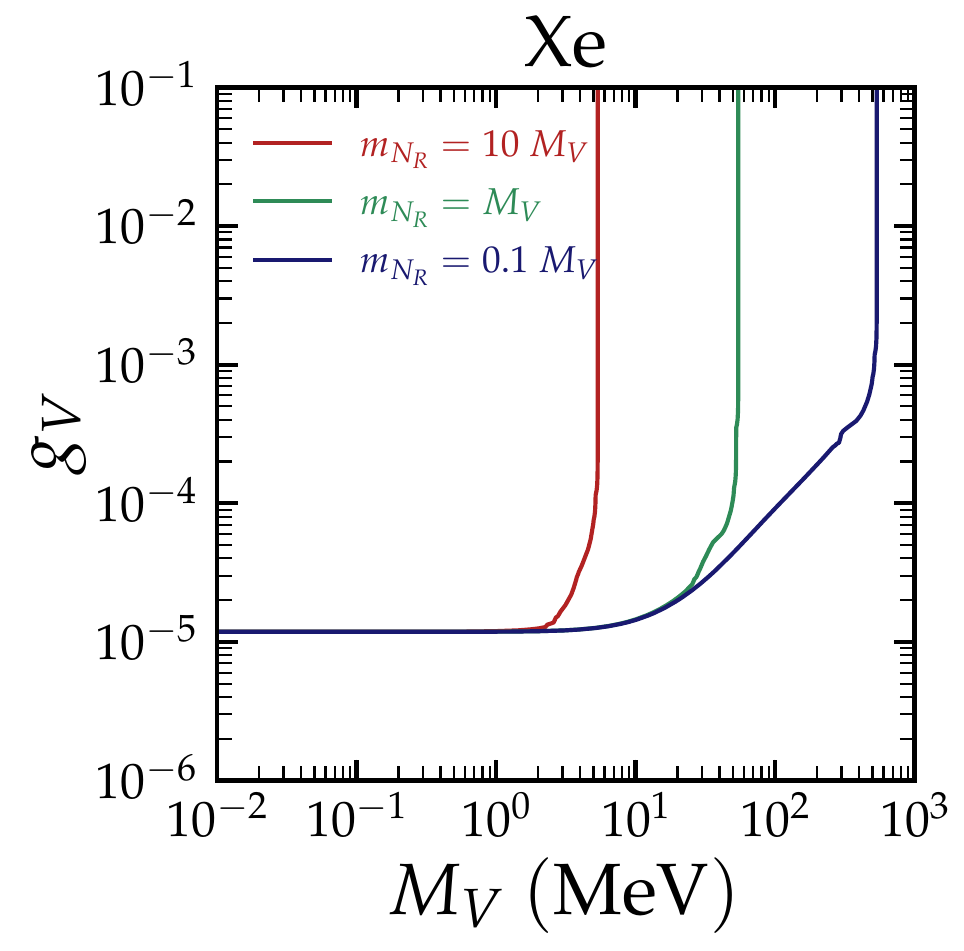}
    \includegraphics[width=0.328\linewidth]{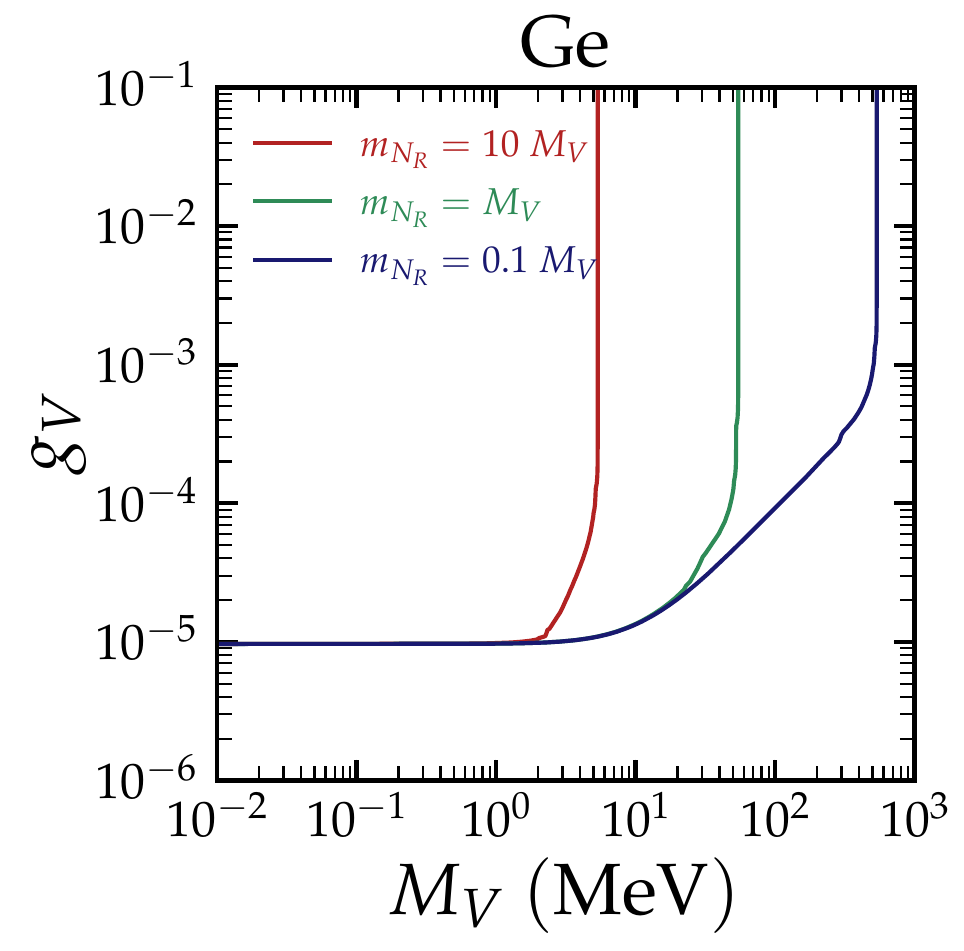}
    \includegraphics[width=0.328\linewidth]{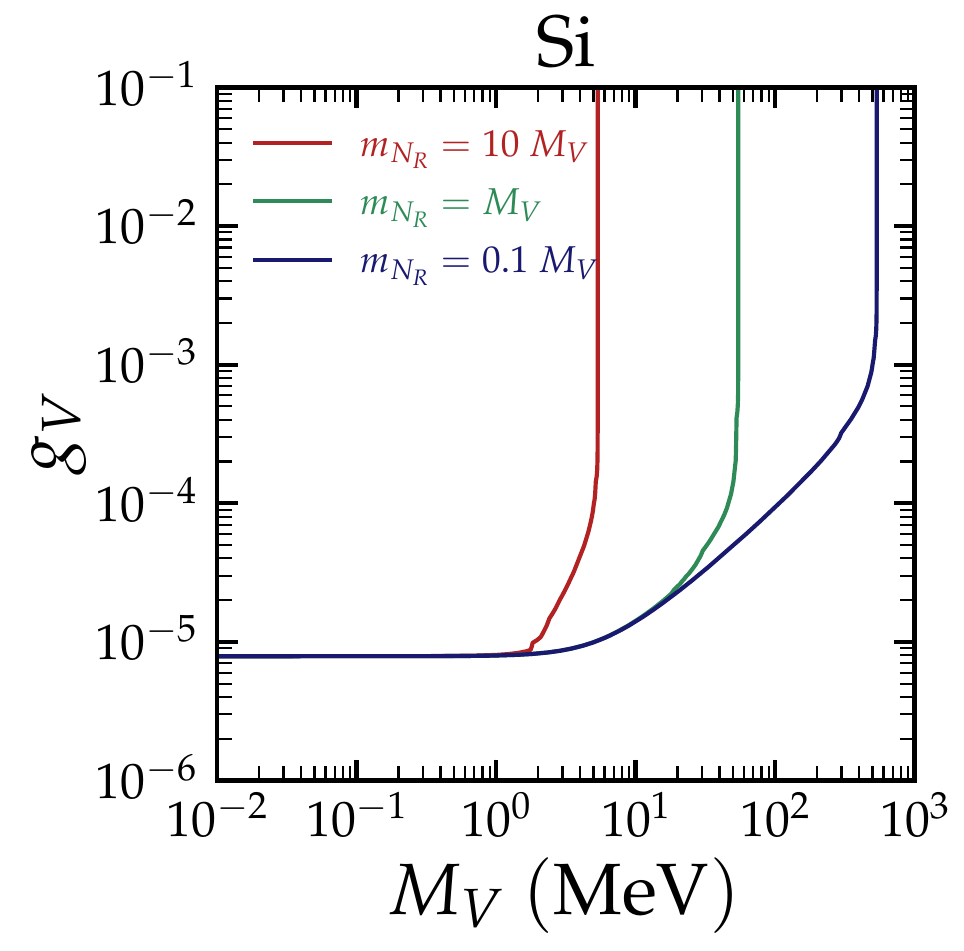}
    \includegraphics[width=0.328\linewidth]{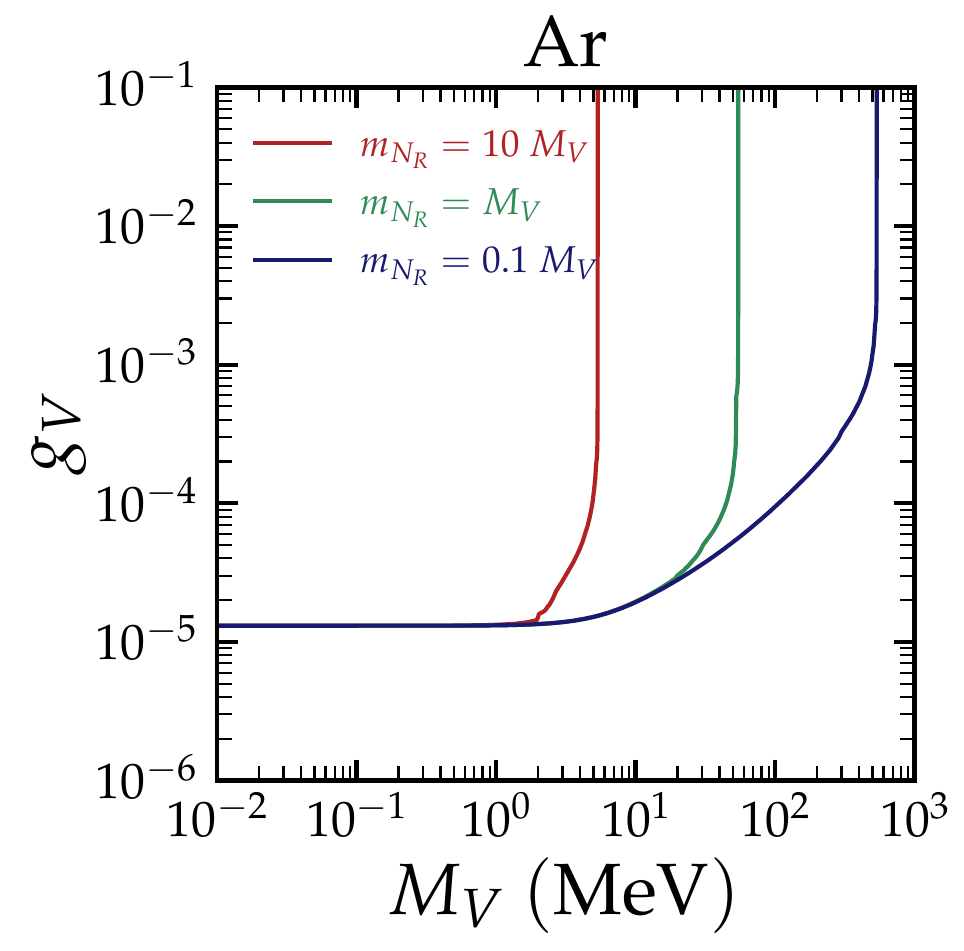}
    \includegraphics[width=0.328\linewidth]{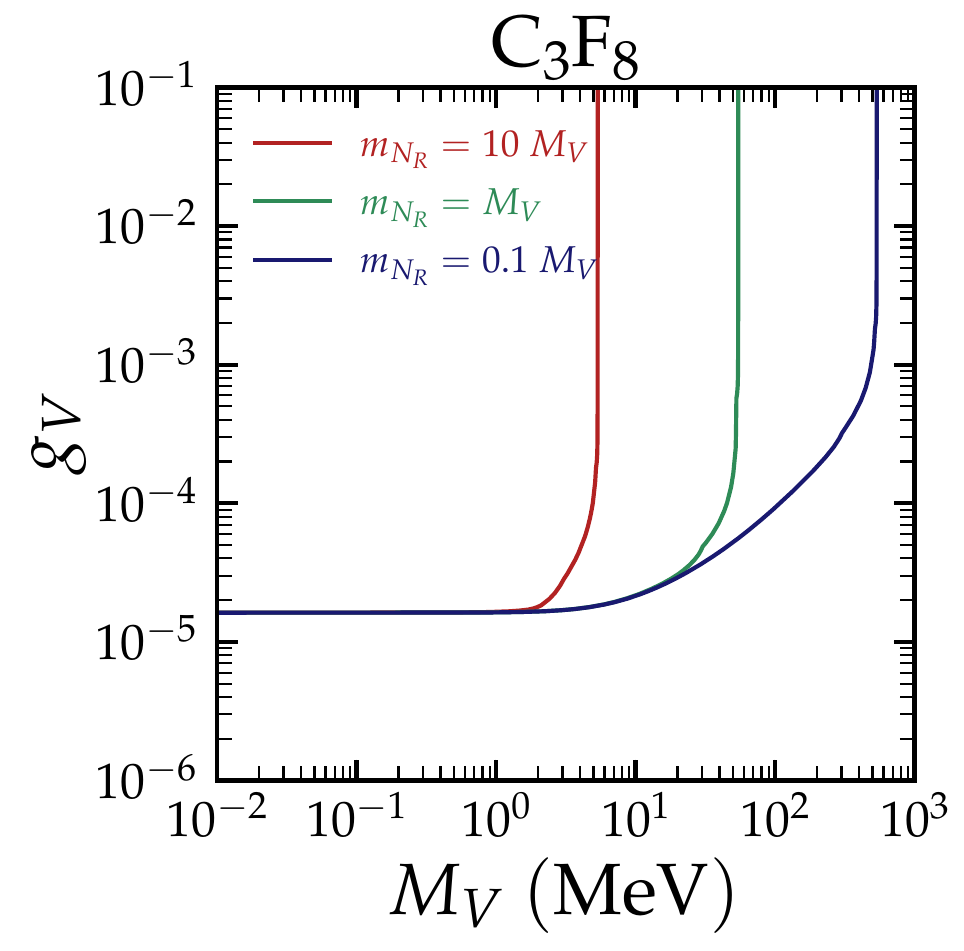}
    \includegraphics[width=0.328\linewidth]{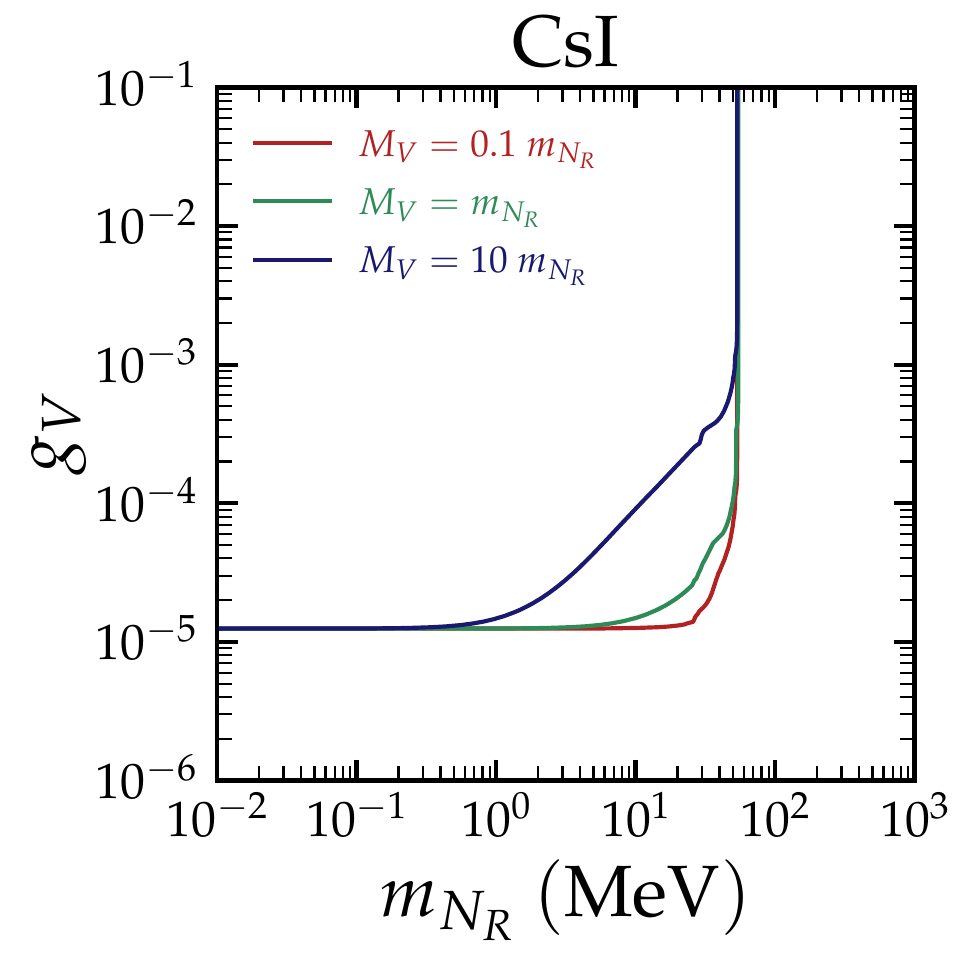}
    \includegraphics[width=0.328\linewidth]{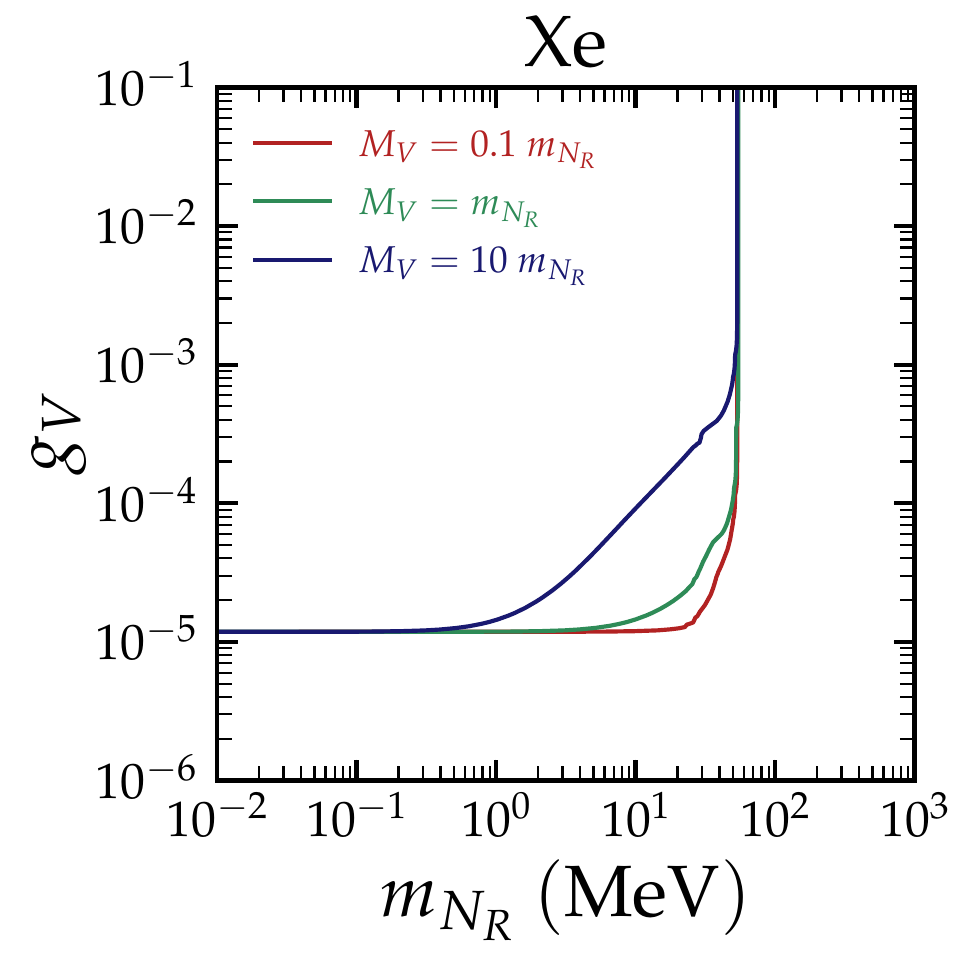}
    \includegraphics[width=0.328\linewidth]{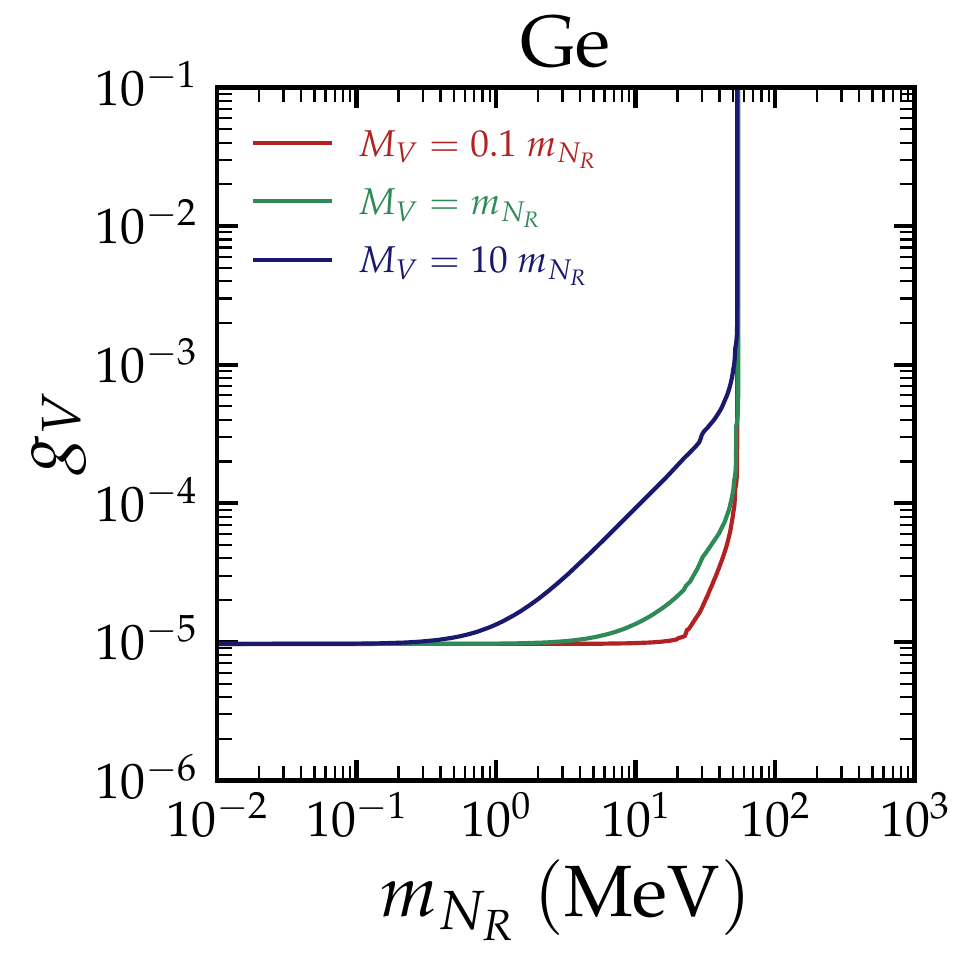}
    \includegraphics[width=0.328\linewidth]{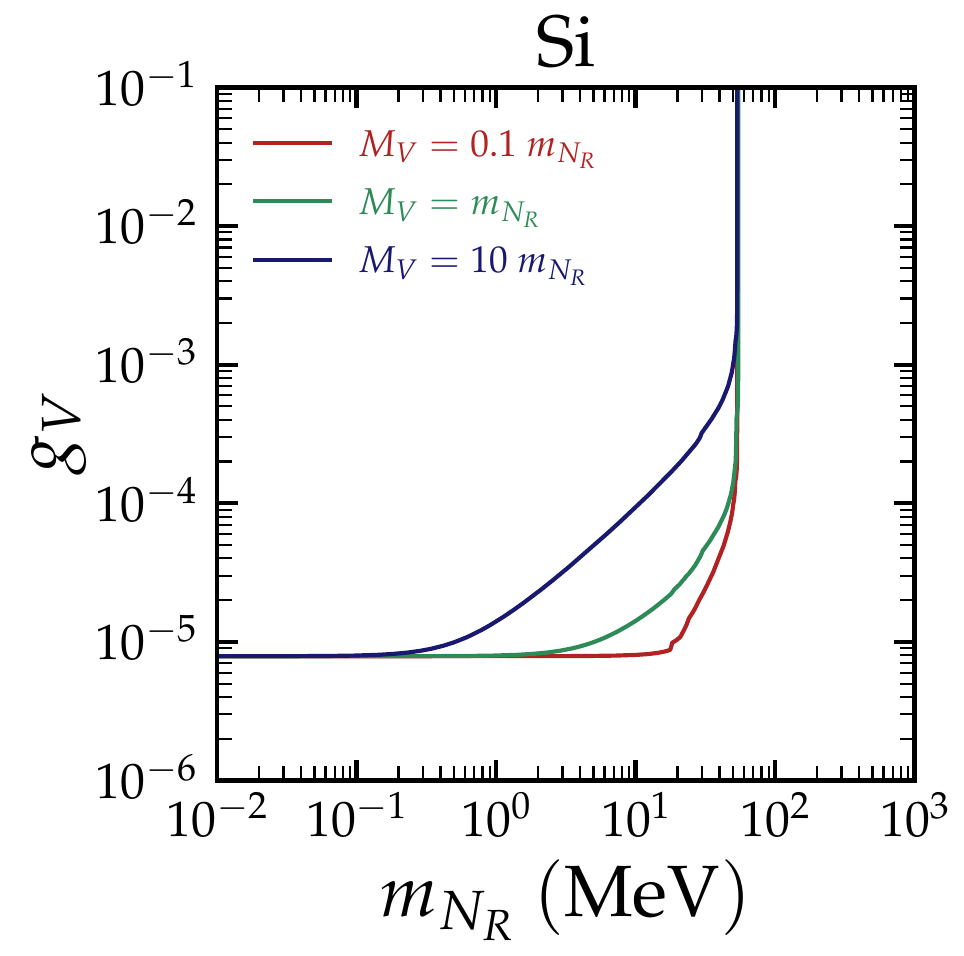}
    \includegraphics[width=0.328\linewidth]{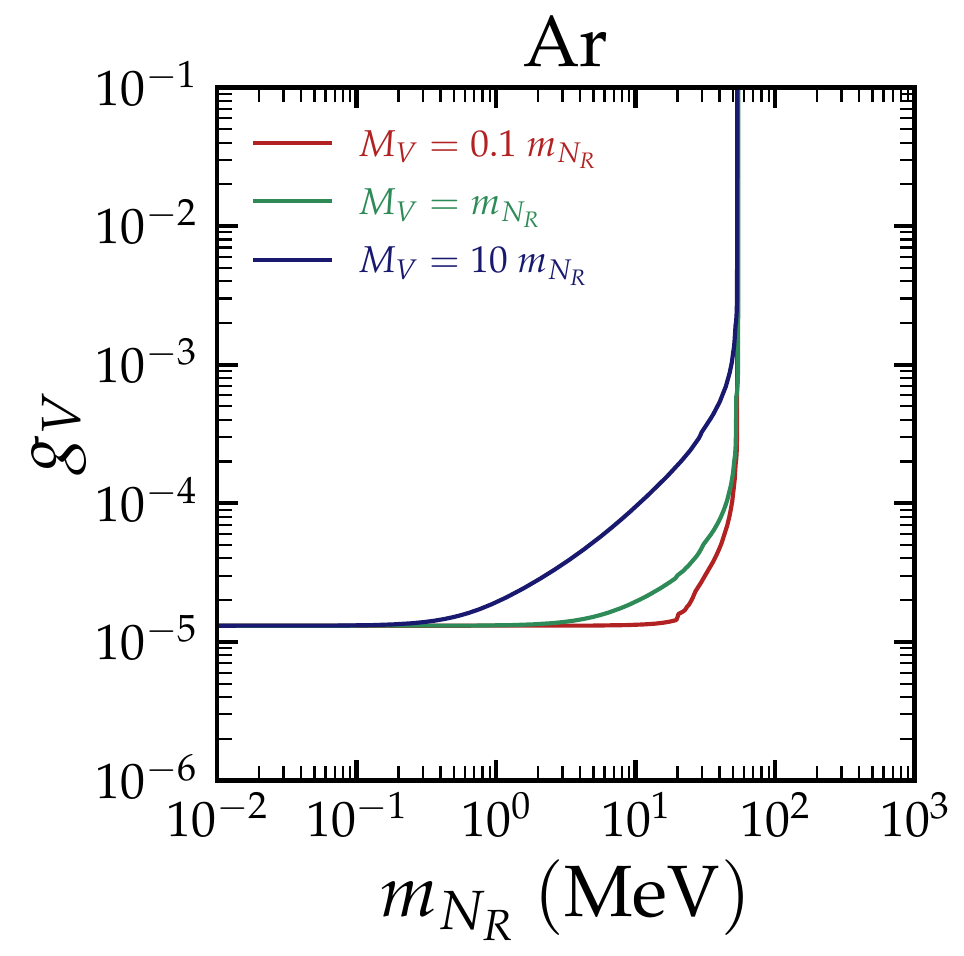}
    \includegraphics[width=0.328\linewidth]{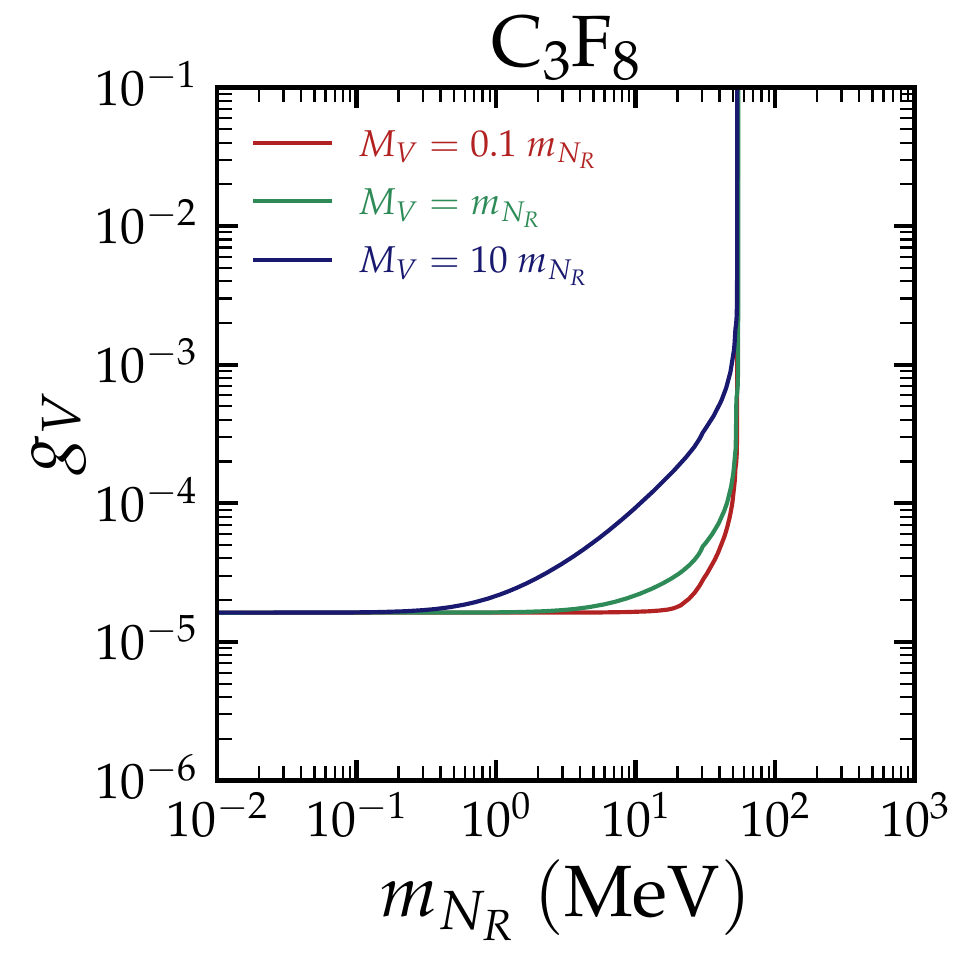}
    \caption{Same as Fig.~\ref{fig:Upscattering_Scalar_different_detector_1}, but for vector-mediated upscattering.}
    \label{fig:Upscattering_Vector_different_detector_1}
\end{figure}

\begin{figure}
    \centering
            \includegraphics[width=0.45\linewidth]{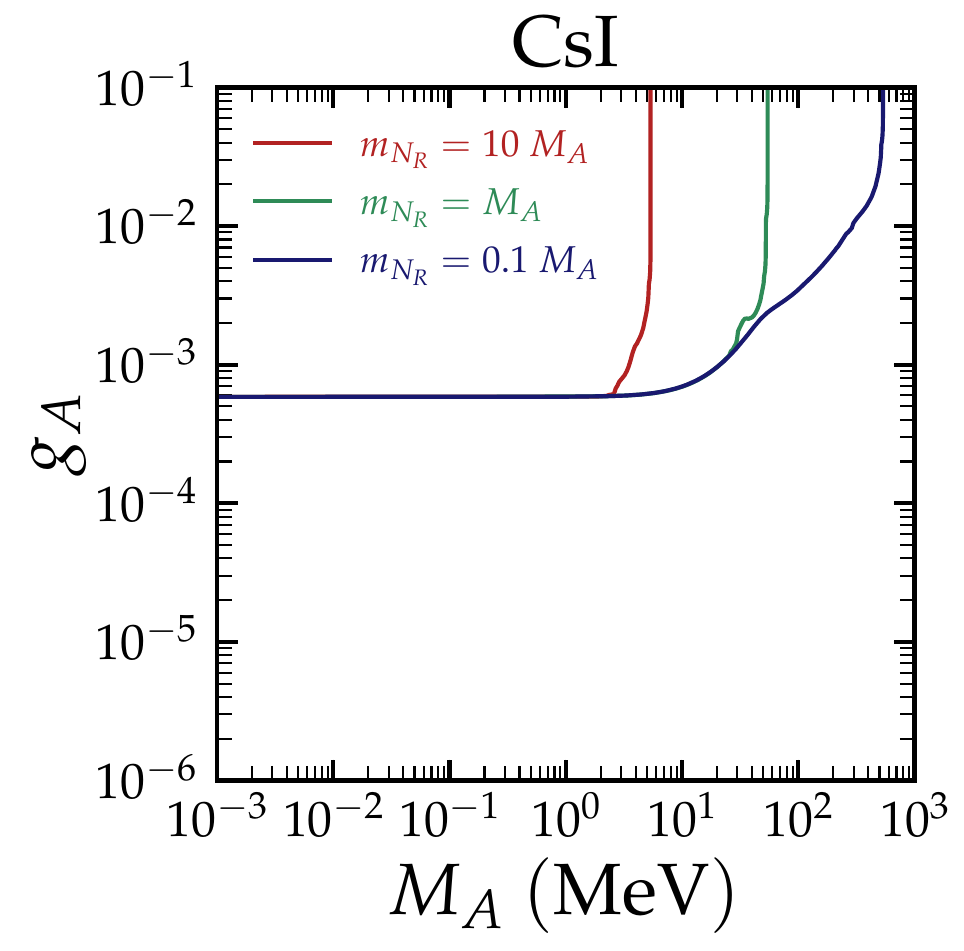}
            \includegraphics[width=0.45\linewidth]{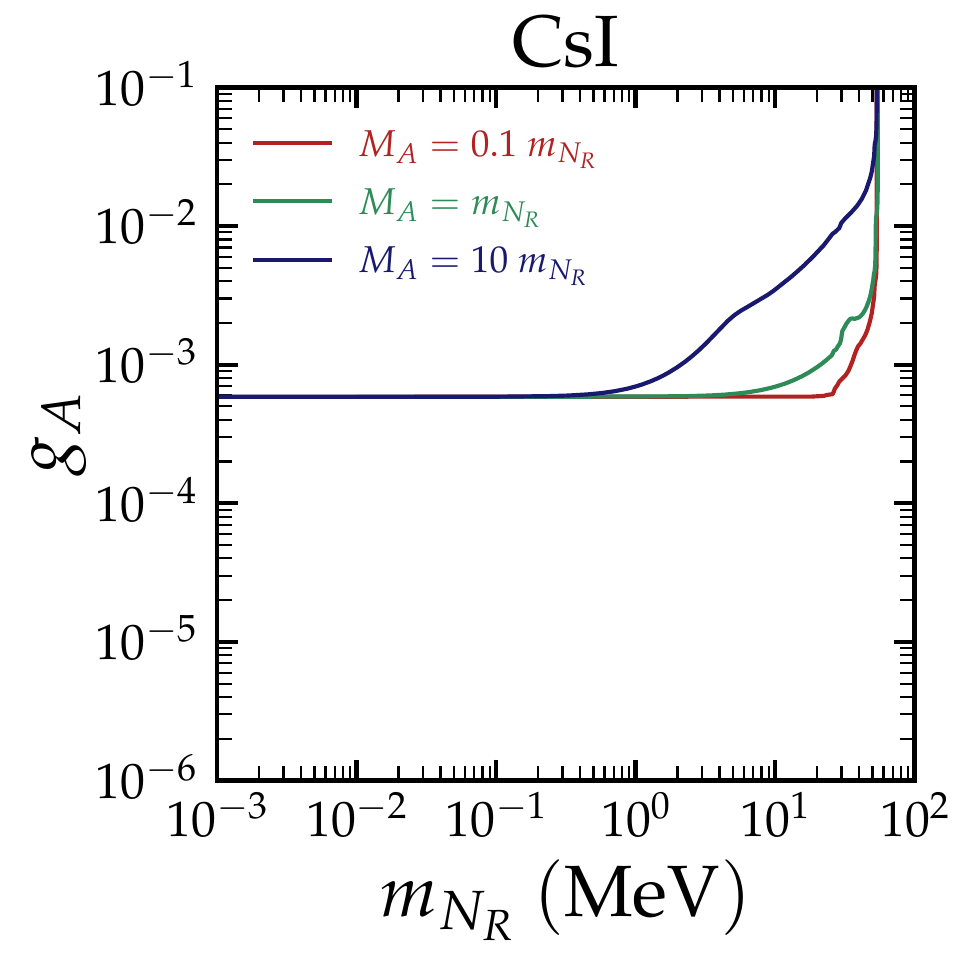}
            \includegraphics[width=0.45\linewidth]{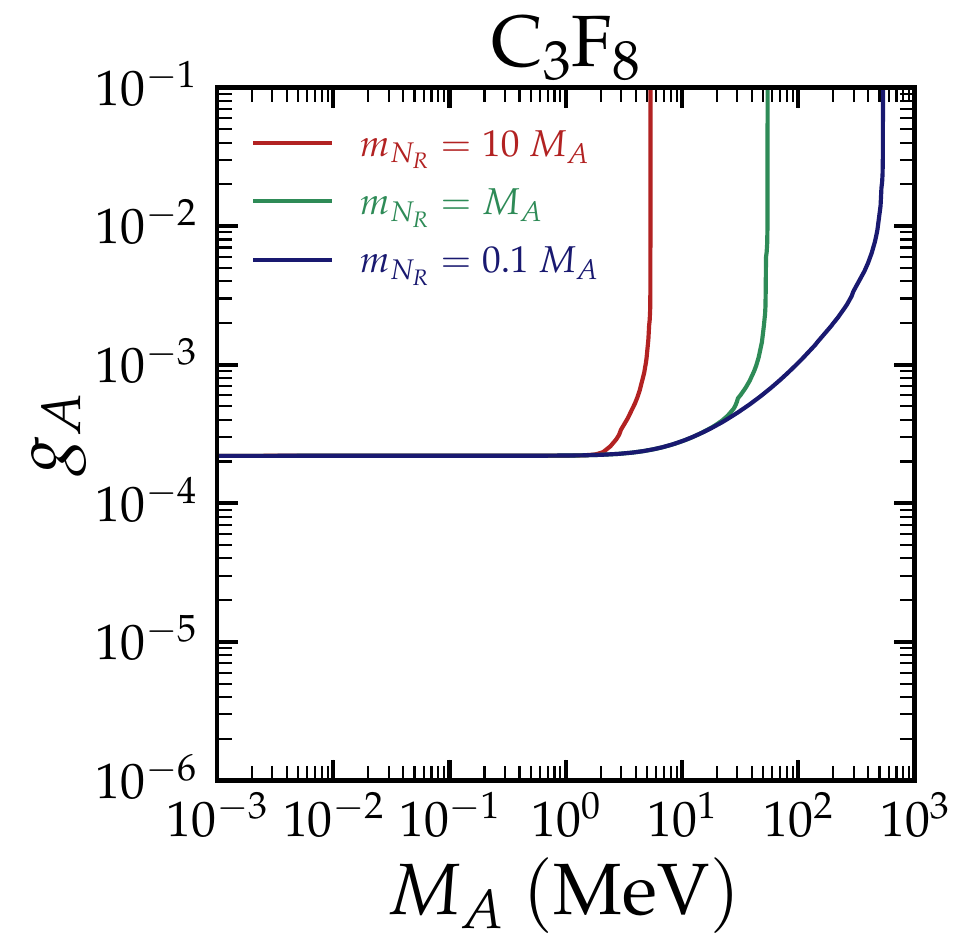}
            \includegraphics[width=0.45\linewidth]{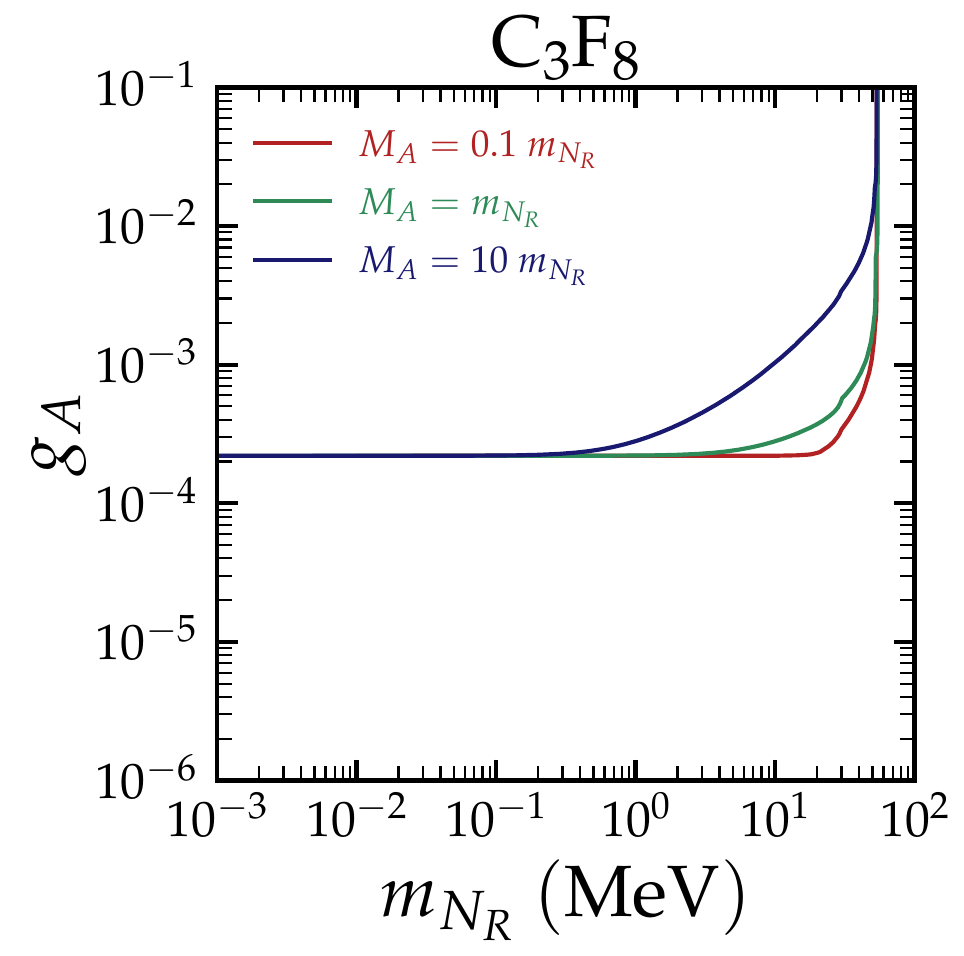}
    \caption{Projected 90\% C.L. limits on the axial vector-mediated scenario for SNL  production via upscattering, for the three considered  benchmark scenarios: $m_{N_R} = \{0.1, 1, 10\} \times M_A$. The left panel shows the projections in $(M_A, g_A)$ plane, while the right panel  shows the projections in the $(m_{N_R}, g_A)$ plane.}
    \label{fig:Upscattering_Axial_Vector_different_detector}
\end{figure}

\begin{figure}
    \centering
            \includegraphics[width=0.45\linewidth]{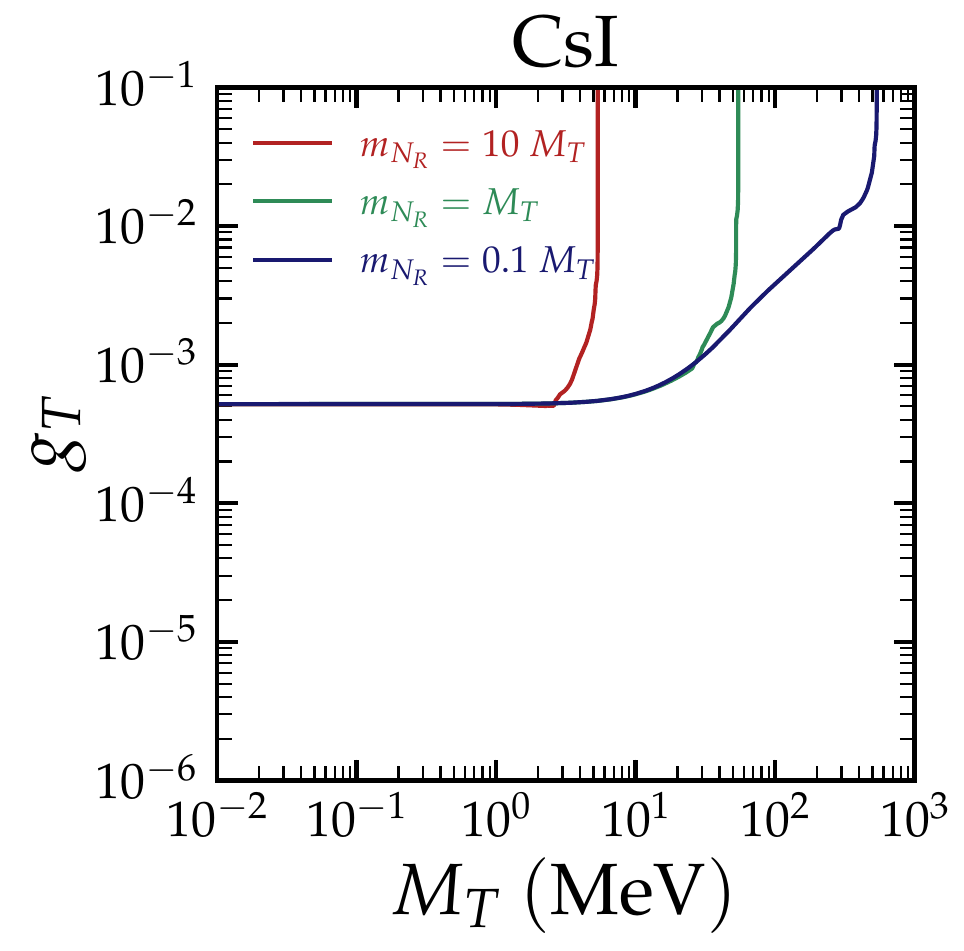}
            \includegraphics[width=0.45\linewidth]{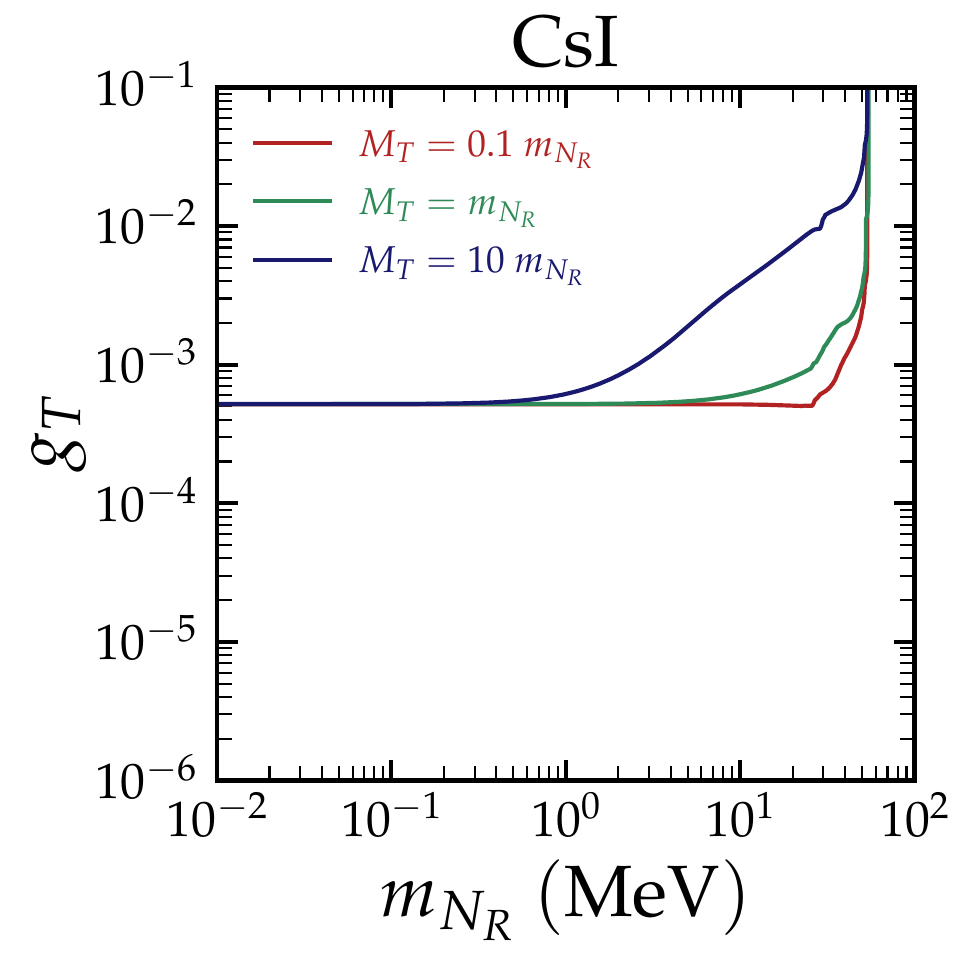}  
            \includegraphics[width=0.45\linewidth]{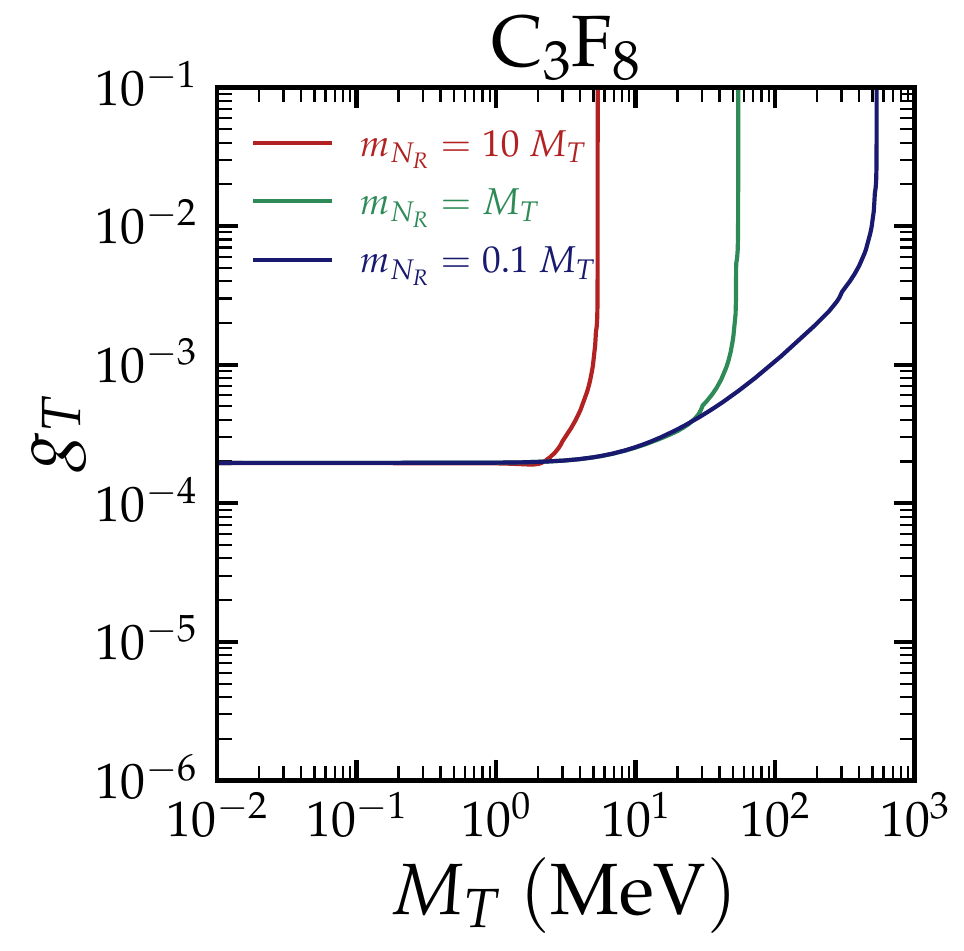}
            \includegraphics[width=0.45\linewidth]{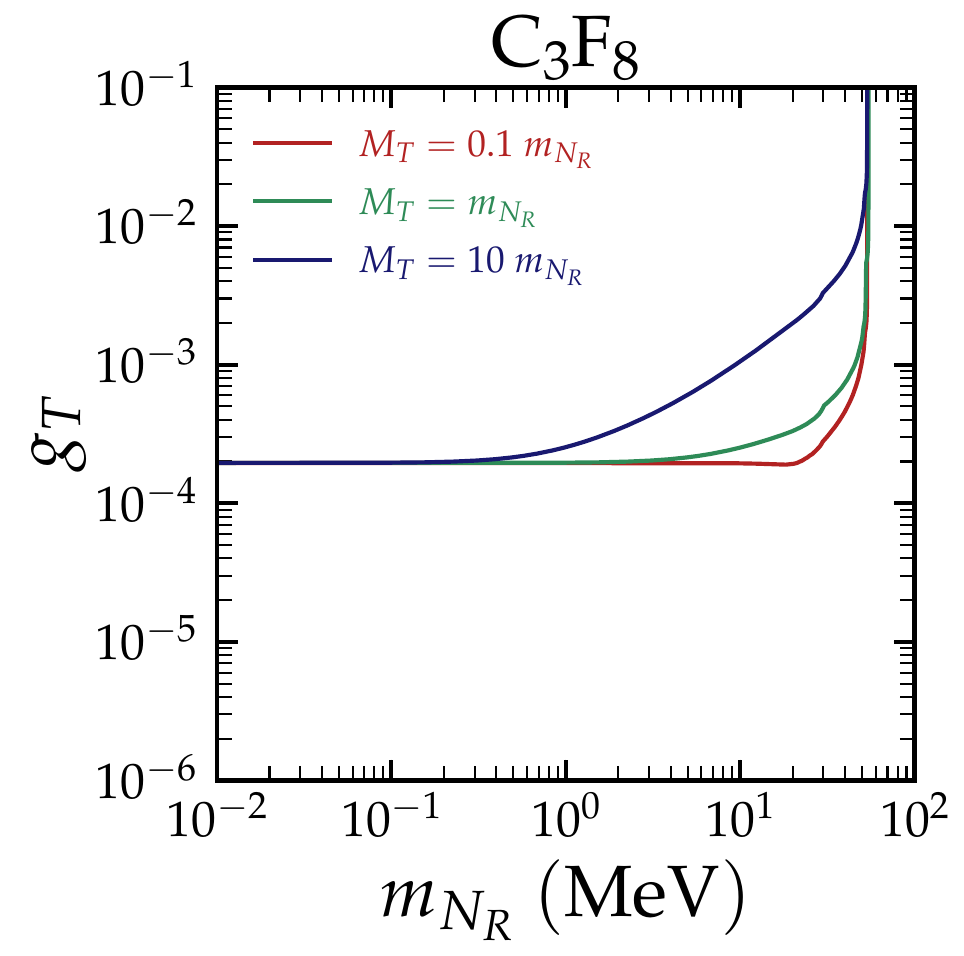}
    \caption{Same as Fig.~\ref{fig:Upscattering_Axial_Vector_different_detector}, but for tensor-mediated upscattering.}
    \label{fig:Upscattering_Tensor_different_detector}
\end{figure}

Finally, we present the 90\% C.L. projected limits regarding the  production of sterile neutral leptons via neutrino upscattering in the presence of NGIs. Again, the results are obtained from the individual analysis of the various proposed ESS detectors. The upper two rows of Fig.~\ref{fig:Upscattering_Scalar_different_detector_1} and Fig.~\ref{fig:Upscattering_Vector_different_detector_1}, illustrate the limits for the scalar and vector-mediated interaction by projecting on the mediator mass. The lower two rows of Fig.~\ref{fig:Upscattering_Scalar_different_detector_1} and Fig.~\ref{fig:Upscattering_Vector_different_detector_1}, depict the corresponding projections on $m_{N_R}$. The spin-dependent axial vector and tensor interactions are shown in Fig.~\ref{fig:Upscattering_Axial_Vector_different_detector} and Fig.~\ref{fig:Upscattering_Tensor_different_detector}, respectively. The left (right) panels show the projections on the mediator (SNL) mass. In each case the limits are demonstrated for three benchmark scenarios: $m_{N_R} = \{0.1, 1, 10\} \times M_X$. Concerning the spin-independent scalar and vector interactions, the CsI, Xe, Ge and Si detectors perform equally well, while the Ar and $\mathrm{C_3F_8}$ ones provide less stringent constraints. As explained previously, this is because a single-bin analysis is performed for these detectors. Regarding the spin-dependent axial vector and tensor interactions, only CsI and $\mathrm{C_3F_8}$ detectors are relevant since the nuclear ground state of Cs, I and F isotopes is different from $0^+$. Among these detectors, the lighter $\mathrm{C_3F_8}$ nuclear target performs significantly better (see the discussion on the NGI in the main text).
%%%%%%%%%%%%%%%%%%%%%%%%%%%%%%%%%%%%%%%%%%%%%%%%%%%%%%

\FloatBarrier

\section{Impact of  detector-specific inputs on the attainable sensitivities}
\label{Appendix_2}

In this Appendix we examine the impact of the various  factors affecting our presented sensitivities. To this end, we perform further analyses by varying the nuclear recoil energy thresholds and background uncertainties, assuming different target nuclei. As representative cases, we focus on the CsI and Si detectors, which serve as examples of heavy and light targets, respectively. Three distinct physics scenarios are considered: (a) The weak mixing angle determination; (b) The neutrino magnetic dipole portal, where the cross-section scales as $T_\mathcal{N}^{-1}$; (c) The light vector $B-L$ mediator case, where the cross-section scales as $T_\mathcal{N}^{-2}$.

For each of these cases, we compute the projected sensitivity curves for the CsI and Si detectors by varying the nuclear recoil energy threshold ($T_\mathcal{N}^{\mathrm{th}}$) and the background uncertainty ($\sigma_\beta$). Specifically, we consider:
\begin{itemize}
    \item CsI: $T_\mathcal{N}^{\mathrm{th}} = 1$ keV$_{nr}$ (same as specified in Table~\ref{Tab:Detectors_Specs}) and 5 keV$_{nr}$, with $\sigma_\beta = 1\%$ (same as considered in this study) and 10\%.
    \item Si: $T_\mathcal{N}^{\mathrm{th}} = 0.16$ keV$_{nr}$ (same as specified in Table~\ref{Tab:Detectors_Specs}) and 5 keV$_{nr}$, with $\sigma_\beta = 1\%$ (same as considered in this study) and 10\%.
\end{itemize}

\begin{figure}[ht!]
    \centering
    \includegraphics[width=0.49\linewidth]{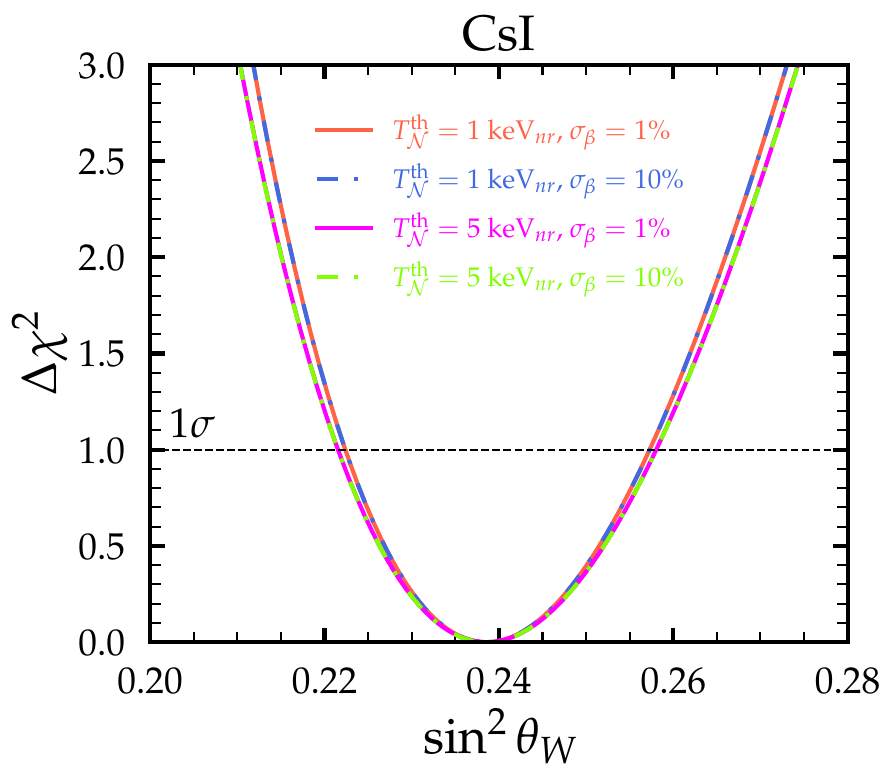}
    \includegraphics[width=0.49\linewidth]{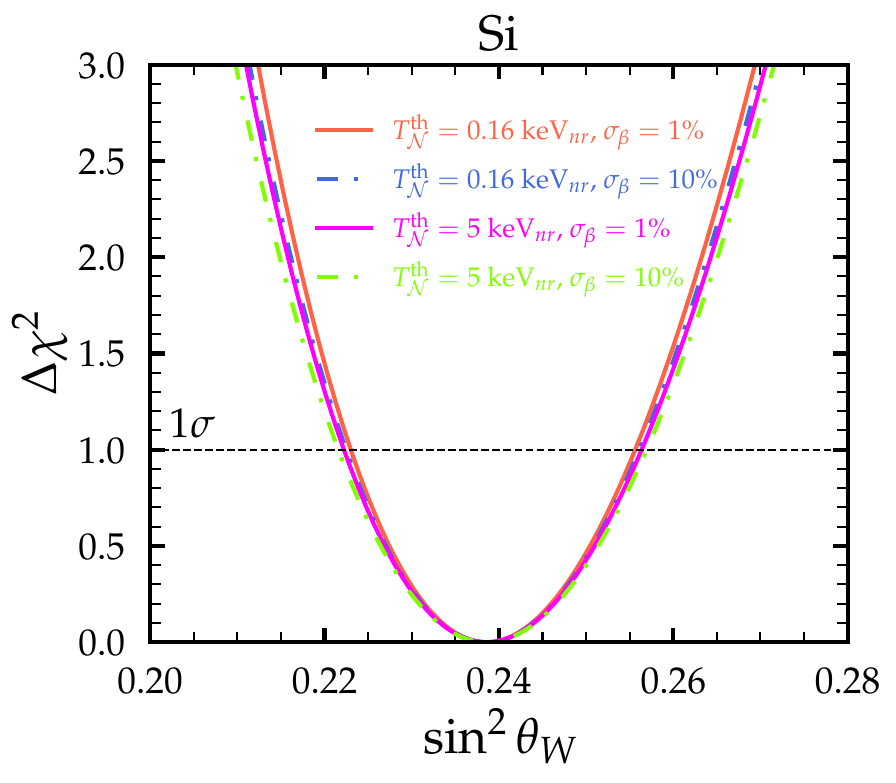}
    \caption{Projected sensitivity to the determination of weak mixing angle for CsI (left) and Si (right) detectors, for different recoil energy thresholds and background uncertainties.}
    \label{fig:Appendix_2_sw2}
\end{figure}

\begin{figure}
    \centering
    \includegraphics[width=0.49\linewidth]{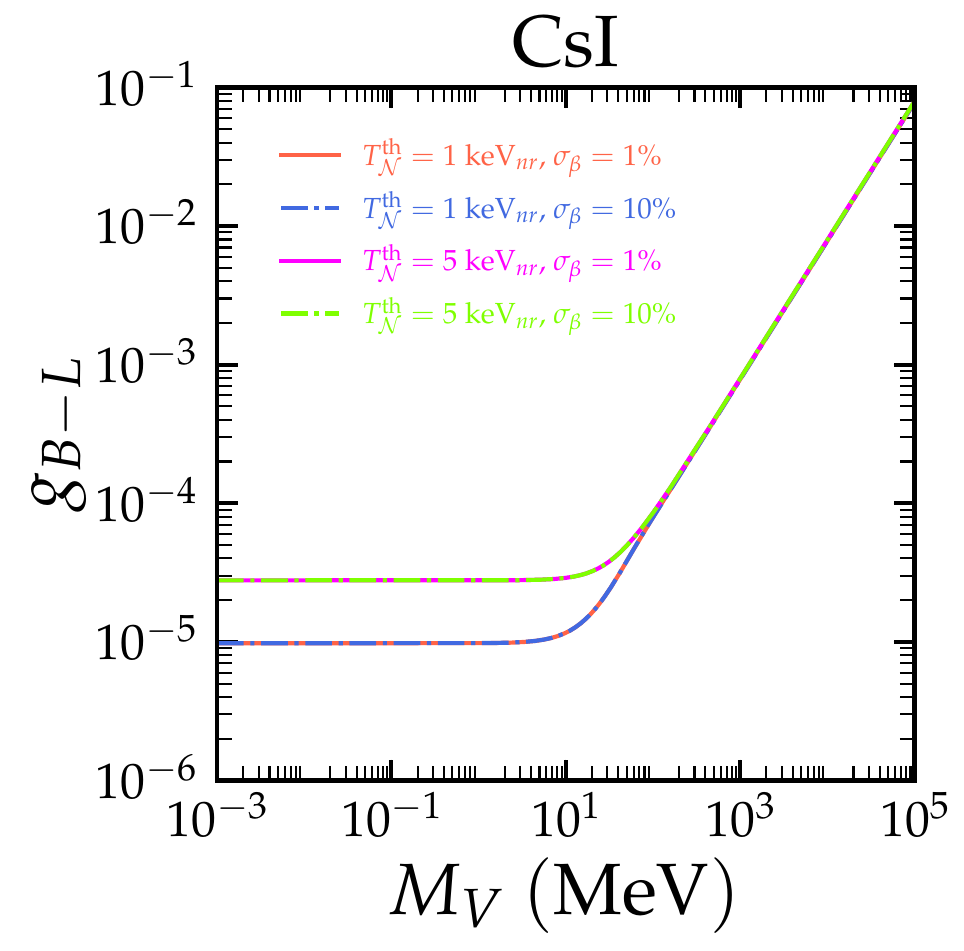}
    \includegraphics[width=0.49\linewidth]{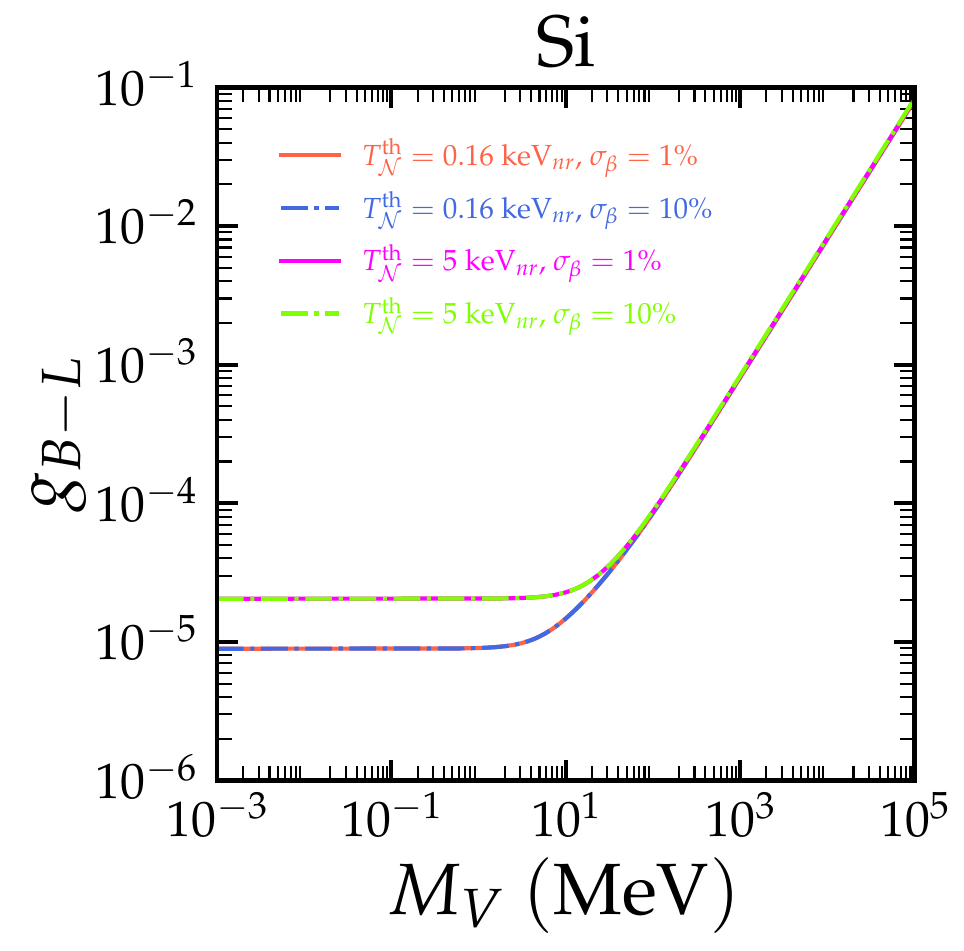}
    \caption{Projected sensitivity in the $(M_V, \textsl{g}_{B-L})$ plane for CsI (left) and Si (right) detectors, for different recoil energy thresholds and background uncertainties.}
    \label{fig:Appendix_2_B_minus_L}
\end{figure}

\begin{figure}
    \centering
    \includegraphics[width=0.49\linewidth]{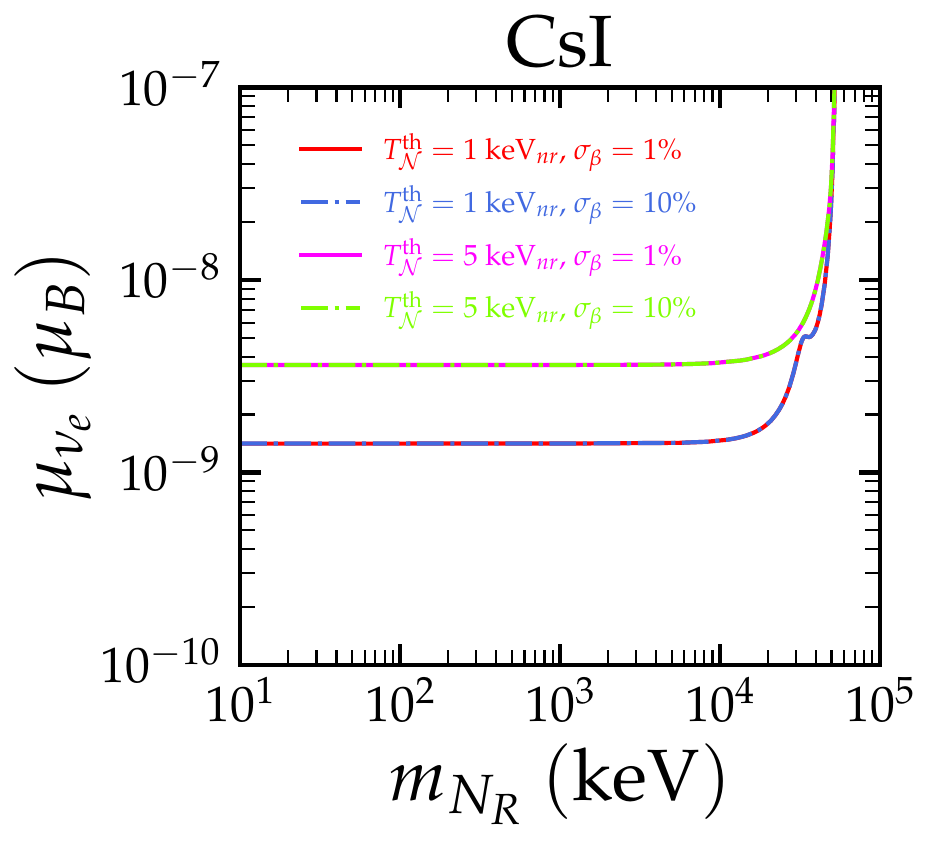}
    \includegraphics[width=0.49\linewidth]{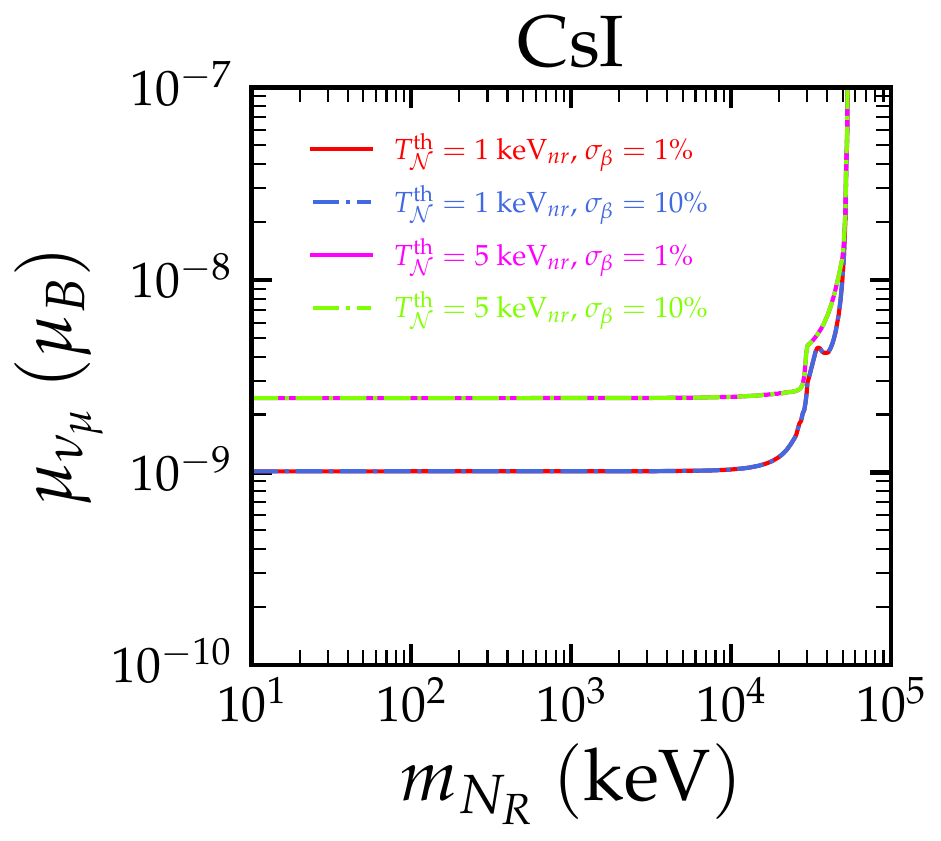}
    \includegraphics[width=0.49\linewidth]{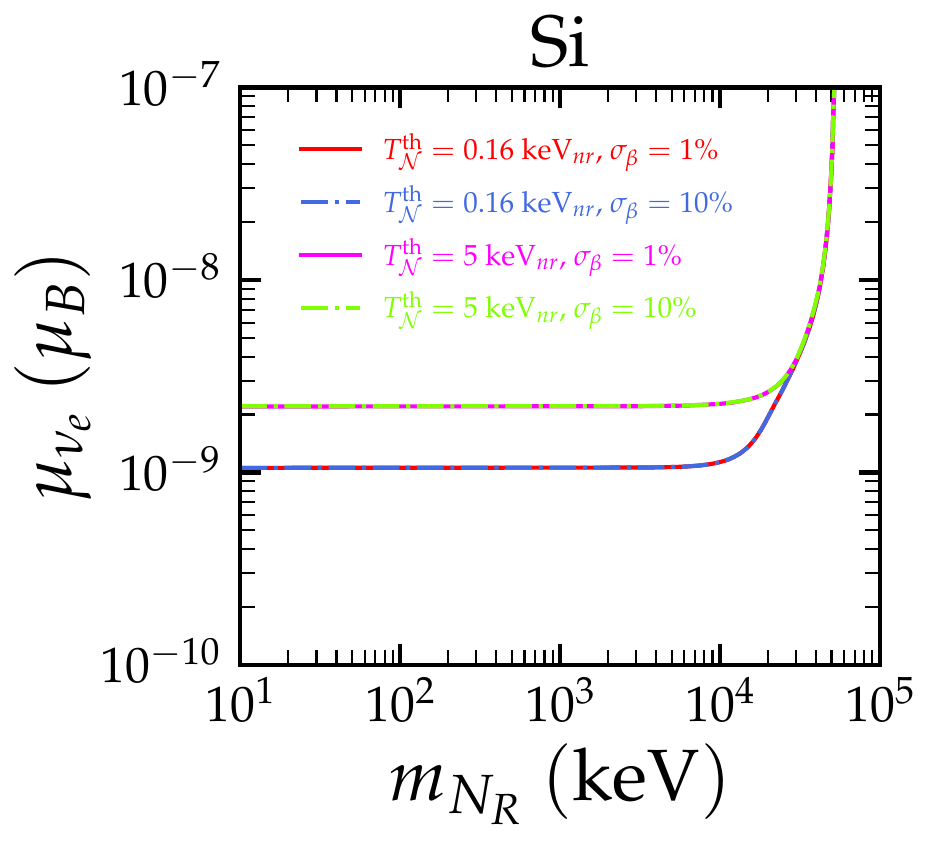}
    \includegraphics[width=0.49\linewidth]{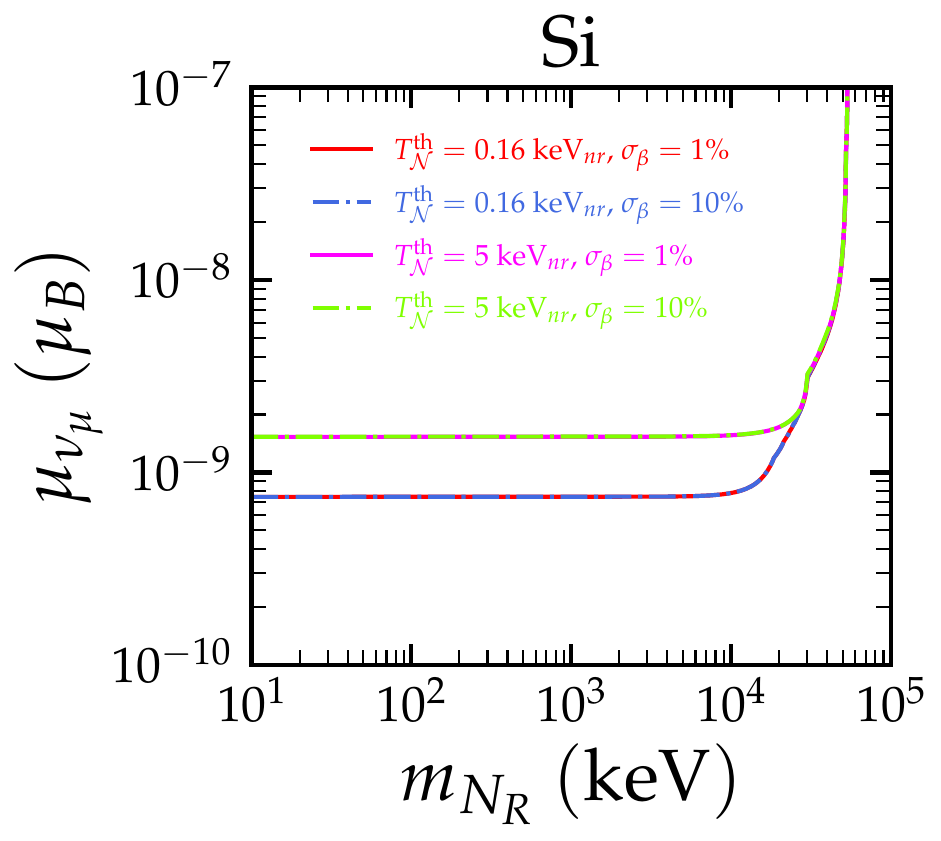}
    \caption{Projected sensitivity in the $(\mu_{\nu_e}, m_{N_R})$ and $(\mu_{\nu_\mu}, m_{N_R})$ planes for CsI (top panel) and Si (lower panel) detectors, for different recoil energy thresholds and background uncertainties.}
    \label{fig:Appendix_2_Dipole_Portal}
\end{figure}

The results of this study are illustrated in Figures \ref{fig:Appendix_2_sw2}, \ref{fig:Appendix_2_B_minus_L}, and \ref{fig:Appendix_2_Dipole_Portal}.

Evidently, the recoil energy threshold has the greatest impact  on the projected sensitivity for all the physics cases, with the case of the light vector $B-L$ mediator scenario being affected the most\footnote{This applies to all the light mediator cases $X=\{S,A,V,T\}$.}. By lowering the threshold the sensitivity gets enhanced mainly from the low recoil bin contributions to the event rates (see also Fig.~\ref{fig:Events_Spectra}), eventually leading to an improvement of a factor   $\sim 3$ for CsI when reducing the recoil threshold from 5 keV$_{nr}$ to 1 keV$_{nr}$, and a factor $\sim 2$ for Si when reducing the recoil threshold from 5 keV$_{nr}$ to 0.16 keV$_{nr}$. In contrast, the weak mixing angle determination remains largely insensitive to threshold variations. For the sterile dipole portal, we find a slightly less important improvement on the  sensitivity, e.g., a factor  $\sim 2$ improvement for CsI in the $\mu_{\nu_e}$ case and $\sim 2.5$ in the $\mu_{\nu_\mu}$ case when reducing the recoil threshold from 5 keV$_{nr}$ to 1 keV$_{nr}$. For Si, the sensitivity improves by a factor of $\sim 2$ for both $\mu_{\nu_e}$ and $\mu_{\nu_\mu}$ when reducing the recoil threshold from 5 keV$_{nr}$ to 0.16 keV$_{nr}$. The impact of varying the background uncertainty and the choice of target nucleus are found to be negligible across all scenarios, indicating that the results are largely driven by the nuclear recoil energy threshold. These findings reinforce the importance of achieving lower nuclear recoil energy thresholds in future \cevns experiments to maximize sensitivity to new physics scenarios.

\FloatBarrier
%%%%%%%%%%%%%%%%%%%%%%%%%%%%%%%%%%%%%%%%%%%%%%%%%%%%%%
\bibliographystyle{utphys}
\bibliography{bibliography}
\end{document}